\documentclass[final, a4paper, 10pt,twocolumn]{IEEEtran}

\usepackage{import}
\usepackage{url}
\usepackage[T1]{fontenc}
\usepackage[latin1]{inputenc}
\usepackage{textcomp}
\usepackage{bm}
\usepackage{graphicx}
\usepackage{cite}
\usepackage{amsmath,amssymb,amsfonts,mathtools}
\DeclareMathOperator{\diag}{diag}
\usepackage{balance}
\usepackage[table]{xcolor}

\usepackage{hyperref}
\hypersetup{colorlinks=true,linkcolor=black,urlcolor=black,}

\usepackage[acronym,style=super,nogroupskip,nonumberlist]{glossaries-extra}
\makeglossaries
\setabbreviationstyle[acronym]{long-short}
\newacronym{PLS}{PLS}{physical~layer~security}
\newacronym{RIS}{RIS}{reconfigurable~intelligent~surface}
\newacronym{IIoT}{IIoT}{industrial~internet~of~things}
\newacronym{LIS}{LIS}{large~intelligent~surface}
\newacronym{DQN}{DQN}{deep~Q-learning}
\newacronym{EM}{EM}{electromagnetic}
\newacronym{ARIS}{ARIS}{aerial~RIS}
\newacronym{DRL}{DRL}{deep~reinforcement~learning}
\newacronym{RL}{RL}{reinforcement~learning}
\newacronym{DDPG}{DDPG}{deep~deterministic~policy~gradient}
\newacronym{DPG}{DPG}{deterministic~policy~gradient}
\newacronym{QoS}{QoS}{quality~of~service}
\newacronym{MIMO}{MIMO}{multiple-input~multiple-output}
\newacronym{SIMO}{SIMO}{single-input~multiple-output}
\newacronym{SISO}{SISO}{single-input~single-output}
\newacronym{MISO}{MISO}{multiple-input~single-output}
\newacronym{RF}{RF}{radio~frequency}
\newacronym{ML}{ML}{machine~learning}
\newacronym{OWC}{OWC}{optical~wireless~communication}
\newacronym{KPIs}{KPIs}{key~performance~indicators}
\newacronym{5G}{5G}{fifth~generation}
\newacronym{6G}{6G}{sixth~generation}
\newacronym{EE}{EE}{energy~efficiency}
\newacronym{PNSC}{PNSC}{probability~of~non-zero~secrecy~capacity}
\newacronym{SCA}{SCA}{successive~convex~approximation}
\newacronym{BCD}{BCD}{block~coordinate~descent}
\newacronym{AO}{AO}{alternating~optimization}
\newacronym{LPWAN}{LPWAN}{low~power~wide~area~network}
\newacronym{SOP}{SOP}{secrecy~outage~probability}
\newacronym{SINR}{SINR}{signal-to-interference-plus-noise~ratio}
\newacronym{BS}{BS}{base~station}
\newacronym{Eav}{Eav}{eavesdropper}
\newacronym{mmWave}{mmWave}{millimeter~wave}
\newacronym{AN}{AN}{artificial~noise}
\newacronym{CSI}{CSI}{channel~state~information}
\newacronym{WPT}{WPT}{wireless~power~transfer}
\newacronym{WPC}{WPC}{wireless~power~communication}
\newacronym{SWIPT}{SWIPT}{simultaneous~wireless~information\quad~and~power~transfer}
\newacronym{MEC}{MEC}{mobile~edge~computing}
\newacronym{SC}{SC}{secrecy~capacity}
\newacronym{SR}{SR}{secrecy~rate}
\newacronym{UAV}{UAV}{unmanned~aerial~vehicle}
\newacronym{DoF}{DoF}{degree~of~freedom}
\newacronym{BER}{BER}{bit~error~ratio}
\newacronym{NOMA}{NOMA}{non-orthogonal~multiple~access}
\newacronym{LoS}{LoS}{line-of-sight}
\newacronym{NLoS}{NLoS}{non-line-of-sight}
\newacronym{THz}{THz}{terahertz}
\newacronym{SNR}{SNR}{signal-to-noise~ratio}
\newacronym{SDR}{SDR}{semidefinite~relaxation}
\newacronym{CCP}{CCP}{convex-concave~procedure}
\newacronym{IoE}{IoE}{internet-of-everything}
\newacronym{AR}{AR}{augmented~reality}
\newacronym{MR}{MR}{mixed~reality}
\newacronym{VR}{VR}{virtual~reality}
\newacronym{VLC}{VLC}{visible~light~communications}
\newacronym{MM}{MM}{majorization-minimization}
\newacronym{QCQP}{QCQP}{quadratically~constrained~quadratic~programming}
\newacronym{IoT}{IoT}{internet~of~things}
\newacronym{V2X}{V2X}{vehicle-to-everything}
\newacronym{V2V}{V2V}{vehicle-to-vehicle}
\newacronym{ASR}{ASR}{average~secrecy~rate}
\newacronym{ASC}{ASC}{average~secrecy~capacity}
\newacronym{SSR}{SSR}{sum~secrecy~rate}
\newacronym{SER}{SER}{symbol~error~rate}
\newacronym{ESC}{ESC}{ergodic~secrecy~capacity}
\newacronym{D2D}{D2D}{Device-to-Device}
\newacronym{RRS}{RRS}{reconfigurable~refractive~surface}
\newacronym{IOS}{IOS}{intelligent~omni-surface}
\newacronym{SOCP}{SOCP}{second~order~cone~programming}
\newacronym{SIC}{SIC}{successive~interference~cancellation}
\newacronym{2D}{2D}{two-dimensional}
\newacronym{DC}{DC}{direct-current}
\newacronym{WBAN}{WBAN}{wireless~body~area~network}
\newacronym{TDD}{TDD}{time~division~duplexing}
\newacronym{CRN}{CRN}{cognitive~radio~network}
\newacronym{AP}{AP}{access~point}
\newacronym{ISAC}{ISAC}{integrated~sensing~and~communications}
\newacronym{MSE}{MSE}{mean~square~error}
\newacronym{IM/DD}{IM/DD}{intensity~modulation~and~direct~detection}
\newacronym{BF}{BF}{beamforming}
\newacronym{LED}{LED}{light~emitting~diode}
\newacronym{SPSC}{SPSC}{strictly~positive~secrecy~capacity}
\newacronym{FSO}{FSO}{free~space~optical}
\newacronym{RGB}{RGB}{red-green-blue}
\newacronym{AI}{AI}{artificial~intelligence}
\newacronym{WPAN}{WPAN}{wireless~personal~area~network}
\newacronym{WSSCE}{WSSCE}{weighted~sum~secrecy~computation~efficiency}
\newacronym{AANET}{AANET}{aeronautical~Ad-hoc~networks}
\newacronym{DNN}{DNN}{deep~neural~network}
\newacronym{SDP}{SDP}{semidefinite~programming}
\newacronym{LC}{LC}{liquid~crystal}
\newacronym{COP}{COP}{connection~outage~probability}
\newacronym{ANT}{ANT}{average~number~of~transmission}
\newacronym{EST}{EST}{effective~secrecy~throughput}
\newacronym{IP}{IP}{intercept~probability}
\newacronym{PPSC}{PPSC}{probability~of~positive~secrecy~capacity}
\newacronym{HARQ}{HARQ}{hybrid~automatic~repeat~request}
\newacronym{IRIS}{RIS}{illegal~reconfigurable~intelligent~surface}
\newacronym{WSNs}{WSNs}{wireless~sensor~networks}
\newacronym{IoMT}{IoMT}{internet~of~medical~things}
\newacronym{CoMP}{CoMP}{coordinate~multi-point}
\newacronym{3D}{3D}{three-dimensional}
\newacronym{PN}{PN}{primary~network}
\newacronym{SN}{SN}{secondary~network}
\newacronym{BackCom}{BackCom}{backscatter~communication}
\newacronym{EH}{EH}{energy~harvesting}
\newacronym{STAR-RIS}{STAR-RIS}{simultaneous~transmitting~and~reflecting\qquad~reconfigurable~intelligent~surface}
\newacronym{SymR}{SymR}{symbiotic~radio}
\newacronym{MA}{MA}{movable~antenna}
\newacronym{FPA}{FPA}{fixed~position~antenna}
\newacronym{RSMA}{RSMA}{rate~splitting~multiple~access}
\newacronym{MWN}{MWN}{multiband~wireless~network}
\newacronym{FA}{FA}{fluid~antenna}
\newacronym{FSS}{FSS}{frequency~selective~surface}
\newacronym{B5G}{B5G}{beyond~fifth-generation}
\newacronym{FD}{FD}{full-duplex}
\newacronym{AF}{AF}{amplify-and-forward}
\newacronym{DF}{DF}{decode-and-forward}
\newacronym{BD-RIS}{BD-RIS}{beyond~diagonal~RIS}
\newacronym{BD-IRS}{BD-IRS}{beyond~diagonal~IRS}
\newacronym{IRS}{IRS}{intelligent~reflecting~surface}
\newacronym{SE}{SE}{spectral~efficiency}
\newacronym{mMIMO}{mMIMO}{massive~MIMO}
\newacronym{LS}{LS}{least~squares}
\newacronym{OFDM}{OFDM}{orthogonal~frequency~division~multiplexing}
\newacronym{DFBS}{DFBS}{dual-function~BS}
\newacronym{SCNR}{SCNR}{signal-to-clutter-plus-noise~ratio}
\newacronym{DFRC}{DFRC}{dual-function~radar-communication}
\newacronym{BC}{BC}{broadcast~channels}
\newacronym{SU}{SU}{single-user}
\newacronym{WMMSE}{WMMSE}{weighted~minimum~mean~square~error}
\newacronym{SAA}{SAA}{sample~average~approximation}
\newacronym{CW-DGC}{CW-DGC}{cell-wise~dynamically~group-connected}
\newacronym{TBF}{TBF}{transmission~beamforming}
\newacronym{MRC}{MRC}{maximal~ratio~combining}
\newacronym{SIM}{SIM}{stacked~intelligent~metasurface}
\newacronym{i.i.d.}{i.i.d.}{independent~and~identically~distributed}
\newacronym{IR}{IR}{information~receiver}
\newacronym{ER}{ER}{energy~receiver}
\newacronym{RHS}{RHS}{right-hand~side}
\newacronym{LHS}{LHS}{left-hand~side}
\newacronym{OIRS}{OIRS}{optical~intelligent~reflecting~surface}
\newacronym{OBD-RIS}{OBD-RIS}{optical~BD-RIS}
\newacronym{QC}{QC}{quantum~computing}
\newacronym{SL}{SL}{supervised~learning}
\newacronym{UL}{UL}{unsupervised~learning}
\newacronym{FL}{FL}{federated~learning}
\newacronym{ITU-R}{ITU-R}{international~telecommunication~union~radio~communication}
\newacronym{IMT}{IMT}{international~mobile~telecommunications}
\newacronym{ISG}{ISG}{industry~specification~group}
\newacronym{ETSI}{ETSI}{European~telecommunications~standards\qquad~institute}
\newacronym{D-RIS}{D-RIS}{diagonal~RIS}
\newacronym{Tx}{Tx}{transmitter}
\newacronym{Rx}{Rx}{receiver}
\newacronym{ZF}{ZF}{zero-forcing}
\newacronym{TS}{TS}{time-switching}
\newacronym{SEP}{SEP}{symbol~error~probability}
\newacronym{CF-mMIMO}{CF-mMIMO}{cell-free~massive~MIMO}
\newacronym{URLLC}{URLLC}{ultra-reliable~low-latency~communication}
\newacronym{UE}{UE}{user~equipment}
\newacronym{OFDMA}{OFDMA}{orthogonal~frequency~division~multiple~access}
\newacronym{SDMA}{SDMA}{space~division~multiple~access}
\newacronym{HAPS}{HAPS}{high~altitude~platform~station}
\newacronym{CRB}{CRB}{Cram$\acute{\textnormal{e}}$r-Rao~bound}
\newacronym{OMA}{OMA}{orthogonal~multiple~access}
\newacronym{LEO}{LEO}{low~earth~orbit}
\newacronym{CNN}{CNN}{convolutional~neural~network}
\newacronym{NTN}{NTN}{non-terrestrial~networks}
\newacronym{SAGIN}{SAGIN}{space-air-ground~integrated~networks}

\glssetcategoryattribute{acronym}{glossdesc}{title}
\glsdisablehyper

\usepackage{enumitem}
\setlist{nosep}

\usepackage{multirow}
\usepackage[export]{adjustbox}
\usepackage{arydshln} 
\usepackage{pifont}
\newcommand{\amark}{\ding{212}} 

\begin{document}

\bstctlcite{IEEEexample:BSTcontrol} 

\title{Beyond Diagonal RIS-Aided Wireless Communications Systems: State-of-the-Art and Future Research Directions}

\author{~Omar~Maraqa,~{Majid~H.~Khoshafa},~\IEEEmembership{Senior~Member,~IEEE}, Olutayo~O.~Oyerinde,~\IEEEmembership{Senior Member,~IEEE}, and Telex~M.~N.~Ngatched,~\IEEEmembership{Senior Member,~IEEE}

\thanks{O. Maraqa, M. H. Khoshafa, and T. M. N. Ngatched are with the Department of Electrical and Computer Engineering, McMaster University, Hamilton, ON L8S 4L8, Canada (e-mail:\{maraqao@mcmaster.ca, khoshafm@mcmaster.ca, and ngatchet@mcmaster.ca\}).

O. O. Oyerinde is with the School of Electrical and Information Engineering, University of the Witwatersrand, Johannesburg, 2020, South Africa (e-mail: Olutayo.Oyerinde@wits.ac.za).

This work has been submitted to the IEEE for possible publication. Copyright may be transferred without notice, after which this version may no longer be accessible.
} }

\markboth{}%
{Shell \MakeLowercase{\textit{et al.}}: Bare Demo of IEEEtran.cls for IEEE Journals}

\maketitle
\begin{abstract}
Integrating beyond diagonal reconfigurable intelligent surface (BD-RIS) into wireless communications systems has attracted significant interest due to its transformative potential in enhancing system performance. This survey provides a comprehensive analysis of BD-RIS technology, examining its modeling, structural characteristics, and network integration while highlighting its advantages over traditional diagonal RIS (D-RIS). Specifically, we review various BD-RIS modeling approaches, including multiport network theory, graph theory, and matrix theory, and emphasize their application in diverse wireless scenarios. The survey also covers BD-RIS's structural diversity, including different scattering matrix types, transmission modes, intercell architectures, and circuit topologies, showing their flexibility in improving network performance. We delve into the potential applications of BD-RIS, such as enhancing wireless coverage, improving physical layer security (PLS), enabling multi-cell interference cancellation, improving precise sensing and localization, and optimizing channel manipulation. Further, we explore BD-RIS architectural development, providing insights into new configurations focusing on channel estimation, optimization, performance analysis, and circuit complexity perspectives. Additionally, we investigate the integration of BD-RIS with emerging wireless technologies, such as millimeter-wave and terahertz communications, integrated sensing and communications, mobile edge computing, and other cutting-edge technologies. These integrations are pivotal in advancing the capabilities and efficiency of future wireless networks. Finally, the survey identifies key challenges, including channel state information estimation, interference modeling, and phase-shift designs, and outlines future research directions. The survey aims to provide valuable insights into BD-RIS's potential in shaping the future of wireless communications systems.
\end{abstract}


\tableofcontents

{\footnotesize{\printglossary[type=\acronymtype,title={List of Abbreviations}]}}

\section{Introduction}\label{sec:intro}

\IEEEPARstart{I}{n} recent years massive research efforts have been directed towards \gls{RIS} technology in both academia and industry. These efforts were motivated by the fact that the \gls{RIS} is considered a revolutionary technology for \gls{B5G} and \gls{6G} wireless communications networks due to its ability to build controllable radio environments and enhance the quality of communication in a cost-effective way~\cite{8766143,8869705,10054381}. Metamaterials are commonly used to build the \glspl{RIS}~\cite{9140329} as a reflective array that is made up of a large number of passive elements. Each of these elements can reflect the incident signals passively using a controllable phase shift~\cite{8811733}. The passive nature of the metamaterials used to build \glspl{RIS} that, in turn, leads to the requirement of low power consumption of the device together with its cost-effective architecture lends credence to why \glspl{RIS} have attracted extensive attention. To be specific, some of the advantages of \glspl{RIS} include~\cite{9424177}: (i) Simplicity of usage: Because \glspl{RIS} are electromagnetic-based, virtually passive devices, and inexpensive, they can be used on a variety of structures, including building facades, interior walls, aerial platforms, roadside billboards, highway polls, car windows, and pedestrian clothing~\cite{9148781}. (ii)  Enhanced spectral efficiency and throughput: By making up for power loss over extended distances, \glspl{RIS} can change the wireless propagation environment. Also, by passively reflecting the incoming radio signals with the aid of \glspl{RIS}, \glspl{BS} and mobile users can establish virtual \gls{LoS} connections. When barriers, like high-rise buildings, obstruct the \gls{LoS} link between \glspl{BS} and users, the throughput enhancement becomes considerable. (iii) High compatibility with existing networks' hardware: Because \glspl{RIS} only reflect electromagnetic waves, they are compatible with many existing communications systems. Furthermore, existing wireless networks' standards are compatible with \gls{RIS}-assisted wireless communications networks~\cite{9079457}. (iv) Environmentally and energy friendly: \glspl{RIS} can shape the incoming signal by adjusting the phase shift of each reflecting element in contrast to the conventional relaying systems like \gls{AF} and \gls{DF}~\cite{9136592}, which control the power amplifier. As a result, using \glspl{RIS} instead of traditional relaying systems is more environment- and energy-friendly~\cite{10598369}.

Generally, in terms of mode of operation, \glspl{RIS} can be divided into three main categories. These include reflective mode, transmissive mode, and hybrid mode~\cite{9122596}. In terms of its architectural design, \glspl{RIS} employ single-connected architecture. In single-connected architecture, each port of the \glspl{RIS} is directly connected to a load and then to the ground, while there are no interconnections between ports. This absence of connections between the \gls{RIS}'s ports results in a diagonal scattering matrix which places constraints on \glspl{RIS}' signal manipulation flexibility, and \gls{BF} capabilities, as well as introduces some potential vulnerabilities, as elaborated in Section~\ref{subsection: Diagonal-RIS (D-RIS)}. It is worth noting that because traditional \glspl{RIS} have diagonal scattering matrices, sometimes they are called \glspl{D-RIS}~\cite{khan2024integration}.

To overcome the above constraints, \glspl{D-RIS} has been scaled up to a more robust version called \glspl{BD-RIS}. There are various proposed \gls{BD-RIS} architectures that feature interconnections between all or a subset of the \gls{BD-RIS} ports. As such the scattering matrix of these architectures is not limited to being diagonal. For example, the scattering matrix becomes a full matrix in the case where each of the ports of the \glspl{BD-RIS} is connected to all the other ports of the \glspl{BD-RIS} (i.e., fully-connected \gls{BD-RIS}), and satisfies the constraints of being unitary and symmetry. The unitary constraint is due to the lossless property of the system, which means that there is energy conservation because there is no loss of energy when the signal waves interact with the scattering surface. The symmetry constraint is due to the possibility of using a reciprocal impedance network to simplify the \glspl{BD-RIS} circuit. As such, based on the scattering matrix being a full matrix, there is an advantage of not only being able to control the phase of each of the \glspl{BD-RIS}' elements, but also the amplitude, polarization, and direction of the impinging signal waves. Different structures of the \glspl{BD-RIS} have been put forward to operate in different transmission modes, such as reflective, transmissive, hybrid, and multi-sector modes. Also, to operate in different circuit topologies, such as fully-/group-/tree-/forest-connected. All of the aforementioned transmission modes and circuit topologies are discussed in Section~\ref{sec: Classification of BD-RIS}.

Consequent to the advantages associated with deploying \glspl{D-RIS} in wireless communications networks, several reviews on \glspl{D-RIS} were documented in the literature, for example,~\cite{9140329,8796365,9122596,9424177,10584518}. As \glspl{BD-RIS} can be considered as a generalization of \glspl{D-RIS}, there is a merit in dedicating a survey to advocate their potential. As far as the authors are aware, there is no existing comprehensive survey conducted about \glspl{BD-RIS}. Therefore, this survey addresses the knowledge gap by providing a comprehensive overview of BD-RIS. Specifically, it highlights BD-RIS modeling, advantages, scattering matrix types, transmission modes, inter-cell architectures, circuit topologies, potential applications, architectural developments, performance evaluation, integration with emerging wireless technologies, challenges, and promising future research directions by emphasizing on the
following aspects:
\begin{itemize}
    \item \textit{Classification of \gls{BD-RIS}}: We first provide a comprehensive study on \gls{BD-RIS}, covering key aspects of their modeling, structural characteristics, and network integration. We review various \gls{BD-RIS} modeling approaches based on multiport network theory, graph theory, and matrix theory, highlighting their applicability to different use cases. The enhanced capabilities of \gls{BD-RIS} over \gls{D-RIS} are discussed, emphasizing their potential to enable more robust and adaptable wireless networks. Furthermore, we explore the structural diversity of \gls{BD-RIS}, including different scattering matrix types (block diagonal, permuted block diagonal, and non-diagonal), transmission modes (reflective, transmissive, hybrid, and multi-sector), inter-cell architectures (single-, fully-, group-, and dynamically group-connected), and circuit topologies (single-, fully-, group-, tree-, and forest-connected). By highlighting these aspects, we provide a thorough insight into \gls{BD-RIS} and its transformative potential in future wireless communications networks.
    \item \textit{BD-RIS Potential Applications}: We provide an in-depth investigation of the diverse potential applications of \gls{BD-RIS} structures in next-generation wireless communications networks. Specifically, we explore how \gls{BD-RIS} can (i) extend wireless coverage by intelligently reflecting and refracting signals to mitigate blockages and enhance connectivity in complex environments, (ii) improve \gls{PLS} by enabling precise control over signal propagation to suppress eavesdropping threats and improve secure transmission, (iii) facilitate multi-cell interference cancellation in dense networks by dynamically reconfiguring the wireless environment to suppress inter-cell interference and enhance spectral efficiency, (iv) improve precise sensing and localization in \gls{ISAC} systems by leveraging advanced wave manipulation to achieve high-resolution target detection and positioning, (v) enable advanced channel manipulation capabilities, allowing for customized signal shaping and adaptive \gls{BF} to optimize communication performance. These applications emphasize the pivotal role of \gls{BD-RIS} in enhancing network efficiency, security, and reliability in future wireless systems.
    \item \textit{\gls{BD-RIS} Architectural Development}: First, we present the details of some new \gls{BD-RIS} architectures, such as (i) the opposite link with horizontally-connected sectors \gls{BD-RIS}, (ii) the full-link with horizontally-connected sectors \gls{BD-RIS}, (iii) the opposite-link with horizontally-connected sectors and vertically connected elements in each sector \gls{BD-RIS}, (iv) full-link with horizontally-connected sectors and vertically connected elements in each sector \gls{BD-RIS}, (v) coordinated \gls{BD-RIS}, (vi) group-connected \gls{BD-RIS} with dynamic and static grouping strategies, (vii) distributed \gls{BD-RIS}, (viii) non-reciprocal \gls{BD-RIS} and (ix) Dual-polarized \gls{BD-RIS}. Second, we review the research efforts that examined the \gls{BD-RIS} structure in practical systems, which include the investigation efforts of \gls{BD-RIS} under mutual coupling, lossy impedance interconnections between \gls{BD-RIS} elements, and with discrete-value scattering matrix. Third, we provide a discussion about the channel estimation solutions that can facilitate the operation for \gls{BD-RIS}-aided systems. Furthermore, We shed some light on the research efforts that evaluated the performance of \gls{BD-RIS}-aided systems from (i) an optimization perspective, (ii) a performance analysis perspective, and (iii) a circuit complexity perspective.
    \item \textit{\gls{BD-RIS} Integration with Emerging Wireless Technologies}: We provide a detailed investigation of how \gls{BD-RIS} can be integrated with diverse emerging wireless technologies, highlighting its potential across various domains. Specifically, we discuss its integration with: (i) \gls{mmWave}/\gls{THz} communications, where \gls{BD-RIS} enhances the performance of high-frequency bands by improving signal propagation; (ii) \gls{ISAC}, enabling the simultaneous use of \gls{BD-RIS} for both communications and precise environmental sensing; (iii) \gls{MEC}, where \gls{BD-RIS} facilitates the offloading of computational tasks to edge servers; (iv) radar communications, where \gls{BD-RIS} can be leveraged to enhance radar signal processing, improving target detection and localization while minimizing interference; (v) \gls{SWIPT} systems, where \gls{BD-RIS} aids in optimizing power transmission and communication, improving energy harvesting capabilities and ensuring efficient use of radio resources; (vi) \gls{URLLC}, where \gls{BD-RIS} contributes to achieving high reliability and low latency by enabling dynamic signal control; (vii) \gls{RSMA} and \gls{NOMA} schemes, where \gls{BD-RIS} enables more efficient spectrum utilization and improves user fairness by facilitating advanced power control and \gls{BF} strategies; (viii) wideband systems, (ix) large-scale systems, and (x) vehicular networks. Through its integration with these cutting-edge technologies, \gls{BD-RIS} is positioned to play a crucial role in advancing the capabilities and efficiency of future wireless communications networks.
    \item \textit{\gls{BD-RIS} Challenges and Future Research Directions}: We highlight the potential of \gls{BD-RIS} in addressing contemporary communications challenges that are related to (i) \gls{CSI} estimation, (ii) joint consideration of non-ideal aspects, (iii) continuous, discrete, and quantized phase-shift designs, (iv) near-field vs. far-field propagation, and (v) modeling the inter-sector interference of multi-sector \gls{BD-RIS}-based systems. Furthermore, we propose several topics for future research consideration. These topics include the investigation of (i) active vs. passive \gls{BD-RIS} structures, (ii) The \gls{BD-RIS} integration with additional emerging wireless technologies, (iii) utilizing advanced signal processing tools to optimize \gls{BD-RIS}-based systems, (iv) deploying a transmissive \gls{BD-RIS} at the receiver, (v) prototyping/experimentation, and (vi) development of standards for the \gls{BD-RIS} technology.
\end{itemize}

The rest of this survey is organized as follows. Section~\ref{sec: Fundamentals of BD-RIS} introduces the fundamentals, architecture and design, modeling, and the main differences between \gls{D-RIS} and \gls{BD-RIS}. Section~\ref{sec: Classification of BD-RIS} presents a detailed classification of \gls{BD-RIS} in terms of its scattering matrix types, modes of transmission, and circuit design. Section~\ref{sec: Potential Areas of Applications of BD-RIS} discusses the potential areas of applications of \gls{BD-RIS}. Section~\ref{sec: BD-RIS-Based Wireless Communications} discusses the various technical contributions that investigated \gls{BD-RIS} from (i) the architectural development perspective, (ii) performance evaluation perspective, and (iii) its integration with the emerging technologies/schemes that are expected to meet the various requirements of \gls{B5G} and \gls{6G} networks. Section~\ref{sec: Challenges And Future Research Directions In BD-RIS} presents challenges and potential future research directions associated with \gls{BD-RIS}-aided systems. Finally, we conclude this survey paper in Section~\ref{sec: Conclusion}.

\section{Fundamentals of RIS} \label{sec: Fundamentals of BD-RIS}

\subsection{RIS Fundamentals}
\Glspl{RIS} are advanced electromagnetic platforms capable of dynamically manipulating signal transmission through reflection, refraction, or diffraction of radio waves. This capability enables \glspl{RIS} to enhance signal quality for intended receivers while minimizing interference for others~\cite{8910627, 8796365, 9140329}. Comprising numerous cost-effective components, \glspl{RIS} modify wireless channels by adjusting both the amplitude and phase of incoming signals, providing an energy-efficient alternative to traditional power-intensive methods such as power amplifiers~\cite{10268023}. As quasi-passive devices, \glspl{RIS} require minimal power to maintain their reconfigurability, with more recent models incorporating active elements in hybrid passive-active \gls{RIS} structures~\cite{10584518}. This technology presents an economically viable and sustainable solution for improving network coverage, data rates, energy usage, and spectral efficiency, particularly in challenging environments and \gls{NLoS} scenarios~\cite{9424177}. Passive \glspl{RIS}, constructed from electromagnetic materials, offer a cost-effective and readily deployable solution across various infrastructures~\cite{9424177,9475160}. Through passive signal reflection, \glspl{RIS} contribute to the reduction of long-distance power losses for both \glspl{BS} and mobile users, presenting an energy-efficient alternative to conventional relays that necessitate power amplifiers~\cite{9140329}. While passive \gls{RIS} facilitates full-duplex transmission through signal reflection, it is subject to multiplicative fading effects. To address this limitation, active \gls{RIS} was developed, incorporating amplifiers to enhance signal strength and convert multiplicative fading into additive gains, albeit at the cost of increased power consumption~\cite{9530403, 9998527}. Furthermore, hybrid passive-active \gls{RIS} designs integrate active elements to facilitate more efficient \gls{CSI} acquisition~\cite{9370097, 9053976}. In contrast to reflective \glspl{RIS} that only support communications on one side, \gls{STAR-RIS} technology enables simultaneous transmission and reflection, providing full $360^{\circ}$ coverage and expanding its range of applications~\cite{9690478, 9722826}. Given the evolving demands of \gls{6G} and future wireless networks, \gls{RIS} and advanced multiple access schemes such as \gls{RSMA}~\cite{khoshafa2025rsma} hold significant promise in enhancing spectral efficiency, ensuring user fairness, and improving interference management.

\subsection{Diagonal-RIS (D-RIS)}
\label{subsection: Diagonal-RIS (D-RIS)}

\Gls{D-RIS}, often simply called \gls{RIS}, represents the earliest design of the \gls{RIS} technology. In this configuration, the main function is to adjust the phase of incoming electromagnetic waves~\cite{9140329}, using passive elements that impose phase shifts on the signals. \Gls{D-RIS} constrains the reflection matrix to a diagonal configuration. The architecture is primarily single-connected, enabling each \gls{D-RIS} element to modify the phase shift of the incoming signal independently, without inter-element connections. The primary objective of \gls{D-RIS} is to achieve a balance between performance and system complexity. By focusing only on phase adjustment, \gls{D-RIS} substantially reduces computational and hardware requirements, making it practical for applications that necessitate low-latency processing and energy efficiency, such as \gls{mMIMO} systems, \gls{mmWave} communications, space-air-ground integrated networks, and prospective \gls{6G} wireless networks~\cite{khoshafa2025ris, 9122596}.
\subsubsection{D-RIS Architecture and Design}
The \gls{D-RIS} architecture is based on a simplified hardware design featuring a diagonal reflection matrix that simplifies control circuits and associated signal processing operations~\cite{8466374}. Each element within a \gls{D-RIS} structure incorporates a phase-shifting component capable of altering the phase of the incoming signal. Moreover, \gls{D-RIS} employs less control hardware, often implemented using adjustable passive components such as varactor diodes or micro-electromechanical systems~\cite{9122596}. From a design perspective, \gls{D-RIS} is particularly advantageous for large-scale implementations where maintaining low energy consumption and ensuring rapid adaptation to changing channel conditions are essential. \gls{D-RIS} also supports scalable deployments in distributed networks, enhancing wireless coverage in dense urban areas and challenging \gls{NLoS} scenarios~\cite{10663469}.

\subsubsection{D-RIS Modeling}
The functioning of a \gls{D-RIS} can be mathematically described by a reflection matrix, represented as $\mathbf{\Theta}_\textnormal{D-RIS} =\diag \big(e^{{j\theta}_1},e^{{j\theta}_2},...,e^{{j\theta}_M} \big)$, where $M$ denotes the number of reflective elements, and $\theta_m, \ \forall m \in [1, 2, ..., M]$, indicates the phase shift applied by the \(m\)-th element. In this diagonal configuration, each \(\theta_m\) is optimized to modify the phase of the incoming signal to improve the received signal at the intended destination~\cite{8910627}. The diagonal constraint simplifies the optimization problem, allowing for more efficient resource allocation and faster convergence in \gls{BF} algorithms. The received signal at a user terminal can be expressed as the product of the incident signal, the channel matrix, and the diagonal reflection matrix~\cite{9140329}. \Gls{D-RIS}-based optimization aims to determine the optimal set of phase shifts \(\theta_m\) that maximizes the received signal power or other performance metrics, such as the \gls{SNR} or data rate. \Gls{D-RIS} simplifies the process of channel estimation and reflection coefficient design, enabling more practical implementation in wireless systems. 
\subsubsection{D-RIS Limitations} \Gls{D-RIS} enhances signal transmission by reflecting signals with controlled phase shifts along diagonal elements of their reflection matrix. However, their functionality is limited by several factors, as summarized below.
\begin{itemize}
    \item \textit{Limited Flexibility in Signal Manipulation}: In a \gls{D-RIS}, each element is constrained to a diagonal reflection matrix, meaning it can only apply phase shifts to the incident signals along the diagonal. This limits the ability to comprehensively control the signal's amplitude, polarization, or direction~\cite{10771739}. As a result, \gls{D-RIS} struggles to achieve complex signal manipulations required for advanced wireless environments.
    \item \textit{Limited \gls{BF} Capabilities}: \Gls{D-RIS} offers fundamental \gls{BF} functionality, primarily reflecting signals along direct \gls{LoS} paths. This leads to limited beam steering flexibility and diminished performance in complex settings, such as urban or dense environments characterized by significant multipath and \gls{NLoS} components~\cite{10499196}. As \gls{D-RIS} operates using only diagonal elements, its degrees of freedom for controlling reflected signals are inherently constrained, reducing its ability to adapt to dynamic variations in user locations or environmental conditions.
    \item \textit{Potential Vulnerabilities}: Due to its constrained \gls{BF} capabilities, \gls{D-RIS} cannot fully exploit advanced physical layer security techniques~\cite{9941040}. Such limitation decreases its effectiveness in countering security threats like eavesdropping and jamming, particularly in space-air-ground networks and multi-user communications scenarios.
\end{itemize}

\subsection{Beyond diagonal RIS (BD-RIS)} \label{sec: Beyond diagonal RIS}
Wireless networks in the first five generations relied on advanced transmitter/receiver designs to handle the unpredictable wireless environment. Future networks (\gls{B5G} and \gls{6G}) are expected to go further by using \gls{RIS} to actively manipulate the wireless environment, creating what's known as a ``smart radio environment''~\cite{10316535,9140329}. Although \gls{D-RIS} features a simple design that enhances signal strength for intended users and reduces network energy consumption, its lack of inter-element coordination limits its \gls{BF} abilities, resulting in reduced performance in more complex communications environments~\cite{khan2024integration}. To bridge this gap and allow for more flexible signal manipulation capabilities, \gls{BD-RIS}, sometimes termed as \gls{RIS} $2.0$, was introduced in~\cite{9514409} to allow for great coordination and interaction among reconfiguring elements, though at the cost of increased circuit complexity. 

\subsubsection{BD-RIS Architecture and Design}
\label{subsubsec: BD-RIS Architecture and Design}
\Gls{D-RIS} features a simple, single-connected architecture where each element operates independently, making \gls{D-RIS} structure easy to control and implement. However, this simplicity limits its ability to manipulate the electromagnetic wavefront. In contrast, \gls{BD-RIS} introduces a more complex design with group-connected or fully-connected architectures, allowing for greater inter-element interaction. This enhanced connectivity enables more precise and dynamic control of electromagnetic waves, significantly improving signal manipulation and expanding \gls{RIS} technology's potential for advanced wireless communications applications. \Gls{BD-RIS} is conceptualized based on the analysis of scattering parameter networks~\cite{9514409}. Its classification is defined by its scattering matrix types, the supported transmission modes, and its inter-cell architecture and/or circuit-topology designs. All the aforementioned classification aspects are discussed in-detail in Section~\ref{sec: Classification of BD-RIS}.

\subsubsection{BD-RIS Modeling}
\label{subsubsec: BD-RIS Modeling}

Only a few works have delved into the \gls{BD-RIS} modeling to provide a convenient way of understanding the role that \gls{BD-RIS} can play in wireless networks. In this regard, three main modeling approaches were investigated in the literature, namely, (i) multiport network theory~\cite{10574199, 10694015, del2024physics}, (ii) graph theory~\cite{10453384}, and (iii) matrix theory~\cite{10499196}. The details of these approaches are elaborated as follows:
\begin{itemize}
    \item \textit{Multiport Network Theory}: Modeling the communications systems using multiport network theory is not new. More than a decade ago \gls{MIMO} systems were analyzed using this approach, for example in~\cite{5446312}. This approach involves analyzing \gls{BD-RIS} based on impedance ($Z$-parameters), admittance ($Y$-parameters), and scattering ($S$-parameters) parameters. The $Z$-parameters are effective for modeling mutual coupling and impedance mismatches, making them suitable for complex systems~\cite{10574199}. The $Y$-parameters, related to the $Z$-parameters through matrix inversion, are advantageous for modeling systems with sparse interconnections, i.e., the \gls{BD-RIS} elements are interconnected by a limited number of adjustable admittance components, such as the modeling of \gls{BD-RIS}-enabled wideband systems~\cite{10694393, 10857964} and tree-/forest-connected \gls{BD-RIS} architectures~\cite{10453384}. The $S$-parameters excel in high-frequency \gls{RF} applications, as they efficiently model wave reflections and transmissions, and are particularly useful for \glspl{RIS} with no mutual coupling and perfect matching. In~\cite{10574199}, a universal framework was introduced to unite the analysis of scattering, impedance, and admittance parameters for modeling \gls{BD-RIS}-aided communications systems. Additionally, detailed discussions were offered therein on the most preferable parameter analyses for various use cases.
    \item \textit{Graph Theory}: In this approach, the \gls{BD-RIS} is represented as a graph where the vertices correspond to \gls{BD-RIS} ports and the edges denote tunable impedance connections between the ports. This graph theoretical approach helps in understanding the system's complexity and performance trade-offs. Specifically, graph theory helps characterize various architectures (e.g., fully-connected, group-connected) and provides insights into how different topologies can achieve the performance upper bound with minimized circuit complexity.
    \item \textit{Matrix theory}: In this approach, \gls{BD-RIS} is modeled through the decomposition of the phase-shift matrix into two components: a power divider (energy allocator) system represented by a real unitary matrix, and a phase-shift system represented by a diagonal matrix. This method, termed real-unitary decomposition, simplifies the representation of complex unitary matrices, facilitating the design and analysis of \gls{BD-RIS} in systems such as \gls{SISO} and \gls{MISO}. The Takagi factorization~\cite{xu2008divide} plays a central role in this decomposition process, ensuring that the matrix is expressed as a product of simpler matrices that still satisfy the necessary constraints for optimal communication performance. This decomposition aids in reducing computational complexity while maintaining system efficiency, especially in large-scale deployments.
\end{itemize} 

\subsubsection{BD-RIS Merits} 
\label{subsubsec: BD-RIS Merits}

Compared to \gls{D-RIS}, \Gls{BD-RIS} provides superior \gls{BF}, improved signal control, greater flexibility, higher spectrum efficiency, better interference management, and increased resilience to environmental changes. These improved features make \gls{BD-RIS} a powerful and adaptable technology for enhancing future wireless communications networks, especially for \gls{B5G} and \gls{6G} networks. The key benefits of \gls{BD-RIS} over \gls{D-RIS} are listed as follows~\cite{10716670,ahmed2024comprehensive,10316535,khan2024integration}:
\begin{itemize}
    \item \textit{Precise \gls{BF} in Full-Space Coverage}: Despite the fact that \gls{D-RIS} and \gls{BD-RIS} are both able to offer a full $360^{\circ}$ coverage (through \gls{STAR-RIS} and hybrid/multi-sector \gls{BD-RIS} structures, respectively), multi-sector \gls{BD-RIS} structure standout by its ability to enable precise \gls{BF} through high-gain elements with narrower beamwidth on each \gls{BD-RIS} sector, improving the channel gain toward all intended receivers.  
    \item \textit{Advanced Interference Management}: \Gls{D-RIS} mitigates interference by adjusting the phase of incoming signals, while \gls{BD-RIS} offers effective interference management features by leveraging advanced signal processing techniques and sophisticated structures. This results in notably lower interference compared to the \Gls{D-RIS} counterpart.
    \item \textit{Improved Adaptability and Flexibility}: \Gls{D-RIS} works well in scenarios with stable and predictable signal paths. In contrast, \gls{BD-RIS} provides greater adaptability to handle dynamic signal conditions such as user movement, mobile blockages, and changing network demands. This flexibility ensures reliable performance even in highly unpredictable environments.
    \item \textit{Improved Signal Manipulation Capabilities}: Unlike \gls{D-RIS}, which can only control the diagonal elements of the phase response matrix, \gls{BD-RIS} offers more flexibility by enabling manipulation of both diagonal and non-diagonal elements. Specifically, \gls{D-RIS} can only manipulate the phase of the signal, while \gls{BD-RIS} is also capable of manipulating the signal's amplitude, polarization, and direction. These advanced manipulation capabilities greatly enhance performance in various wireless communications scenarios, particularly in \gls{6G} networks.
\end{itemize}
A comprehensive summary of the comparison between \gls{D-RIS} and \gls{BD-RIS} is provided in Table~\ref{tab: D-RIS-comparison}.

\begin{table*}[!t]
\vspace{-2em}
\caption{Comparison between \gls{D-RIS} and \gls{BD-RIS}}
\vspace{-1em}
    \centering
    \begin{tabular}{|l|p{7cm}|p{7cm}|}
        \hline
        \textbf{Feature} & \multicolumn{1}{c|}{\textbf{Diagonal-RIS (D-RIS)}}& \multicolumn{1}{c|}{\textbf{Beyond Diagonal RIS (BD-RIS)}} \\
        \hline
        \textbf{Architecture} & Reflects signals with diagonal phase shifts only & Capable of reflecting and scattering signals with non-diagonal phase shifts \\\hline
        \textbf{Connection} & Single-connected scheme & Fully-connected and group-connected schemes \\\hline
        \textbf{Signal Control} & Only controls amplitude and phase shift of the reflected signal& Provides more control over signal phase, amplitude, polarization, and waveform shaping \\\hline
         \textbf{Complexity} & Simpler hardware and signal processing & More complex due to non-diagonal scattering patterns, requiring advanced control mechanisms  \\\hline
         \textbf{Flexibility} & Limited to specific reflection angles and paths & Flexible in \gls{BF}, allowing multi-path reflections  \\\hline
         \textbf{Beamforming} & Traditional \gls{BF} with limited degrees of freedom & Enhanced \gls{BF} with full control over electromagnetic wavefronts  \\\hline
         \textbf{Channel Model} & Suitable for simple channel models (e.g., diagonal reflection matrix) & Requires more advanced channel models due to non-diagonal elements and complex scattering behaviors  \\\hline \textbf{Power Consumption} & Low power consumption due to simpler design and passive reflection & High power consumption due to additional complex hardware and signal processing  \\
        \hline \textbf{SNR} & Moderate improvement in SNR due to limited reflection options & Significant improvement in SNR through complex scattering and multi-path propagation
        \\\hline\textbf{Applications} & Suitable for basic signal enhancement and scenarios & Suitable for more sophisticated applications such as multi-user communications and MIMO systems  \\
        \hline\textbf{Costs} & Low cost due to simple hardware components and signal processing&   High cost due to complex hardware components and signal processing\\\hline
    \end{tabular}    
    \label{tab: D-RIS-comparison}
\end{table*}

\section{Classification of BD-RIS} \label{sec: Classification of BD-RIS}

\begin{figure*}[!t]
\centering
\vspace{-1em}
\includegraphics[width=\textwidth]{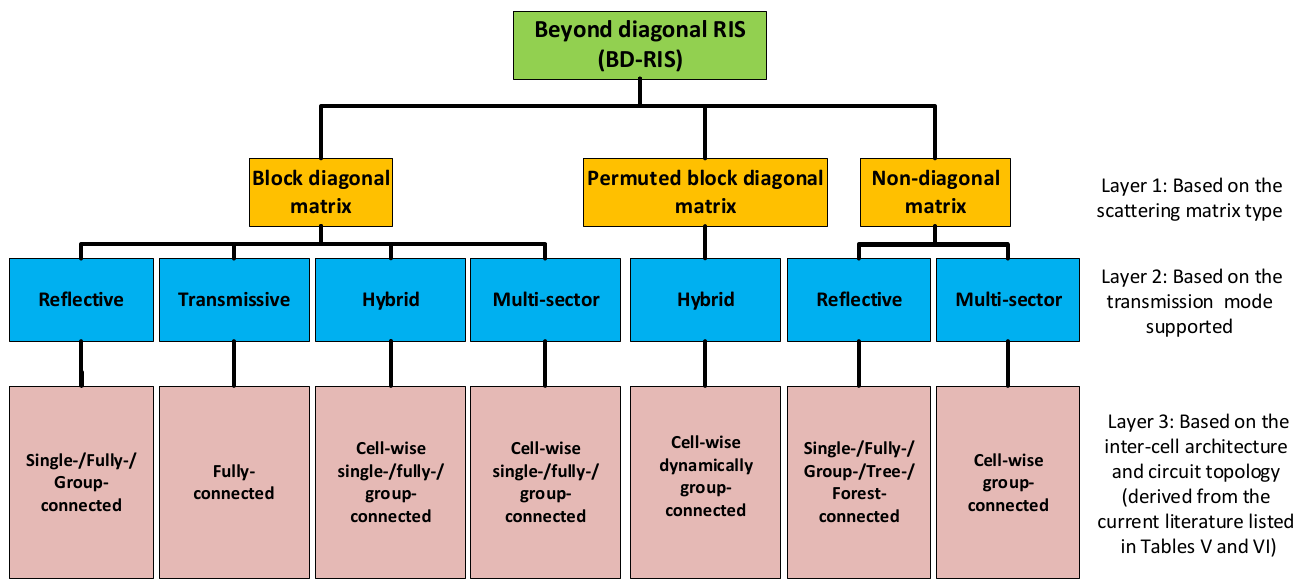}
\caption{\gls{BD-RIS} classification tree.}
\label{fig: 1_Classification}
\vspace{-2em}
\end{figure*}

\Gls{BD-RIS} structure can be classified in terms of its scattering matrix, modes of transmission, inter-cell architecture, and circuit topology. Specifically, its scattering matrix can be either block diagonal, permuted block diagonal, or non-diagonal. Its modes of transmission can be either reflective, transmissive, hybrid, or multi-sector. Its inter-cell architecture can be either cell-wise single-/fully-/group-/dynamically group-connected architectures. Its circuit topology can be either single-/fully-/group-/tree-/forest-connected topologies. The pictorial classification of the aforementioned \gls{BD-RIS} structures is illustrated in Fig.~\ref{fig: 1_Classification} as a three-layer classification tree. All these layers are discussed next in detail.

\begin{table*}[!tbph]
\centering
\vspace{-2em}
\caption{\gls{BD-RIS} classification (a detailed look)}
\vspace{-1em}
\label{tab: Layer charts}
\resizebox{\textwidth}{!}{%
\begin{tabular}{|l|c|c|}
\hline

\multicolumn{1}{|c|}{\textbf{Layer 1: Scattering Matrix (With Constraints)}} & \textbf{Layer 2: Transmission Mode} & \textbf{Layer 3: Inter-Cell Architecture and Circuit Topology} \\ \hline

\rowcolor{lightgray}\multicolumn{3}{|c|}{\textbf{(Reflective - Transmissive) D-RIS}} \\ \hline
\begin{tabular}[l]{@{}l@{}}
\amark \ Single-connected:~\cite{8910627} \\[1mm]
\(\displaystyle
\mathbf{\Theta_{\textnormal{D-RIS}}} = 
\begin{bmatrix} \label{1}
e^{j\theta_1}& 0            &0            &0\\
0            & e^{j\theta_2}&0            &0\\
0            & 0            &e^{j\theta_3}&0\\
0            & 0            &0            & e^{j\theta_4}
\end{bmatrix}
\)
\end{tabular}
&
\includegraphics[width=0.22\textwidth,valign=c]{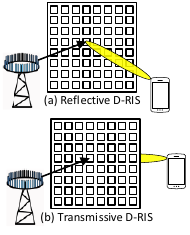}
& \begin{tabular}[l]{@{}l@{}}  Note: In here, only reflective architecture is presented.\\ Transmissive architecture is straightforward. \\[1mm] \includegraphics[width=0.4\textwidth,valign=c]{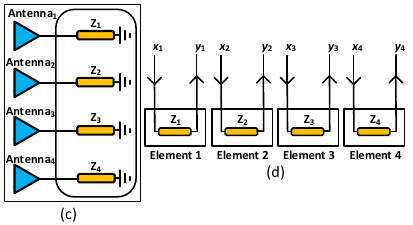} \end{tabular} \\ 
\hline
\rowcolor{lightgray}\multicolumn{3}{|c|}{\textbf{(Reflective - Distributed Reflective) BD-RIS}} \\ \hline
\begin{tabular}[l]{@{}l@{}}
\amark \ Fully-connected:~\cite{9913356} \\[1mm]
\(\displaystyle
\mathbf{\Theta} = 
\begin{bmatrix}
e^{j\theta_{1,1}}&e^{j\theta_{1,2}}&e^{j\theta_{1,3}}&e^{j\theta_{1,4}}\\
e^{j\theta_{2,1}}&e^{j\theta_{2,2}}&e^{j\theta_{2,3}}&e^{j\theta_{2,4}}\\
e^{j\theta_{3,1}}&e^{j\theta_{3,2}}&e^{j\theta_{3,3}}&e^{j\theta_{3,4}}\\
e^{j\theta_{4,1}}&e^{j\theta_{4,2}}&e^{j\theta_{4,3}}&e^{j\theta_{4,4}}
\end{bmatrix} 
\) \\[1mm]
Constraints: $\mathbf{\Theta}=\mathbf{\Theta}^T, \mathbf{\Theta}^H \mathbf{\Theta} = \mathbf{I}_M$. \\[2mm] \hdashline \\

\amark \ Group-connected:~\cite{9913356} \\[1mm] 
\(\displaystyle
\mathbf{\Theta} = 
 \begin{bmatrix}
e^{j\theta_{1,1}}&e^{j\theta_{1,2}}&0                &0\\
e^{j\theta_{2,1}}&e^{j\theta_{2,2}}&0                &0\\
0                &0                &e^{j\theta_{3,3}}&e^{j\theta_{3,4}}\\
0                &0                &e^{j\theta_{4,3}}&e^{j\theta_{4,4}}
\end{bmatrix}
\) \\[1mm]
Constraints:
$\mathbf{\Theta} = \diag (\mathbf{\Theta}_1,\mathbf{\Theta}_2,\ldots,\mathbf{\Theta}_G),$ \\ \qquad \qquad \quad \ $ \mathbf{\Theta}_g=\mathbf{\Theta}_g^T, \ \mathbf{\Theta}_g^H \mathbf{\Theta}_g = \mathbf{I}_{\bar{M}}, \ \forall g, \ \bar{M} = \frac{M}{G} $. \\[2mm] \hdashline \\ 

\amark \ Tree-connected (Tridiagonal):~\cite{10453384} \\[1mm]
\(\displaystyle
\mathbf{\Theta} = 
 \begin{bmatrix}
e^{j\theta_{1,1}}&e^{j\theta_{1,2}}&0                &0\\
e^{j\theta_{2,1}}&e^{j\theta_{2,2}}&e^{j\theta_{2,3}}&0\\
0                &e^{j\theta_{3,2}}&e^{j\theta_{3,3}}&e^{j\theta_{3,4}}\\
0                &0                &e^{j\theta_{4,3}}&e^{j\theta_{4,4}}
\end{bmatrix} \) \\[1mm]
\amark \ Tree-connected (Arrowhead):~\cite{10453384} \\[1mm]
\(\displaystyle
\mathbf{\Theta} = 
 \begin{bmatrix}
e^{j\theta_{1,1}}&e^{j\theta_{1,2}}&e^{j\theta_{1,3}}&e^{j\theta_{1,4}}&e^{j\theta_{1,5}}\\
e^{j\theta_{2,1}}&e^{j\theta_{2,2}}&0                &0&0\\
e^{j\theta_{3,1}}&0                &e^{j\theta_{3,3}}&0&0\\
e^{j\theta_{4,1}}&0                &0                &e^{j\theta_{4,4}}&0\\
e^{j\theta_{5,1}}&0                &0                &0&e^{j\theta_{5,5}}
\end{bmatrix}  
\) \\[1mm]
Constraints: 
$\boldsymbol{\Theta}=\left(\mathbf{I}+jZ_{0}\mathbf{B}\right)^{-1}\left(\mathbf{I}-jZ_{0}\mathbf{B}\right),$ \\ \qquad \qquad \quad \ $\mathbf{B}=\mathbf{B}^{T}, \mathbf{B}\in\mathcal{B}_{\mathcal{G}}.$ \\[2mm] \hdashline \\

\amark \ Forest-connected (Tridiagonal):~\cite{10453384} \\[1mm]
\(\displaystyle
\mathbf{\Theta} = \left[
\begin{array}{cccc}
e^{j\theta_{1,1}}&e^{j\theta_{1,2}}&0                &0\\
e^{j\theta_{2,1}}&e^{j\theta_{2,2}}&e^{j\theta_{2,3}}&0\\
0                &e^{j\theta_{3,2}}&e^{j\theta_{3,3}}&e^{j\theta_{3,4}}\\
0                &0                &e^{j\theta_{4,3}}&e^{j\theta_{4,4}}\\
\hline
e^{j\theta_{5,1}}&e^{j\theta_{5,2}}&0                &0\\
e^{j\theta_{6,1}}&e^{j\theta_{6,2}}&e^{j\theta_{6,3}}&0\\
0                &e^{j\theta_{7,2}}&e^{j\theta_{7,3}}&e^{j\theta_{7,4}}\\
0                &0                &e^{j\theta_{8,3}}&e^{j\theta_{8,4}}\\
\end{array} \right] \) \\[1mm]
\amark \ Forest-connected (Arrowhead):~\cite{10453384} \\[1mm]
\(\displaystyle
\mathbf{\Theta} = \left[
\begin{array}{ccccc}
e^{j\theta_{1,1}}&e^{j\theta_{1,2}}&e^{j\theta_{1,3}}&e^{j\theta_{1,4}}&e^{j\theta_{1,5}}\\
e^{j\theta_{2,1}}&e^{j\theta_{2,2}}&0                &0&0\\
e^{j\theta_{3,1}}&0                &e^{j\theta_{3,3}}&0&0\\
e^{j\theta_{4,1}}&0                &0                &e^{j\theta_{4,4}}&0\\
e^{j\theta_{5,1}}&0                &0                &0&e^{j\theta_{5,5}}
\\ \hline
e^{j\theta_{6,1}}&e^{j\theta_{6,2}}&e^{j\theta_{6,3}}&e^{j\theta_{6,4}}&e^{j\theta_{6,5}}\\
e^{j\theta_{7,1}}&e^{j\theta_{7,2}}&0                &0&0\\
e^{j\theta_{8,1}}&0                &e^{j\theta_{8,3}}&0&0\\
e^{j\theta_{9,1}}&0                &0                &e^{j\theta_{9,4}}&0\\
e^{j\theta_{10,1}}&0                &0                &0&e^{j\theta_{10,5}}
\end{array}  
\right] \) \\[1mm]
Constraints: 
$\boldsymbol{\Theta}=\left(\mathbf{I}+jZ_{0}\mathbf{B}\right)^{-1}\left(\mathbf{I}-jZ_{0}\mathbf{B}\right)$, \\ \qquad \qquad \quad \
$\mathbf{B}=\mathrm{diag}\left(\mathbf{B}_{1},\ldots,\mathbf{B}_{G}\right)$,\\ \qquad \qquad \quad \
$\mathbf{B}_{g}=\mathbf{B}_{g}^{T},\:\mathbf{B}_{g}\in\mathcal{B}_{\mathcal{G},g},\:\forall g$.\\[1mm]
\end{tabular}

& \includegraphics[width=0.22\textwidth,valign=c]{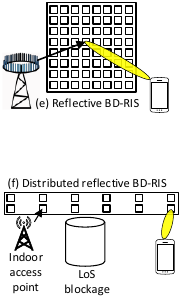} 

& \includegraphics[width=0.4\textwidth,valign=c]{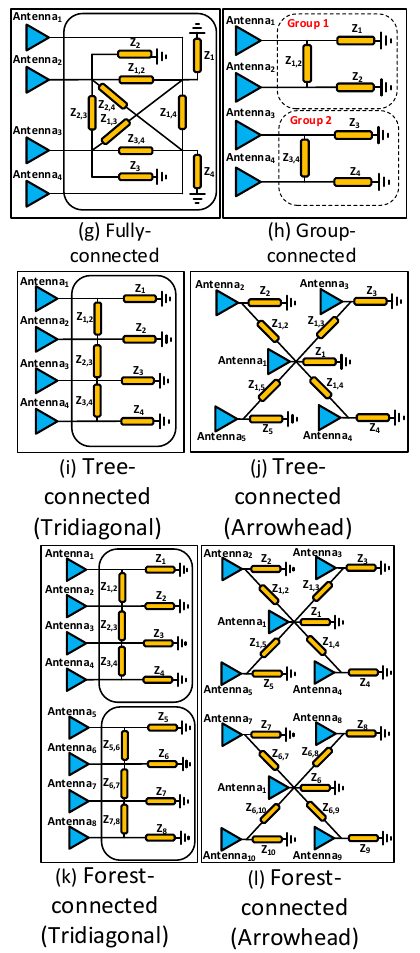} \\ \hline
 
\end{tabular}%
}
\end{table*}

\begin{table*}[!t]
\centering
\vspace{-2em}
\caption{\gls{BD-RIS} classification (a detailed look) (Cont.).}
\vspace{-1em}
\label{tab: Layer charts 2}
\resizebox{\textwidth}{!}{%
\begin{tabular}{|l|c|c|}
\hline

\multicolumn{1}{|c|}{\textbf{Layer 1: Scattering Matrix (With Constraints)}} & \textbf{Layer 2: Transmission Mode} & \textbf{Layer 3: Inter-Cell Architecture and Circuit Topology} \\ \hline

\rowcolor{lightgray}\multicolumn{3}{|c|}{\textbf{(Transmissive - Hybrid - Multi-sector) BD-RIS}} \\ \hline
\begin{tabular}[l]{@{}l@{}}
\amark \ Fully-connected:~\cite{Wali2025transmissive,10530995} \\[1mm]
\(\displaystyle
\mathbf{\Theta} = 
\begin{bmatrix}
e^{j\theta_{1,1}}&e^{j\theta_{1,2}}&e^{j\theta_{1,3}}&e^{j\theta_{1,4}}\\
e^{j\theta_{2,1}}&e^{j\theta_{2,2}}&e^{j\theta_{2,3}}&e^{j\theta_{2,4}}\\
e^{j\theta_{3,1}}&e^{j\theta_{3,2}}&e^{j\theta_{3,3}}&e^{j\theta_{3,4}}\\
e^{j\theta_{4,1}}&e^{j\theta_{4,2}}&e^{j\theta_{4,3}}&e^{j\theta_{4,4}}
\end{bmatrix} 
\) \\[1mm]
Constraints: $\mathbf{\Theta}=\mathbf{\Theta}^T, \mathbf{\Theta}^H \mathbf{\Theta} = \mathbf{I}_M$.
\end{tabular}
& 
\includegraphics[width=0.235\textwidth,valign=c]{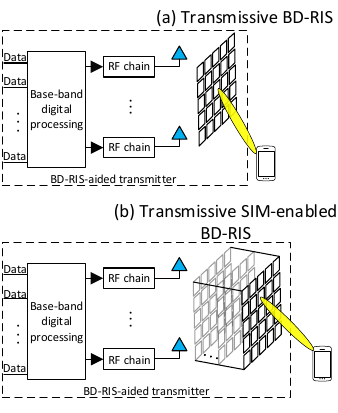}
&  
\includegraphics[width=0.25\textwidth,valign=c]{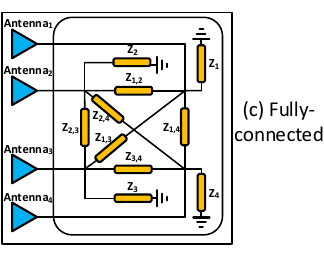}
\\ \hline

\begin{tabular}[l]{@{}l@{}}
\amark \ Cell-wise single-connected:~\cite{9913356} \\[1mm]
$\mathbf{\Theta}_{\mathrm{r}}=\mathsf{diag}(e^{{j\theta}_{1,\mathrm{r}}},\ldots,e^{{j\theta}_{M,\mathrm{r}}})$, \\[1mm] $\mathbf{\Theta}_{\mathrm{t}}=\mathsf{diag}(e^{{j\theta}_{1,\mathrm{t}}},\ldots,e^{{j\theta}_{M,\mathrm{t}}})$, \\[1mm] 
Constraint:  $|e^{{j\theta}_{m,\mathrm{r}}}|^{2}+|e^{{j\theta}_{m,\mathrm{t}}}|^{2}=1, \ \forall m \in M.$ \\[2mm] \hdashline \\

\amark \ Cell-wise fully-connected:~\cite{9913356} \\[1mm]
$\mathbf{\Theta}_{\mathrm{r}}=\mathsf{Full}(e^{{j\theta}_{1,\mathrm{r}}},\ldots,e^{{j\theta}_{M,\mathrm{r}}})$, \\[1mm] $\mathbf{\Theta}_{\mathrm{t}}=\mathsf{Full}(e^{{j\theta}_{1,\mathrm{t}}},\ldots,e^{{j\theta}_{M,\mathrm{t}}})$, \\[1mm] 
Constraint:  $\mathbf{\Theta}_{\mathrm{r}}^{H}\mathbf{\Theta}_{\mathrm{r}}+\mathbf{\Theta}_{\mathrm{t}}^{H}\mathbf{\Theta}_{\mathrm{t}}=\mathbf{I}_{\frac{M}{2}}.$ \\[2mm] \hdashline \\

\amark \ Cell-wise group-connected:~\cite{9913356} \\[1mm]
$\mathbf{\Theta}_{\mathrm{r}}=\mathsf{blkdiag}(\mathbf{\Theta}_{\mathrm{r},1},\ldots,\mathbf{\Theta}_{\mathrm{r},G})$, \\ $\mathbf{\Theta}_{\mathrm{t}}=\mathsf{blkdiag}(\mathbf{\Theta}_{\mathrm{t},1},\ldots,\mathbf{\Theta}_{\mathrm{t},G})$, \\[1mm]
Constraint:  
$\mathbf{\Theta}_{\mathrm{r},g}^{H}\mathbf{\Theta}_{\mathrm{r},g}+\mathbf{\Theta}_{\mathrm{t},g}^{H}\mathbf{\Theta}_{\mathrm{t},g}=\mathbf{I}_{\bar{M}},\forall g\in\mathcal{G}.$  \\[2mm] \hdashline \\

\amark \ Cell-wise dynamically group-connected:~\cite{10159457} \\[1mm]
$\mathbf{\Theta}_{\mathrm{r}}=\mathsf{blkdiag}(\mathbf{\Theta}_{\mathrm{r},1},\ldots,\mathbf{\Theta}_{\mathrm{r},G})$, \\ $\mathbf{\Theta}_{\mathrm{t}}=\mathsf{blkdiag}(\mathbf{\Theta}_{\mathrm{t},1},\ldots,\mathbf{\Theta}_{\mathrm{t},G})$, \\[1mm]
Constraints: $[\mathbf{\Theta}_\mathrm{t}]_{m,n} = 0, \ \forall m \in \mathcal{D}_p, \ \forall n \in \mathcal{D}_q, \ p \ne q$, 
\\[1mm] \qquad \qquad \quad \
$[\mathbf{\Theta}_\mathrm{r}]_{m,n} = 0, \ \forall m \in \mathcal{D}_p, \ \forall n \in \mathcal{D}_q, \ p \ne q$, 
\\[1mm] \qquad \qquad \quad \ $\mathbf{\Theta}_{\mathrm{t},\mathcal{D}_g}^H\mathbf{\Theta}_{\mathrm{t}, \mathcal{D}_g} + \mathbf{\Theta}_{\mathrm{r},\mathcal{D}_g}^H\mathbf{\Theta}_{\mathrm{r},\mathcal{D}_g} = \mathbf{I}_{|\mathcal{D}_g|}, \ \forall g$, 
\\[1mm] \qquad \qquad \quad \
$\mathcal{D}_p \cap \mathcal{D}_q = \varnothing, \ \forall p \ne q$,
\\[1mm] \qquad \qquad \quad \
$\mathcal{D}_g \ne \varnothing, \ \forall g$,
\\[1mm] \qquad \qquad \quad \
$\cup_{g = 1}^G \mathcal{D}_g = M$.
\end{tabular} 
 &  
\includegraphics[width=0.25\textwidth,valign=c]{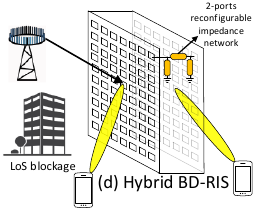} 
 &  
\includegraphics[width=0.4\textwidth,valign=c]{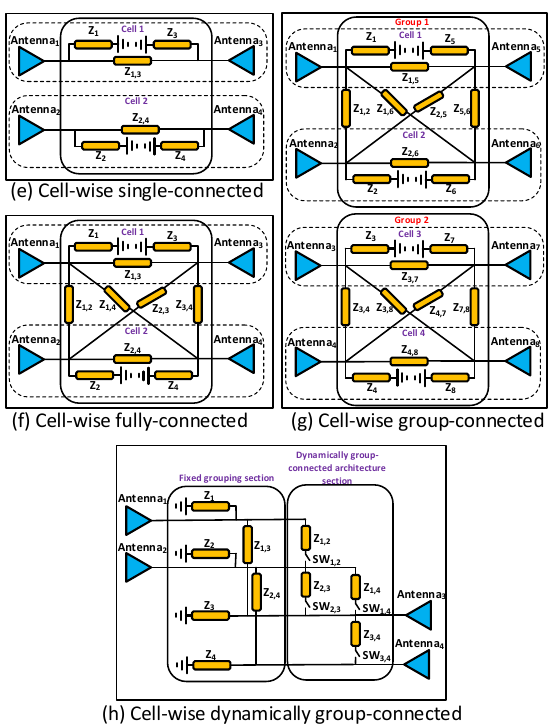} 
 
 \\ \hline

\begin{tabular}[l]{@{}l@{}}
\amark \ Cell-wise single-connected:~\cite{10158988} \\[1mm]
$\mathbf{\Theta}_l = \mathsf{diag}(e^{{j\theta}_{(l-1)M+1}},\ldots,e^{{j\theta}_{lM}})$, \\[1mm] 
Constraint:  $\sum_{i\in \mathcal{L}_m} |e^{{j\theta}_i}|^2 = 1$, $\forall m\in M $ \\[2mm] \hdashline \\

\amark \ Cell-wise fully-connected:~\cite{10158988} \\[1mm]
$\mathbf{\Theta}_l =\mathsf{Full}(e^{{j\theta}_{(l-1)M+1}},\ldots,e^{{j\theta}_{lM}})$,  \\[1mm] 
Constraint:  $\sum_{\forall l \in \mathcal{L}} \mathbf{\Theta}_{l}^H\mathbf{\Theta}_{l} = \mathbf{I}_\frac{M}{\mathcal{L}}$. \\[2mm] \hdashline \\

\amark \ Cell-wise group-connected:~\cite{10158988} \\[1mm]
$\mathbf{\Theta}_l = \mathsf{blkdiag}(\mathbf{\Theta}_{l,1}, \ldots, \mathbf{\Theta}_{l,G})$, \\[1mm]
Constraint:  
$\sum_{\forall l \in \mathcal{L}}\mathbf{\Theta}_{l,g}^H\mathbf{\Theta}_{l,g} = \mathbf{I}_{\bar{M}}$, $\forall g\in\mathcal{G}$, \\ \qquad \qquad \quad \textnormal{where} $\mathcal{G} = \{1,\ldots,G\}$, $\bar{M} = M/G$.  \\[2mm] 
\hdashline \\ 

\amark \ Opposite-link with horizontally-connected sectors BD-RIS:~\cite{10643263} \\ \quad \ (Depending on the switch array setup, a possible $\mathbf{\Theta}_l$ can be)\\[1mm]
\(\displaystyle
\mathbf{\Theta}_l = 
 \begin{bmatrix}
0&0&0&e^{j\theta_{1,4}}&0                &0\\
0&0&0&0                &e^{j\theta_{2,5}}&0\\
0&0&0&0                &0                &e^{j\theta_{3,6}}
\end{bmatrix}  
\) \\[1mm]
Constraint: $\sum_{\forall l \in \mathcal{L}}\mathbf{\Theta}_{l,g}^H\mathbf{\Theta}_{l,g} = \mathbf{I}_{\bar{M}}$, $\forall g\in\mathcal{G}$, \\ \qquad \qquad \quad \textnormal{where} $\mathcal{G} = \{1,\ldots,G\}$, $\bar{M} = M/G$.  \\[1mm]
\hdashline \\ 

\amark \ Full-link with horizontally-connected sectors BD-RIS:~\cite{10643263} \\ \quad \ (Depending on the switch array setup, a possible $\mathbf{\Theta}_l$ can be) \\[1mm]
\(\displaystyle
\mathbf{\Theta}_l = 
 \begin{bmatrix}
0&0&e^{j\theta_{1,3}}&0                &0&0\\
0&0&0                &e^{j\theta_{2,4}}&0&0\\
0&0&0                &0                &e^{j\theta_{3,5}}&0
\end{bmatrix}  
\) \\[1mm]
Constraint: $\sum_{\forall l \in \mathcal{L}} \mathbf{\Theta}_{l}^H\mathbf{\Theta}_{l} = \mathbf{I}_M$. \\[2mm] 
\end{tabular}

 &
\includegraphics[width=0.25\textwidth,valign=c]{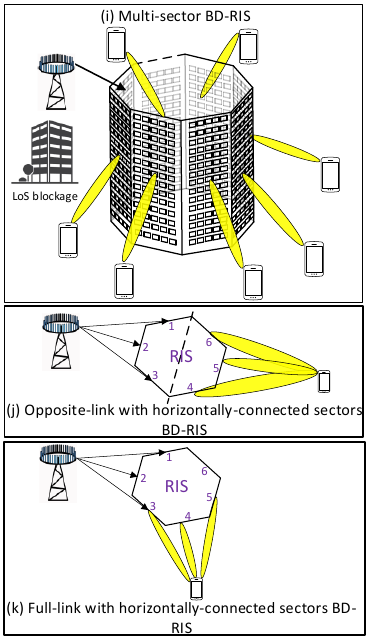} 
 &
\begin{tabular}[c]{@{}c@{}}  Note: In here, only cell-wise single-connected architecture \\ is presented. Cell-wise group-/full-connected architectures \\ are straightforward. \\ \includegraphics[width=0.4\textwidth,valign=c]{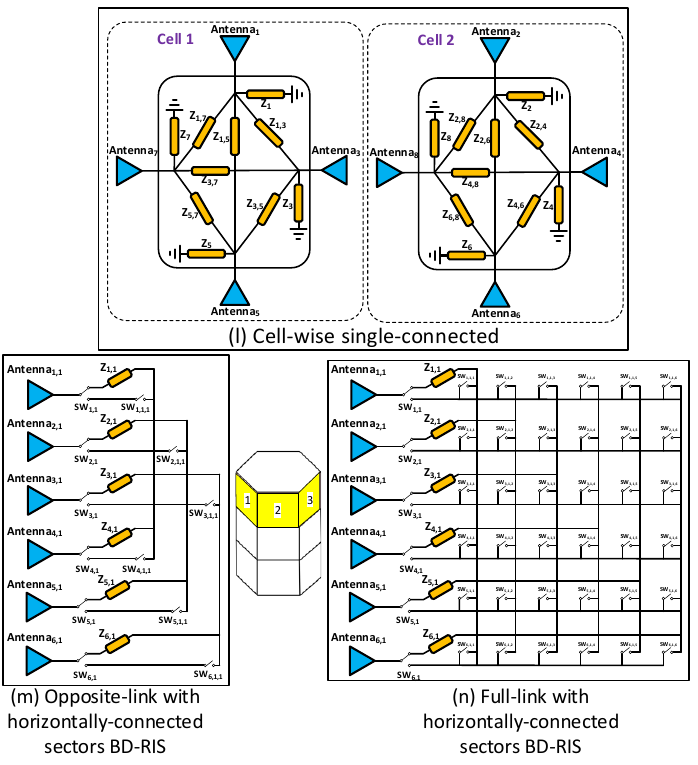} \end{tabular}
 \\ \hline

\end{tabular}%
}
\end{table*}

\begin{table*}[!t]
\centering
\vspace{-2em}
\caption{\gls{BD-RIS} classification (a detailed look) (Cont.).}
\vspace{-1em}
\label{tab: Layer charts 3}
\resizebox{\textwidth}{!}{%
\begin{tabular}{|l|c|c|}
\hline

\multicolumn{1}{|c|}{\textbf{Layer 1: Scattering Matrix (With Constraints)}} & \textbf{Layer 2: Transmission Mode} & \textbf{Layer 3: Inter-Cell Architecture and Circuit Topology} \\ \hline

\rowcolor{lightgray}\multicolumn{3}{|c|}{\textbf{(Reflective - Multi-sector) non-diagonal BD-RIS}} \\ \hline

\begin{tabular}[l]{@{}l@{}}
\amark \ Non-diagonal (ND) BD-RIS:~\cite{9737373} \\[1mm]
\(\displaystyle
\mathbf{\breve{\Theta}} = 
\begin{bmatrix}
0                &0                &0              &e^{j\theta_{1,4}}\\
e^{j\theta_{2,1}}&0                &0                & 0        \\
0                &0                &e^{j\theta_{3,3}}& 0         \\
0                &e^{j\theta_{4,2}}&0                & 0
\end{bmatrix} 
\) \\
Constraint: $|e^{{j\theta}_m}|=1, \ \forall m \in [1, 2, ..., M]$. \\[2mm] 
\hdashline \\ 

\amark \ Opposite-link with horizontally-connected sectors and vertically \\ \quad \ connected elements:~\cite{10643263} \\ \quad \ (Depending on the switch array setup, a possible $\mathbf{\breve{\Theta}}_l$ can be) \\[2mm]
\(\displaystyle
\mathbf{\breve{\Theta}}_l = 
 \begin{bmatrix}
0&e^{j\theta_{1,2}}&0&0&0&0\\
e^{j\theta_{2,1}}&0&0&0&0&0\\
0&0&0&e^{j\theta_{3,4}}&0&0\\
0&0&e^{j\theta_{4,3}}&0&0&0\\
0&0&0&0&0&e^{j\theta_{5,6}}\\
0&0&0&0&e^{j\theta_{6,5}}&0
\end{bmatrix}  
\) \\[1mm]
Constraint: $\sum_{\forall l \in \mathcal{L}}\mathbf{\breve{\Theta}}_{l,g}^H\mathbf{\breve{\Theta}}_{l,g} = \mathbf{I}_{\bar{M}}$, $\forall g\in\mathcal{G}$, \\ \qquad \qquad \quad \textnormal{where} $\mathcal{G} = \{1,\ldots,G\}$, $\bar{M} = M/G$.  \\
\hdashline \\ 

\amark \ Full-link with horizontally-connected sectors and vertically \\ \quad \ connected elements in each sector BD-RIS:~\cite{10643263}  \\ \quad \ (Depending on the  switch array setup, a possible $\mathbf{\breve{\Theta}}_l$ can be)  \\[2mm]
\(\displaystyle
\mathbf{\breve{\Theta}}_l = 
 \begin{bmatrix}
0&e^{j\theta_{1,2}}&0&0&0&0\\
e^{j\theta_{2,1}}&0&0&0&0&0\\
0&0&e^{j\theta_{3,3}}&0&0&0\\
0&0&0&0&e^{j\theta_{4,5}}&0\\
0&0&0&e^{j\theta_{5,4}}&0&0\\
0&0&0&0&0&e^{j\theta_{6,6}}
\end{bmatrix}  
\) \\[1mm]
Constraint: $\sum_{\forall l \in \mathcal{L}} \mathbf{\breve{\Theta}}_{l}^H\mathbf{\breve{\Theta}}_{l} = \mathbf{I}_M$. \\[2mm] 
\end{tabular}
 &
\includegraphics[width=0.25\textwidth,valign=c]{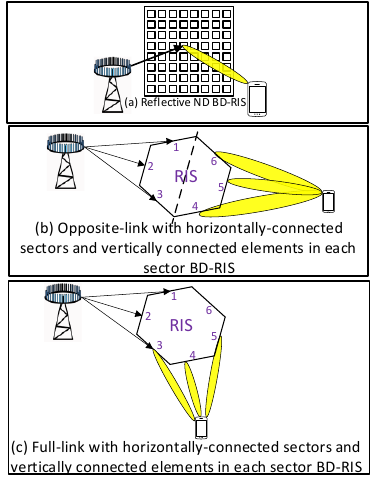}  
 &
\includegraphics[width=0.4\textwidth,valign=c]{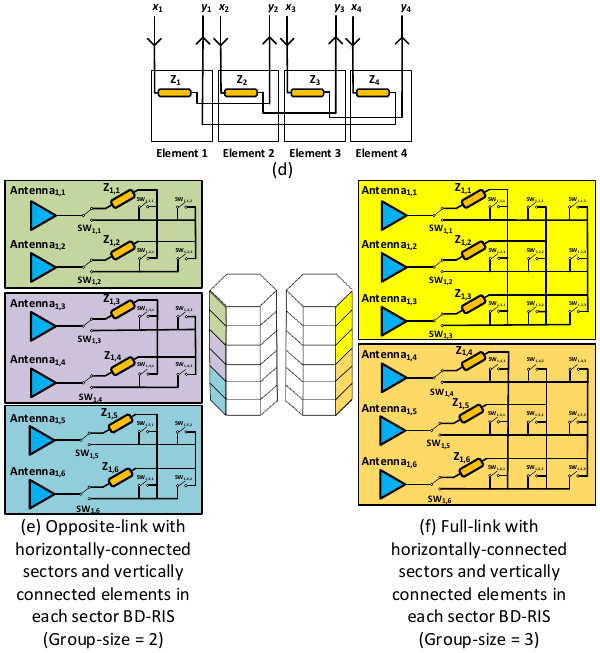}   
 \\ \hline
 
\end{tabular}%
}
\end{table*}

\subsection{Layer~\texorpdfstring{$1$}{1}: Based on the Scattering Matrix Type}

The scattering matrix $\mathbf{\Theta}$ of \gls{BD-RIS} structures is beyond diagonal and, under a lossless assumption, satisfies the following two constraints (i) the unitary constraint, $\mathbf{\Theta}^H \mathbf{\Theta} = \mathbf{I}$, where $(.)^H$ is the conjugate transpose operator. This means that there is energy conservation when the impinging waves interact with the scattering surface; and (ii) symmetry/reciprocal constraint, $\mathbf{\Theta}=\mathbf{\Theta}^T$, where $(.)^T$ is the transpose operator. This constraint allows for using a reciprocal impedance network to simplify the circuit of the \gls{BD-RIS} system~\cite{9514409}.  The scattering matrix of \gls{BD-RIS} structures can be either block diagonal, permuted block diagonal, or non-diagonal as explained in the following.

\subsubsection{Block Diagonal Matrix} 

In this scattering matrix type, the $M$ antennas of the \gls{BD-RIS} structure are uniformly divided into $G$ groups. As illustrated in Table~\ref{tab: Layer charts}(h), group-connected \gls{BD-RIS} structure, a $4$-element \gls{BD-RIS} structure is divided into two groups of two antenna ports each, where two antennas are paired together in each group. This \gls{BD-RIS} structure leads to a scattering matrix with a block diagonal structure, for which the number of blocks corresponds to the number of groups in the structure. That is, $\mathbf{\Theta}$ is a block diagonal scattering matrix, with each block being unitary~\cite{10316535}. In the case where there is only one group as shown in Table~\ref{tab: Layer charts}(g), fully-connected \gls{BD-RIS} structure, each of the $M$ antenna ports is connected to all the remaining $M-1$ antenna ports. This makes $\mathbf{\Theta}$ a one-block unitary scattering matrix. 

\subsubsection{Permuted Block Diagonal Matrix} 

In this scattering matrix type, the way the $M$ antennas of the \gls{BD-RIS} structure are grouped can change based on the current channel conditions. In other words, the antennas' grouping strategy is dynamic and adapted to the channel environment. This dynamic grouping allows for more flexible beam control compared to systems with fixed groupings~\cite{10316535}. In this case, the resulting scattering matrix is called permuted block diagonal~\cite{10159457} and the whole \gls{BD-RIS} structure is called dynamically group-connected, as shown in Table~\ref{tab: Layer charts 2}(h). To create an $M$-element permuted block diagonal scattering matrix for the dynamically group-connected \gls{BD-RIS} structure, an array of switches is needed. The locations of non-zero elements in the \gls{RIS} scattering matrix define the ON/OFF state of these switches. To be more precise, if the matching element in the \gls{RIS} scattering matrix is non-zero, the switches are turned ON; otherwise, they remain OFF.

\subsubsection{Non-diagonal Matrix} 

In this scattering matrix type, the antenna ports of the \gls{BD-RIS} structure have unique interconnections. Specifically, the signal received by one antenna is reflected off another, leading to an asymmetric non-diagonal scattering matrix~\cite{9737373}. This architecture allows for precise control over the direction of the reflected signal, enhancing the overall performance and efficiency of the antenna system. To put this in perspective, Table~\ref{tab: Layer charts}(d) and Table~\ref{tab: Layer charts 3}(d) are used to illustrate the \gls{D-RIS} structure with a diagonal matrix and the \gls{BD-RIS} structure with a non-diagonal matrix, respectively. In Table~\ref{tab: Layer charts}(d), each \gls{RIS} element directly reflects the impinging signal waves after phase shift adjustment. On the other hand, in Table~\ref{tab: Layer charts 3}(d), the signal received by one antenna element is reflected by another antenna element after phase shift adjustment. For example, in Table~\ref{tab: Layer charts 3}(d), by denoting the incident signal impinging on the $1$-st element as $x_1$ and denoting the reflected signal, after phase shift modification, reflected from the $2$-nd element as $y_2$, the relationship between $x_1$ and $y_2$ is given as $y_2 = x_1 e^{j\theta_{2,1}}$. In the non-diagonal scattering matrix $\mathbf{\breve{\Theta}}$, each row and each column of $\mathbf{\breve{\Theta}}$ has exactly one non-zero element, as shown in the scattering matrices provided in Table~\ref{tab: Layer charts 3}.

\subsection{Layer~\texorpdfstring{$2$}{2}: Based on the Transmission Mode} 
\label{subsec: Layer2: Based on the Transmission Mode}

The modes of transmission of \gls{BD-RIS} structures can be either reflective, transmissive, hybrid, or multi-sector as explained in the following.

\subsubsection{Reflective Mode} 

In this category, both the transmitter and receiver are located on the same side of the \gls{BD-RIS} structure as illustrated in Table~\ref{tab: Layer charts}(e). Further, all the $M$ antenna ports are oriented in the same direction. Hence, an impinging signal wave on one side of the \gls{BD-RIS} structure is reflected on the same side. For this mode of transmission, the scattering matrix $\mathbf{\Theta}$ satisfies the unitary and symmetry constraints~\cite{10316535}.

\subsubsection{Transmissive Mode} 
In this category, a \gls{BD-RIS} is deployed at the transmitter side to reduce the extensive power consumption typical of wireless transmitters that rely on fully digital \gls{RF} chains to achieve high \gls{BF} gains. Specifically, \gls{BD-RIS} removes the requirement for extensive \gls{RF} signal processing at the transmitter, while carrying out the digital \gls{BF}, by modifying the phase of incoming signals with minimal power consumption~\cite{10693959}. Also, at the transmitter side, the \gls{BD-RIS} is placed a few wavelengths away from the active transmitter antennas (as depicted in Table~\ref{tab: Layer charts 2}(a)) to reduce the feed blockages~\cite{10693959}.

\subsubsection{Hybrid Mode} 

In this category, some of the impinging signal energy is reflected by the \gls{BD-RIS} elements and some portion of the energy passes through to the other side of the \gls{BD-RIS} structure. This mode enables coverage at the front and the back of the \gls{BD-RIS} structure employing both reflective and transmissive modes, respectively, as illustrated in Table~\ref{tab: Layer charts 2}(d). To ensure the hybrid reflective and transmissive mode, two uni-directional radiation pattern antennas are connected back-to-back to form a cell as illustrated in Table~\ref{tab: Layer charts 2}(e)-(g). Two matrices are associated with this mode, one for the reflective mode and the other for the transmissive mode, denoted respectively as $\mathbf{\Theta}_r$ and $\mathbf{\Theta}_t$. These two matrices satisfy the following constraint:  $\mathbf{\Theta}_r^H \mathbf{\Theta}_r + \mathbf{\Theta}_t^H \mathbf{\Theta}_t = \mathbf{I}_{\frac{M}{2}}$, where $\mathbf{\Theta}_r \neq 0 $ and $\mathbf{\Theta}_t \neq 0 $~\cite{9913356}.

\subsubsection{Multi-Sector Mode} 

In this category, the antennas are divided into multiple $\mathcal{L} \geq$ $2$-sectors, where each sector covers a limited region of space as shown in Table~\ref{tab: Layer charts 2}(i). In this mode, when signals impinge on one sector of the \gls{BD-RIS} structure, they can be partially reflected in different directions. Specifically, some of the signals may be partially reflected towards the same sector, while others are partially scattered towards the other $\mathcal{L}-1$ sectors. To achieve this mode, each cell is made to have $M$ antennas that are positioned at the edges of an $l$-th sided \gls{BD-RIS} structure. These antennas have unidirectional radiation patterns that cover $1/\mathcal{L}$ to prevent overlap between sectors. The $M$ antennas are connected to an $M$-port single-connected reconfigurable impedance network, as depicted in Table~\ref{tab: Layer charts 2}(l) for versatile and efficient signal transmission. Consequently, the multi-sector mode can enable full-space coverage and finds applications in coverage extension in instances where there are multiple obstacles. The multi-sector mode \gls{BD-RIS} has $\mathcal{L}$-sub-matrices, $\mathbf{\Theta}_l, \ \forall l \in [1,2,...,\mathcal{L}]$, that satisfied the following constraint, $\sum_{\forall l \in \mathcal{L}} \mathbf{\Theta}_{l}^H\mathbf{\Theta}_{l} = \mathbf{I}_{\frac{M}{\mathcal{L}}}$~\cite{10158988}. In~\cite{10694505}, the authors show that the increase in the number of sectors in the multi-sector \gls{BD-RIS}, while maintaining the same total number of elements, results in substantial gains in \gls{SE} and \gls{EE} compared to having fewer sectors with more elements per sector.

\subsection{Layer~\texorpdfstring{$3$}{3}: Based on the Inter-Cell Architecture and Circuit Topology}
\label{Subsec: Layer 3: Based on the Inter-Cell Architecture and Circuit Topology}

Layer $3$ classification is mainly hinged on two forms. The first is the inter-cell architecture of \gls{BD-RIS} structure, which is the way the cells are connected to one another in hybrid/multi-sector transmission modes. In other words, in hybrid/multi-sector transmission modes multiple uni-directionally radiating elements are arranged back-to-back to form cells~\cite{10716670}. The second is the circuit topology of the \gls{BD-RIS} structure, which is how the antenna elements are connected to one another in reflective/transmissive modes.

\subsubsection{Inter-Cell Architecture} 

In this architecture, there are cell-wise single-/fully-/group-/dynamically group-connected inter-cell architectures, as shown for the hybrid transmission mode in Table~\ref{tab: Layer charts 2}(e)-(h). It is demonstrated in~\cite{9913356} that the cell-wise group-/fully-connected architecture performs better than the cell-wise single-connected architecture. In a bid to improve the performance of the cell-wise connected architecture, the inter-cell grouping technique is made adaptable to channel conditions in~\cite{10159457} to form what is named cell-wise dynamically group-connected architecture which is illustrated in Table~\ref{tab: Layer charts 2}(h). The scattering matrices of this architecture can be expressed as $\mathbf{\Theta}_{\mathrm{r}}=\mathsf{blkdiag}(\mathbf{\Theta}_{\mathrm{r},1},\ldots,\mathbf{\Theta}_{\mathrm{r},G})$, $\mathbf{\Theta}_{\mathrm{t}}=\mathsf{blkdiag}(\mathbf{\Theta}_{\mathrm{t},1},\ldots,\mathbf{\Theta}_{\mathrm{t},G})$, and their corresponding constraints are~\cite{10159457}
\begin{align*}
\textnormal{(C1):} & \quad [\mathbf{\Theta}_\mathrm{t}]_{m,n} = 0, \ \forall m \in \mathcal{D}_p, \ \forall n \in \mathcal{D}_q, \ p \ne q, \\ 
\textnormal{(C2):} & \quad [\mathbf{\Theta}_\mathrm{r}]_{m,n} = 0, \ \forall m \in \mathcal{D}_p, \ \forall n \in \mathcal{D}_q, \ p \ne q, \\ \textnormal{(C3):} & \quad \mathbf{\Theta}_{\mathrm{t},\mathcal{D}_g}^H\mathbf{\Theta}_{\mathrm{t}, \mathcal{D}_g} + \mathbf{\Theta}_{\mathrm{r},\mathcal{D}_g}^H\mathbf{\Theta}_{\mathrm{r},\mathcal{D}_g} = \mathbf{I}_{|\mathcal{D}_g|}, \ \forall g, \\ \textnormal{(C4):} & \quad \mathcal{D}_p \cap \mathcal{D}_q = \varnothing, \ \forall p \ne q, \\ \textnormal{(C5):} & \quad
\mathcal{D}_g \ne \varnothing, \ \forall g, \\ \textnormal{(C6):} & \quad \cup_{g = 1}^G \mathcal{D}_g = M,    
\end{align*}
\noindent where $\mathcal{D}=[\mathcal{D}_1,...,\mathcal{D}_G]$ is an array that stores the indexes of \gls{BD-RIS} cells for each group, $(m,n), \ m\neq n$ denotes a pair of inter-connected cells connected by reconfigurable impedance components, $\mathbf{\Phi}_{\mathrm{t/r},\mathcal{D}_g}$ is a sub-matrix of $\mathbf{\Phi}_{\mathrm{t/r}}$ which selects columns and rows of $\mathbf{\Phi}_{\mathrm{t/r}}$ according to indexes in the set $\mathcal{D}_g$,$ \ \forall g$. (C1) and (C2) denote the non-zero entries in $\mathbf{\Theta}_{\mathrm{t}}$ and $\mathbf{\Theta}_{\mathrm{r}}$, respectively. (C3) represents a constraint that relates $\mathbf{\Theta}_{\mathrm{t}}$ and $\mathbf{\Theta}_{\mathrm{r}}$ for each set $\mathcal{D}_g$,$ \ \forall g$. (C4) ensures that every group includes a minimum of one cell. (C5) ensures that each \gls{BD-RIS} cell can belong to only a single group. (C6) ensures that the different groups are mutually exclusive.

\subsubsection{Circuit Topology} 

There are several proposed circuit topologies for the reflective transmission mode, namely, single-/fully-/group-/tree-/forest-connected topologies, as shown in Table~\ref{tab: Layer charts 2}(g)-(m). If the \gls{BD-RIS} structure is divided into $G$ groups of antenna ports, while the antenna ports in each group are paired together, this refers to the group-connected topology. The scattering matrix of this topology is $\mathbf{\Theta} = \diag (\mathbf{\Theta}_1,\mathbf{\Theta}_2,\ldots,\mathbf{\Theta}_G)$, and satisfies both the unitary and symmetry constraints, which are respectively represented as $ \mathbf{\Theta}_g=\mathbf{\Theta}_g^T, \ \mathbf{\Theta}_g^H \mathbf{\Theta}_g = \mathbf{I}_{\bar{M}}, \ \forall g,$ for $\bar{M} = M/G$. $\bar{M}$ denotes the number of antenna ports in each group~\cite{9514409}. In the case where there is only one group and each of the $M$ antenna ports is connected to all the remaining $M-1$ antenna ports, this refers to the fully-connected topology. In single-connected topology, each element operates independently and this topology can be considered as a special case of group-connected topology~\cite{10316535}. The work in~\cite{9514409} demonstrated that the use of fully and group-connected topologies can both (i) boost the received signal power by as much as 62\% compared to a single-connected architecture, (ii) achieve the same received signal power of a single-connected topology while minimizing the number of \gls{BD-RIS} elements by up to 21\%.

The circuit topologies for both tree-connected and forest-connected structures are modeled in~\cite{10453384} using graph theoretical concept and are defined as $\mathbb{G} = (\mathbb{V},\mathbb{E})$, where $\mathbb{G}$ is the graph, $\mathbb{V}$ is the \textit{vertex set} of $\mathbb{G}$ and they are expressed by the set of indices of the \gls{BD-RIS} ports given as $\mathbb{V}=\{1,2,...,M\}$, and $\mathbb{E}$ denotes the edge set of $\mathbb{G}$, which corresponds to the $\dot{M}$ number of admittances connecting one port to another port in the topology. It should be noted that the number of tunable admittance components in a \gls{BD-RIS} architecture's circuit topology determines its circuit complexity. Because there are $M$ tunable admittance components connecting each element to the ground and $\dot{M}$ tunable admittance components connecting the elements to each other, the circuit complexity of a \gls{BD-RIS} represented by a graph with $\dot{M}$ edges is therefore given by $M + \dot{M}$. A tree can be defined as a linked graph $\mathbb{G}$ with $M$ vertices and $M-1$ edges. Thus, a tree-connected \gls{BD-RIS} has a total of $2M-1$ admittance components, which is significantly fewer than the $(M(M +1))/2$ admittance components in a fully-connected \gls{BD-RIS}. This is because a tree-connected \gls{BD-RIS} has $M$ admittance components connecting each port to the ground and $M-1$ admittance components (edges) connecting the ports to each other. The scattering matrix of the tree-connected topology can be expressed as $\boldsymbol{\Theta}=\left(\mathbf{I}+jZ_{0}\mathbf{B}\right)^{-1}\left(\mathbf{I}-jZ_{0}\mathbf{B}\right)$, where $Z_0$ denotes the reference impedance for calculating scattering parameters and $\mathbf{B}\in\mathcal{B}_{\mathcal{G}}$ denotes the $M$-port reconfigurable impedance network susceptance matrix. $\mathcal{B}_{\mathcal{G}}$ is the set of all possible susceptance matrices for \gls{BD-RIS} defined by the graph $\mathbb{G}$~\cite{10453384}. 

In~\cite{10453384}, two specific examples of tree-connected topology were proposed, namely, tridiagonal and arrowhead, as illustrated in Table~\ref{tab: Layer charts}(i) and (j), respectively. The tridiagonal \gls{BD-RIS} structure is represented by a tree-connected topology whose graph $\mathbb{G}$ is a path graph. On the other hand, the arrowhead \gls{BD-RIS} structure is represented by a tree-connected topology whose graph $\mathbb{G}$ is a star graph. It is worth noting that the tridiagonal \gls{BD-RIS} structure is a simple and practical architecture for tree-connected topology and can be practically realized using uniform linear arrays or radio stripes. This is because it minimizes interconnection length by connecting only adjacent \gls{BD-RIS} elements. However, arrowhead \gls{BD-RIS} is better suited for developing tree-connected topology and can be practically realized using uniform planar arrays~\cite{10453384}.

Forest-connected \gls{BD-RIS} was proposed in~\cite{10453384} as a way to further minimize the circuit complexity of tree-connected \gls{BD-RIS}, particularly for large-scale situations. A \gls{BD-RIS} that has its $M$ elements split up into $\bar{M} =  M/G$  groups with group size $G$ and uses the tree-connected architecture, i.e., each group's graph is a tree on $G$ vertices, is known as a forest-connected \gls{BD-RIS}. As a result, a forest is represented by the graph $\mathbb{G}$ connected to the forest-connected \gls{BD-RIS}. Each group in the forest-connected \gls{BD-RIS} may have an arrowhead or tridiagonal tree-connected topologies~\cite{10453384}. Forest-connected \gls{BD-RIS}, or one with $\bar{M}=1$, is a special case of the tree-connected \gls{BD-RIS}. Furthermore, both the single-connected and tree-connected \gls{BD-RIS} are two distinct examples of the forest-connected \gls{BD-RIS}, with $G=1$ and $G=M$, respectively. Furthermore, when $G=2$, forest-connected \gls{BD-RIS} is equivalent to a group-connected \gls{BD-RIS}. In a forest-connected \gls{BD-RIS} with group size $G$, the number of tunable admittance components is  $M ( (2-1)/G )$. The scattering matrix of the forest-connected topology can be expressed as $\boldsymbol{\Theta}=\left(\mathbf{I}+jZ_{0}\mathbf{B}\right)^{-1}\left(\mathbf{I}-jZ_{0}\mathbf{B}\right)$, where $\mathbf{B}=\mathrm{diag}\left(\mathbf{B}_{1},\ldots,\mathbf{B}_{G}\right)$.

\subsection{Lessons Learnt}
The following points summarize the key takeaways regarding the classification of the \gls{BD-RIS} architectures documented in the section.
\begin{itemize}
    \item Due to the evolution of the \gls{RIS} technology, the scattering matrix $\mathbf{\Theta}$ is no longer limited to being diagonal. The scattering matrix $\mathbf{\Theta}$ can be a block diagonal matrix where the antenna ports in each group are disconnected across groups. Further, the scattering matrix $\mathbf{\Theta}$ can be permuted block diagonal matrix, where grouping arrangement is adapted to the channel environment. Lastly, the scattering matrix $\mathbf{\Theta}$ can be an asymmetric non-diagonal scattering matrix, where the signal received by one antenna port is reflected off another antenna port.
    \item The operation mode of the \gls{RIS} is no longer restricted to the reflective, transmissive, and hybrid transmission modes. Now the operation mode can be also multi-sector (multiple transmissive, serving multiple users located at any place around the \gls{BD-RIS} structure). Full-space coverage, enhanced performance, and simplified circuit design can be achieved with the multi-sector \gls{BD-RIS}-based structure in comparison with the other \gls{BD-RIS} structures.
    \item Performance comparison between the group-connected topology and the fully-connected topology shows that as the number of \gls{BD-RIS} elements increases, the \gls{SNR} gain of the fully-connected topology rises quickly. However, for the group-connected topology, the larger the group sizes the higher the \gls{SNR} gain, and the performance of a very large group size can approach the performance of the fully-connected topology.   
    \item When the number of antenna ports increases significantly for the fully-connected reconfigurable impedance network, the number of reconfigurable impedance components also increases (growing quadratically with the number of antenna ports) and the circuit topology gets more complex.
\end{itemize}

\section{Potential Areas of Applications of BD-RIS} \label{sec: Potential Areas of Applications of BD-RIS}
\gls{BD-RIS} presents a promising technology in wireless communications, facilitating advancements across various domains. This section explores the potential applications of \gls{BD-RIS}, which include coverage extension and enabling comprehensive signal reflection and transmission to enhance network performance in poor coverage areas. The \gls{PLS} is significantly improved through the advanced \gls{BF} capabilities of \gls{BD-RIS}, which mitigate risks of eavesdropping and jamming. The architecture also plays a crucial role in multi-cell interference cancellation, allowing for precise signal management in dense networks. Additionally, \gls{BD-RIS} enhances sensing and localization, providing accurate positioning and tracking in challenging environments. Emerging RIS types such as \gls{SIM} and \gls{STAR-RIS} further expand the applications of \gls{BD-RIS}, enabling innovative solutions for next-generation wireless networks. The above-mentioned applications highlight the potential of \gls{BD-RIS} in addressing contemporary communications challenges.

\subsection{Coverage Extension}
Conventional \gls{D-RIS} models are constrained to reflecting signals on a single side of the surface, resulting in limited coverage due to the uniform arrangement of antenna arrays. To address this limitation, \gls{STAR-RIS} and \gls{IOS} were introduced in~\cite{9200683} and~\cite{9690478}, respectively, enabling reflection and transmission to achieve full-space coverage. \gls{STAR-RIS}, however, relies on a specific configuration with back-to-back antenna arrays and a group-connected reconfigurable impedance network~\cite{10158988}, reducing flexibility in signal control. \gls{BD-RIS} offers significant potential for enhancing network coverage. \gls{BD-RIS} extends the coverage capability by enabling signal reflections and transmissions on multiple sides of the surface. \gls{BD-RIS} is particularly advantageous for cellular networks, where signal degradation often limits coverage at the cell edges. By deploying \gls{BD-RIS} at strategic locations, such as cell boundaries or areas with poor coverage, these surfaces can reflect and redirect signals to users outside the conventional coverage area while simultaneously enhancing the quality of service for users within the cell. Furthermore, the flexible configuration of \gls{BD-RIS} allows for adaptive optimization of signal propagation paths, significantly improving network performance in dense urban, rural, or obstructed environments. In~\cite{10158988}, the authors proposed a multi-sector \gls{BD-RIS} which extends the functionality of \gls{STAR-RIS} and \gls{IOS} models to provide comprehensive full-space coverage. Multi-sector \gls{BD-RIS} design surpasses traditional \gls{RIS} and \gls{STAR-RIS} performance by utilizing group-connected reconfigurable impedance networks and optimized antenna array arrangements. With its ability to deliver full-space coverage, generate highly directional beams, and adapt to varying sector configurations, the multi-sector \gls{BD-RIS} is ideal for diverse applications, particularly in \gls{mmWave}, \gls{THz}, and cell-free network deployments.

\subsection{Physical Layer Security}
Conventional \gls{D-RIS} provides a valuable opportunity to improve \gls{PLS} in wireless communications systems by leveraging its capability to manipulate the characteristics of the wireless channel intelligently~\cite{10736549}. Integrating~\gls{D-RIS} with advanced technologies such as \gls{D2D} communications \cite{9305710}, \gls{UAV}~\cite{10778572}, \gls{IoT}~\cite{10285357}, and \gls{CRN}~\cite{10219539} highlights its flexibility in enhancing \gls{PLS}. However, the limited \gls{BF} capabilities of \gls{D-RIS} restrict the ability to fully leverage advanced \gls{PLS} techniques. Such limitation reduces its effectiveness in mitigating security threats, such as eavesdropping and jamming, especially in complex environments like multi-user communications scenarios. Consequently, \gls{D-RIS} could not be suitable for applications that require robust protection against unauthorized access. On the other hand, \gls{BD-RIS} offers a promising enhancement for \gls{PLS} in wireless communications networks by enhancing the \gls{BF} capabilities, which could significantly improve security measures by allowing for more precise control over signal propagation and maintaining robust and secure transmission. For example, by utilizing its advanced multi-sector and hybrid reflection-transmission capabilities, \gls{BD-RIS} can direct confidential signals toward intended receivers while introducing artificial noise to confuse potential eavesdroppers. The architecture allows for dynamic \gls{BF} across multiple sectors, enabling secure communications paths that exploit the randomness of fading channels and noise to limit unauthorized interception. Unlike conventional \gls{D-RIS}, \gls{BD-RIS} can flexibly manage amplitude and phase shifts, ensuring that confidential information is transmitted to legitimate users while reducing signal leakage toward eavesdroppers. Such adaptability significantly enhances \gls{PLS}, particularly in complex wireless environments such as \gls{mmWave} and \gls{THz} networks, where secure and efficient communication is paramount.

\subsection{Multi-Cell Interference Cancellation}

Future cellular networks are anticipated to feature densely deployed small cells with extensive overlapping coverage areas, significantly intensifying multi-cell interference~\cite{9330587}. Managing and mitigating such interference is a critical challenge in dense wireless environments, and \gls{BD-RIS} could be a potential solution. One of the key applications of \gls{BD-RIS}  is the ability to effectively reduce multi-cell interference by intelligently controlling the propagation of signals. Specifically, by leveraging a hybrid or multi-sector \gls{BD-RIS} architecture, the system can be configured to manipulate the transmitted and reflected signals more precisely. The \gls{BD-RIS} architecture allows for the selective nullification of signals directed toward unintended cell-edge users, minimizing the interference experienced by neighboring cells. Simultaneously, \gls{BD-RIS} can enhance signal transmission by directing and reflecting signals specifically toward the intended cell-edge users, improving signal quality and overall system performance.

The advanced control mechanisms of \gls{BD-RIS}, utilizing a non-diagonal scattering matrix, enable effective management of complex signal interactions across multiple sectors or cells. \gls{BD-RIS} optimizes coverage, mitigates interference, and enhances spectral efficiency in densely populated cellular environments by dynamically adjusting signal transmission and reflection characteristics. Consequently, the deployment of \gls{BD-RIS} signifies a substantial advancement in tackling the challenges of multi-cell interference in the next-generation wireless communications networks. Deploying \gls{BD-RIS} for passive MU \gls{BF} in MU-MISO systems was explored in~\cite{10771739}. The design of the \gls{BD-RIS} scattering matrix was utilized to either maximize the aggregate received signal power at user terminals or reduce interference at users through the zero-forcing method. A generalized practical frequency-dependent reflection model was proposed in~\cite{10766364} as a framework for configuring fully-connected and group-connected \gls{BD-RIS} in multi-band \gls{MIMO} networks. The findings highlighted potential harmful interference from insufficient synchronization between \gls{BD-RIS} and adjacent \glspl{BS}.

\subsection{Sensing and Localization}
\Gls{ISAC} has become a key enabler for next-generation wireless networks due to its ability to share spectrum, hardware, and signal processing between communications and sensing functions~\cite{9737357, 9606831}. The integration of \gls{mmWave} technology enhances \gls{ISAC} systems by offering high data rates for communications and high resolution for target sensing. However, \gls{mmWave}'s short wavelength leads to significant path loss, requiring numerous transmit antennas and fully digital \gls{RF} chains to achieve high \gls{BF} gains, which results in high power consumption~\cite{9705498}. Therefore, cost-effective solutions are needed for \gls{mmWave}-based \gls{ISAC} systems. The advanced \gls{BF} capabilities of BD-RIS substantially enhance the network's ability to achieve ultra-precise device localization. These features are critical for creating a new generation of indoor positioning systems that operate independently of additional infrastructure, costs, or energy requirements~\cite{10716670}.

The emerging \gls{BD-RIS} architecture has improved \gls{BF} capabilities compared to \gls{D-RIS}. Using fully-connected \gls{BD-RIS} to enhance throughput in \gls{ISAC} systems while ensuring quality sensing was investigated in~\cite{10495009}, demonstrating the benefits of integrating \gls{BD-RIS} into \gls{ISAC} networks. The authors of~\cite{10493847} investigated using \gls{BD-RIS} to minimize the \gls{BS}'s transmit power in an \gls{ISAC} network while ensuring communication and sensing quality. The effectiveness of the proposed system model demonstrated the advantages of \gls{BD-RIS} deployment. Integrating fully connected \gls{BD-RIS} into \gls{ISAC} systems was investigated in~\cite{10777522}, with its additional degrees of freedom leveraged through non-zero off-diagonal elements in the scattering matrix. By jointly optimizing the weighted sum of SNR for both radar receivers and communications users, substantial performance gains for \gls{ISAC} systems were obtained. In~\cite{10902602}, distributed \gls{BD-RIS} was examined as a solution to the limited coverage of conventional and localized \gls{BD-RIS} systems restricted to serving nearby users. \Gls{BD-RIS} elements were distributed across a wide area, and significant performance gains over both distributed \gls{D-RIS} and localized BD-\gls{RIS} were revealed due to the interconnections facilitating signal propagation. The \gls{BD-RIS}-aided \gls{DFRC} system was presented in~\cite{10643599}, differing from \gls{D-RIS}-aided \gls{DFRC} systems that utilized a diagonal scattering matrix, which limited signal reflection to half-space. The \gls{BD-RIS} was designed to support hybrid reflecting and transmitting modes, allowing for flexible architectures and full-space coverage, enhancing system performance and highlighting the superior communications and sensing capabilities of the \gls{BD-RIS}-aided \gls{DFRC} system compared to \gls{D-RIS}-aided \gls{DFRC} systems. A localization approach for next-generation communications systems was proposed in~\cite{raeisi2024efficient} using BS-assisted passive \gls{BF} with \gls{BD-RIS}s. More control over phase and amplitude was obtained by \gls{BD-RIS}s compared to \gls{D-RIS}s, leading to enhanced localization accuracy. These findings highlighted the potential of \gls{BD-RIS}s as a promising solution for high-accuracy positioning in future wireless networks.

\subsection{Channel Manipulation}
Channel shaping utilizes \gls{D-RIS} to modify the wireless environment, improving various aspects such as mitigating Doppler effects, enhancing \gls{MIMO} channel rank, and introducing artificial channel diversity for multiple access schemes. It separates \gls{BF} into two stages: channel shaping and transceiver design, offering versatile solutions for wireless applications. Channel shaping metrics fall into two categories: singular value metrics~\cite{10187717}, which relate to performance but are sensitive to channel perturbations, and power metrics~\cite{10453384}, which are easier to analyze but less insightful in \gls{MIMO} systems.

Most works~\cite{9200661, 9734015, 10314137, 10187717} have focused on a \gls{D-RIS} model, limiting control to simple phase shifts without inter-element coupling. Using a passive \gls{RIS} to optimize the singular values of point-to-point \gls{MIMO} channels was investigated in~\cite{zhao2024channel}, enhancing power efficiency and data rates. A \gls{BD-RIS} architecture providing greater flexibility in wave manipulation compared to the \gls{D-RIS} was introduced. The achievable channel singular value regions for both RIS models were analyzed. The results showed that \gls{BD-RIS} significantly improved the dynamic range of singular values, leading to better channel power and higher achievable rates, particularly as the number of RIS elements and \gls{MIMO} dimensions increases. The \gls{BD-RIS} was investigated in~\cite{alegria2024channel} for their ability to create orthogonal channels in multiuser \gls{MIMO} scenarios. Practical \gls{RIS} implementations with passive components and constraints on their reconfigurability were proposed. Techniques for optimal channel selection without active amplification at the \gls{RIS} were introduced. Efficient channel estimation and \gls{RIS} configuration techniques, to be handled at the \gls{BS}, were presented. The effect of channel aging due to \gls{UE} mobility on multi-sector \gls{BD-RIS} was investigated in~\cite{10680138}. Unlike \gls{D-RIS}, \gls{BD-RIS} technology provides expanded coverage. The average sum-rate maximization problem under aging conditions was addressed by jointly designing the \gls{BD-RIS} matrix and transmit precoder. Numerical results revealed the impacts of channel aging on system performance.

\subsection{Stacked Intelligent Metasurface (SIM)}
\Gls{SIM} and \gls{BD-RIS} represent sophisticated advances in metasurface technology that significantly enhance the ability to manipulate electromagnetic waves in wireless communications systems. These technologies extend the performance of \gls{D-RIS}. \Gls{SIM} employs a multi-layered or stacked configuration, enabling precise control over transmitted and reflected signals. Leveraging the interaction between multiple metasurface layers allows simultaneous manipulation of signal parameters across various spatial dimensions~\cite{10379500}. In contrast, \gls{BD-RIS} introduces a non-diagonal scattering matrix, which provides additional degrees of freedom for signal control, enabling more dynamic transmission and reflection of electromagnetic waves. Both \gls{SIM} and \gls{BD-RIS} overcome the limitations of \gls{D-RIS} by offering enhanced flexibility in wavefront control, improving wireless coverage, increasing signal quality, and optimizing spectral and energy efficiency~\cite{10534211}.

The \gls{SIM} achieves a non-diagonal scattering matrix involving leveraging a multi-layer \gls{RIS} architecture, as proposed in~\cite{10515204, 10379500}. The \gls{SIM} architecture consists of multiple closely spaced refracting metasurfaces, each designed to manipulate the incident signal progressively as it propagates through the layers~\cite{10515204}. Upon illumination by the incoming signal, the first metasurface layer refracts the signal toward the subsequent layer, and this process continues until the signal reaches the final layer. At this stage, the wave is radiated toward the receiver. During the propagation through the layers, the signal is dynamically shaped by tunable meta-atoms embedded in each layer, resulting in a non-diagonal end-to-end scattering matrix. This effect arises due to the inherent broadcast nature of wireless signals as they propagate from one layer to the next. A non-diagonal scattering matrix can be realized by employing dynamic refracting metasurfaces, each equipped with electronics capable of tuning the signal's phase and amplitude across layers. The scattering matrix comprehensively describes the scattering behavior of an $M$-port reconfigurable impedance network independent of specific circuit designs~\cite{10515204}. The resulting scattering matrix is diagonal in traditional \gls{RIS} implementations, where each port is only connected to its respective reconfigurable impedance.

\subsection{STAR-RIS}
The \gls{STAR-RIS} or \gls{IOS} offer a more versatile approach to wireless communications systems. This hybrid mode, characterized by a mathematical model that extends beyond the conventional diagonal scattering matrix, was explored in~\cite{9722826,9200683,9437234,9754364}. In~\cite{9200683}, the physical properties of IOS were detailed, along with a model capable of programmable electromagnetic responses on both sides of the surface. The study formulated an \gls{SE} maximization problem for downlink communication, where the \gls{IOS} was designed to enhance the \gls{SE} of a multiuser system through optimal phase shift configuration. A branch-and-bound algorithm was proposed to determine the optimal phase shifts in a finite set, providing an effective solution for \gls{IOS} configuration. Simulation results demonstrated significant improvements in \gls{SE} and coverage, showing that IOS-assisted systems greatly expand coverage areas compared to \gls{RIS} systems that rely solely on reflection.

In~\cite{9437234}, a comprehensive hardware model for \gls{STAR-RIS} was introduced, allowing for an in-depth comparison of diversity gain between \gls{STAR-RIS} and conventional RIS. This analysis used channel models tailored for near-field and far-field scenarios. The work in~\cite{9754364} outlined four promising techniques for independent control of transmitted and reflected signals, essential for the practical implementation of \gls{STAR-RIS}. Additionally, three hardware models were presented, each quantifying the tuning capabilities of \gls{STAR-RIS} with varying degrees of precision. The corresponding channel models were categorized into five groups: three for physics-based representations, one for near-field channel modeling, and one for far-field gain calculation. The benefits and limitations of these modeling techniques were thoroughly discussed. Further advancements were reported in~\cite{9722826}, where \gls{IOS} was explored for enabling full-dimensional wireless communications. Based on this concept, a hybrid \gls{BF} technique was proposed for \gls{IOS}-based communications systems. Additionally, a prototype of an \gls{IOS}-aided wireless system was developed, with experimental results confirming the ability of \gls{IOS} to simultaneously support users on both sides of the surface, demonstrating the practicality and effectiveness of the technology in real-world scenarios.

\subsection{Lessons Learnt}
This section explores the various applications of \gls{BD-RIS}. Here are the key takeaways:
\begin{itemize}
    \item \gls{BD-RIS} enhances network coverage by enabling signal reflections and transmissions from multiple sides of the surface. Such capability addresses the limitations of \gls{D-RIS} models, which often struggle with \gls{LoS} restrictions in urban environments. \gls{BD-RIS} allows for flexible optimization of signal propagation paths, thus improving coverage in challenging areas such as dense urban landscapes and rural zones. By dynamically adapting the signal paths, \gls{BD-RIS} can fill coverage gaps and ensure reliable connectivity for users in various locations.
    \item \gls{BD-RIS} is crucial in enhancing \gls{ISAC} capabilities. It enables ultra-precise localization and positioning by using reflective surfaces to interact with multiple signals and gather data from the environment. Such ability is particularly beneficial for applications in autonomous vehicles and smart cities, where accurate positioning is essential. 
    \item \gls{BD-RIS} effectively addresses multi-cell interference by intelligently controlling the propagation of signals. \gls{BD-RIS} enhances spectral efficiency in densely populated cellular environments by optimizing coverage and minimizing interference between adjacent cells. The advanced control mechanisms employed by \gls{BD-RIS} allow for the coordination of signals across different cells, reducing the impact of interference and improving the service quality for users in congested areas.
    \item \gls{BD-RIS} employs channel-shaping techniques to improve various aspects of wireless communication, such as mitigating Doppler effects caused by moving objects and enhancing the rank of \gls{MIMO} channels. Such channel manipulation allows for the optimization of signal transmission, increasing the reliability and efficiency of data delivery. \gls{BD-RIS}'s versatility in channel manipulation offers diverse solutions for applications ranging from mobile communications to \gls{IoT} devices, ensuring robust performance across different scenarios.
    \item \gls{SIM} and \gls{BD-RIS} multi-layered approach provides greater flexibility in wavefront control, enabling improved wireless coverage, signal quality, and energy efficiency. \gls{SIM}'s ability to adjust the phase and amplitude of signals across different layers allows for optimized signal propagation and better performance in complex environments.
\end{itemize}

\begin{table*}[!p]
\centering
\vspace{-2em}
\caption{SUMMARY OF EXISTING WORKS IN BD-RIS-BASED SYSTEMS}
\vspace{-1em}
\label{tab: SUMMARY 1}
\resizebox{0.99\textwidth}{!}{%
\begin{tabular}{|c|ccc|ccc|c|}
\hline
\multirow{2}{*}{\textbf{{[}\#{]}}} & \multicolumn{3}{c|}{\textbf{System related content}} & \multicolumn{3}{c|}{\textbf{BD-RIS related content}} & \multirow{2}{*}{\textbf{\begin{tabular}[c]{@{}c@{}}Performance\\ metric(s)\end{tabular}}} \\ \cline{2-7}

& \multicolumn{1}{c|}{\textbf{\begin{tabular}[c]{@{}c@{}}Technical\\ content\end{tabular}}} & \multicolumn{1}{c|}{\textbf{\begin{tabular}[c]{@{}c@{}}Main focus \end{tabular}}} & \multicolumn{1}{c|}{\textbf{\begin{tabular}[c]{@{}c@{}}System  model\end{tabular}}} & \multicolumn{1}{c|}{\textbf{\begin{tabular}[c]{@{}c@{}}Scattering \\ matrix type\end{tabular}}} & \multicolumn{1}{c|}{\textbf{\begin{tabular}[c]{@{}c@{}}Transmission\\ mode\end{tabular}}} & \textbf{Architecture type} &  \\ \hline


\cite{10574199}  & \multicolumn{1}{c|}{\multirow{13}{*}{\begin{tabular}[c]{@{}c@{}} BD-RIS \\  modeling \\(covered in \\ Sec. II) \end{tabular}}}  & \multicolumn{1}{c|}{\begin{tabular}[c]{@{}c@{}} Multiport network\\ theory BD-RIS\\ modeling  \end{tabular}} & \multicolumn{1}{c|}{SU-MIMO-DL} & \multicolumn{1}{c|}{\begin{tabular}[c]{@{}c@{}} Block diagonal \\ and \\ non-diagonal \end{tabular}} & \multicolumn{1}{c|}{Reflective} & \multicolumn{1}{c|}{\begin{tabular}[c]{@{}c@{}} Single-/Fully-/Group-\\/Tree-/Forest-connected \end{tabular}} & Received power \\ \cline{1-1} \cline{3-8}

\cite{10694015} & \multicolumn{1}{c|}{}  & \multicolumn{1}{c|}{\begin{tabular}[c]{@{}c@{}} Multiport network\\ theory BD-RIS\\ modeling \end{tabular}} & \multicolumn{1}{c|}{SU-MIMO-DL} & \multicolumn{1}{c|}{Block diagonal} & \multicolumn{1}{c|}{Reflective} & \multicolumn{1}{c|}{-} & - \\ \cline{1-1} \cline{3-8}

\cite{del2024physics} & \multicolumn{1}{c|}{}  & \multicolumn{1}{c|}{\begin{tabular}[c]{@{}c@{}} Multiport network\\ theory BD-RIS\\ modeling  \end{tabular}} & \multicolumn{1}{c|}{SU-MIMO-DL} & \multicolumn{1}{c|}{Block diagonal} & \multicolumn{1}{c|}{Reflective} & \multicolumn{1}{c|}{Group-connected} & \begin{tabular}[c]{@{}c@{}} Received signal \\ strength indicator \end{tabular}\\ \cline{1-1} \cline{3-8}

\cite{10453384} & \multicolumn{1}{c|}{} & \multicolumn{1}{c|}{\begin{tabular}[c]{@{}c@{}} Graph theory \\ BD-RIS \\ modeling \end{tabular} } & \multicolumn{1}{c|}{SU-MISO-DL} & \multicolumn{1}{c|}{Non-diagonal} & \multicolumn{1}{c|}{Reflective} & \multicolumn{1}{c|}{Tree/Forest-connected} & Received power \\ \cline{1-1} \cline{3-8}

\cite{10499196}  & \multicolumn{1}{c|}{} & \multicolumn{1}{c|}{\begin{tabular}[c]{@{}c@{}} Matrix theory \\ BD-RIS \\modeling  \end{tabular}} & \multicolumn{1}{c|}{\begin{tabular}[c]{@{}c@{}} SU-SISO-DL \\ SU-MISO-DL \\ MU-MISO-DL \end{tabular}} &  \multicolumn{1}{c|}{Block diagonal} & \multicolumn{1}{c|}{Hybrid} & \multicolumn{1}{c|}{\begin{tabular}[c]{@{}c@{}} Cell-wise \\ fully-connected \end{tabular}} & Sum-rate \\ \hline


\cite{10445725} & \multicolumn{1}{c|}{\multirow{19}{*}{\begin{tabular}[c]{@{}c@{}} BD-RIS \\  applications \\(covered in \\ Sec. IV) \end{tabular}}} & \multicolumn{1}{c|}{\begin{tabular}[c]{@{}c@{}} Physical layer\\ security \end{tabular}} & \multicolumn{1}{c|}{MU-MISO-DL} & \multicolumn{1}{c|}{Non-diagonal} & \multicolumn{1}{c|}{Reflective} & \multicolumn{1}{c|}{Fully-connected} & Ergodic sum-rate \\ \cline{1-1} \cline{3-8}

\cite{10694006} & \multicolumn{1}{c|}{} & \multicolumn{1}{c|}{\begin{tabular}[c]{@{}c@{}} Physical layer\\ key generation \end{tabular}} & \multicolumn{1}{c|}{SU-SISO-DL} & \multicolumn{1}{c|}{Non-diagonal} & \multicolumn{1}{c|}{Reflective} & \multicolumn{1}{c|}{Group-connected} & Secret key rate \\ \cline{1-1} \cline{3-8}

\cite{zhang2024full}  & \multicolumn{1}{c|}{} & \multicolumn{1}{c|}{\begin{tabular}[c]{@{}c@{}} Wireless sensing \end{tabular}} & \multicolumn{1}{c|}{MU-SISO-DL/UL} & \multicolumn{1}{c|}{Block diagonal} & \multicolumn{1}{c|}{Multi-sector} & \multicolumn{1}{c|}{\begin{tabular}[c]{@{}c@{}} Cell-wise \\ fully-connected \end{tabular}} & CRB \\ \cline{1-1} \cline{3-8}

\cite{raeisi2024efficient} & \multicolumn{1}{c|}{} & \multicolumn{1}{c|}{\begin{tabular}[c]{@{}c@{}} Localization \end{tabular}} & \multicolumn{1}{c|}{SU-SISO-DL} & \multicolumn{1}{c|}{Block diagonal} & \multicolumn{1}{c|}{Transmissive} & \multicolumn{1}{c|}{\begin{tabular}[c]{@{}c@{}} Fully-connected \end{tabular}} & CRLB \\ \cline{1-1} \cline{3-8}

\cite{10771739} & \multicolumn{1}{c|}{} & \multicolumn{1}{c|}{\begin{tabular}[c]{@{}c@{}} Interference nulling \end{tabular}} & \multicolumn{1}{c|}{MU-MISO-DL} & \multicolumn{1}{c|}{Block diagonal} & \multicolumn{1}{c|}{Reflective} & \multicolumn{1}{c|}{Fully-connected} & Sum-rate \\ \cline{1-1} \cline{3-8}

\cite{10766364} & \multicolumn{1}{c|}{} & \multicolumn{1}{c|}{Multi-cell} & \multicolumn{1}{c|}{MU-MIMO-DL} & \multicolumn{1}{c|}{Block diagonal} & \multicolumn{1}{c|}{Reflective} & \multicolumn{1}{c|}{\begin{tabular}[c]{@{}c@{}} Fully-/Group-connected \end{tabular}} & Received power \\ \cline{1-1} \cline{3-8}

\cite{10530995} & \multicolumn{1}{c|}{} & \multicolumn{1}{c|}{\begin{tabular}[c]{@{}c@{}} SIM-enabled \\ BD-RIS \end{tabular}} & \multicolumn{1}{c|}{\begin{tabular}[c]{@{}c@{}} SU-SISO-DL \\ SU-MIMO-DL \end{tabular}} & \multicolumn{1}{c|}{Non-diagonal} & \multicolumn{1}{c|}{Transmissive} & \multicolumn{1}{c|}{Fully-connected} & Channel gain \\ \cline{1-1} \cline{3-8}

\cite{9200683} & \multicolumn{1}{c|}{} & \multicolumn{1}{c|}{STAR-RIS} & \multicolumn{1}{c|}{SU-SISO-DL} & \multicolumn{1}{c|}{Block diagonal} & \multicolumn{1}{c|}{Hybrid} & \multicolumn{1}{c|}{Cell-wise single-connected} & SE \\ \cline{1-1} \cline{3-8} 

\cite{zhao2024channel}  & \multicolumn{1}{c|}{} & \multicolumn{1}{c|}{\begin{tabular}[c]{@{}c@{}} Channel shaping\\  capability \end{tabular}} & \multicolumn{1}{c|}{SU-MIMO-DL} & \multicolumn{1}{c|}{Block diagonal} & \multicolumn{1}{c|}{Reflective} & \multicolumn{1}{c|}{Fully-connected} & \begin{tabular}[c]{@{}c@{}} Channel gain, \\ rate \end{tabular} \\ \cline{1-1} \cline{3-8}

\cite{10680138} & \multicolumn{1}{c|}{} & \multicolumn{1}{c|}{\begin{tabular}[c]{@{}c@{}} BD-RIS under \\ channel aging \\ effect \end{tabular}} & \multicolumn{1}{c|}{MU-MISO-DL} & \multicolumn{1}{c|}{Block diagonal} & \multicolumn{1}{c|}{Multi-sector} & \multicolumn{1}{c|}{\begin{tabular}[c]{@{}c@{}} Cell-wise\\ single-connected \end{tabular}} & Sum-SE \\ \cline{1-1} \cline{3-8}

\cite{alegria2024channel}  & \multicolumn{1}{c|}{} & \multicolumn{1}{c|}{\begin{tabular}[c]{@{}c@{}} Channel \\ orthogonalization \end{tabular}} & \multicolumn{1}{c|}{MU-MIMO-DL} & \multicolumn{1}{c|}{Block diagonal} & \multicolumn{1}{c|}{Reflective} & \multicolumn{1}{c|}{Fully-connected} & Channel gain \\ \hline


\cite{9913356} & \multicolumn{1}{c|}{\multirow{48}{*}{\begin{tabular}[c]{@{}c@{}} BD-RIS \\  architectural \\ development  \\(covered in \\ Sec. V-A) \end{tabular}}} & \multicolumn{1}{c|}{\begin{tabular}[c]{@{}c@{}} BD-RIS \\different \\architectures  \end{tabular}} & \multicolumn{1}{c|}{MU-MIMO-DL} & \multicolumn{1}{c|}{Block diagonal} & \multicolumn{1}{c|}{Hybrid} & \multicolumn{1}{c|}{\begin{tabular}[c]{@{}c@{}} Cell-wise\\ group-/fully-connected  \end{tabular}} & Sum-rate \\ \cline{1-1} \cline{3-8}

\cite{10158988} & \multicolumn{1}{c|}{} & \multicolumn{1}{c|}{\begin{tabular}[c]{@{}c@{}} Multi-sector \\ BD-RIS \\ structure \end{tabular}} & \multicolumn{1}{c|}{\begin{tabular}[c]{@{}c@{}} SU-SISO-DL \\ MU-MISO-DL \end{tabular}} & \multicolumn{1}{c|}{Block diagonal} & \multicolumn{1}{c|}{Multi-sector} & \multicolumn{1}{c|}{\begin{tabular}[c]{@{}c@{}} Cell wise\\single-/group-/fully-\\connected  \end{tabular}} & \begin{tabular}[c]{@{}c@{}} Received power, \\ sum-rate \end{tabular} \\ \cline{1-1} \cline{3-8}

\cite{10308579} & \multicolumn{1}{c|}{} & \multicolumn{1}{c|}{\begin{tabular}[c]{@{}c@{}} Transmitter \\side BD-RIS \end{tabular}} & \multicolumn{1}{c|}{MU-MIMO-DL} & \multicolumn{1}{c|}{Block diagonal} & \multicolumn{1}{c|}{Reflective} & \multicolumn{1}{c|}{Fully-connected}  & SE \\ \cline{1-1} \cline{3-8}

\cite{10643263}  & \multicolumn{1}{c|}{} & \multicolumn{1}{c|}{\begin{tabular}[c]{@{}c@{}} Multi-sector \\ structures \end{tabular}} & \multicolumn{1}{c|}{\begin{tabular}[c]{@{}c@{}} SU-MISO-DL \\ MU-MIMO-DL \\ SU-SISO-DL \end{tabular}} & \multicolumn{1}{c|}{\begin{tabular}[c]{@{}c@{}} Block diagonal\\ and\\ non-diagonal \end{tabular}} & \multicolumn{1}{c|}{Multi-sector} & \multicolumn{1}{c|}{ \begin{tabular}[c]{@{}c@{}} Cell-wise \\ group-connected \end{tabular}} & Achievable rate \\ \cline{1-1} \cline{3-8}

\cite{10670007}  & \multicolumn{1}{c|}{} & \multicolumn{1}{c|}{Multi-BD-RIS} & \multicolumn{1}{c|}{MU-MISO-DL} & \multicolumn{1}{c|}{Block diagonal} & \multicolumn{1}{c|}{Reflective} & \multicolumn{1}{c|}{Fully-connected} & Sum-rate \\ \cline{1-1} \cline{3-8}

\cite{10472097} & \multicolumn{1}{c|}{} & \multicolumn{1}{c|}{\begin{tabular}[c]{@{}c@{}} Coordinated \\ BD-RIS \end{tabular}} & \multicolumn{1}{c|}{SU-MISO-DL} & \multicolumn{1}{c|}{Non-diagonal} & \multicolumn{1}{c|}{Reflective} & \multicolumn{1}{c|}{Group-connected} & Power gain \\ \cline{1-1} \cline{3-8}

\cite{10159457} & \multicolumn{1}{c|}{} & \multicolumn{1}{c|}{\begin{tabular}[c]{@{}c@{}} Dynamic \\ grouping \\ strategy  \end{tabular}} & \multicolumn{1}{c|}{MU-MISO-DL} & \multicolumn{1}{c|}{\begin{tabular}[c]{@{}c@{}} Permuted \\ block diagonal \end{tabular}} & \multicolumn{1}{c|}{Hybrid} & \multicolumn{1}{c|}{\begin{tabular}[c]{@{}c@{}} Cell-wise\\ dynamically \\group-connected \end{tabular}} & Sum-rate \\ \cline{1-1} \cline{3-8}

\cite{10526360}  & \multicolumn{1}{c|}{} & \multicolumn{1}{c|}{\begin{tabular}[c]{@{}c@{}} Static grouping \\ strategy \end{tabular}} & \multicolumn{1}{c|}{\begin{tabular}[c]{@{}c@{}} SU-MISO-DL \\ MU-MISO-DL \end{tabular}} & \multicolumn{1}{c|}{Block diagonal} & \multicolumn{1}{c|}{Reflective} & \multicolumn{1}{c|}{Group-connected} & \begin{tabular}[c]{@{}c@{}} Channel gain, \\ sum-rate \end{tabular} \\ \cline{1-1} \cline{3-8}

\cite{10902602}  & \multicolumn{1}{c|}{} & \multicolumn{1}{c|}{\begin{tabular}[c]{@{}c@{}} Lossy \\ interconnections \\ and distributed \\ BD-RIS \end{tabular}} & \multicolumn{1}{c|}{SU-MIMO-DL} & \multicolumn{1}{c|}{Block diagonal} & \multicolumn{1}{c|}{Reflective} & \multicolumn{1}{c|}{Single-/Fully-connected} & Received power \\ \cline{1-1} \cline{3-8}

\cite{10418928} & \multicolumn{1}{c|}{} & \multicolumn{1}{c|}{Mutual coupling} & \multicolumn{1}{c|}{SU-MIMO-DL} & \multicolumn{1}{c|}{Block diagonal} & \multicolumn{1}{c|}{Hybrid} & \multicolumn{1}{c|}{\begin{tabular}[c]{@{}c@{}} Cell-wise\\ group-/fully-connected  \end{tabular}} & Channel gain \\ \cline{1-1} \cline{3-8}

\cite{nerini2024global} & \multicolumn{1}{c|}{} & \multicolumn{1}{c|}{Mutual coupling} & \multicolumn{1}{c|}{SU-SISO-DL} & \multicolumn{1}{c|}{Block diagonal} & \multicolumn{1}{c|}{Reflective} & \multicolumn{1}{c|}{Fully-/Tree-connected} & Channel gain \\ \cline{1-1} \cline{3-8}

\cite{li2024non} & \multicolumn{1}{c|}{} & \multicolumn{1}{c|}{\begin{tabular}[c]{@{}c@{}} Non-reciprocal \\ BD-RIS \end{tabular}} & \multicolumn{1}{c|}{MU-MIMO-DL/UL} & \multicolumn{1}{c|}{Block diagonal} & \multicolumn{1}{c|}{Reflective} & \multicolumn{1}{c|}{Fully-/Group-connected} & Rate \\ \cline{1-1} \cline{3-8}

\cite{liu2024non} & \multicolumn{1}{c|}{} & \multicolumn{1}{c|}{\begin{tabular}[c]{@{}c@{}} Non-reciprocal \\ BD-RIS \end{tabular}} & \multicolumn{1}{c|}{MU-MIMO-DL/UL} & \multicolumn{1}{c|}{Block diagonal} & \multicolumn{1}{c|}{Reflective} & \multicolumn{1}{c|}{Fully-/Group-connected} & Sum-rate \\ \cline{1-1} \cline{3-8}

\cite{10197228}  & \multicolumn{1}{c|}{} & \multicolumn{1}{c|}{\begin{tabular}[c]{@{}c@{}} Discrete-value \\ scattering matrix \end{tabular}} & \multicolumn{1}{c|}{\begin{tabular}[c]{@{}c@{}} SU-SISO-DL \\ SU-MIMO-DL \end{tabular}} & \multicolumn{1}{c|}{Block diagonal} & \multicolumn{1}{c|}{Reflective} & \multicolumn{1}{c|}{\begin{tabular}[c]{@{}c@{}} Fully-/Group-connected \end{tabular}} & Received power \\ \cline{1-1} \cline{3-8}

\cite{9737373} & \multicolumn{1}{c|}{} & \multicolumn{1}{c|}{\begin{tabular}[c]{@{}c@{}} Non-diagonal \\scattering matrix \end{tabular}} & \multicolumn{1}{c|}{\begin{tabular}[c]{@{}c@{}} SU-SISO-DL \\ SU-MISO-DL \\ MU-MIMO-DL \end{tabular} } & \multicolumn{1}{c|}{Non-diagonal} & \multicolumn{1}{c|}{Reflective} & Single-connected & \begin{tabular}[c]{@{}c@{}} Achievable rate, \\ channel gain, \\ OP, and BER \end{tabular}\\ \cline{1-1} \cline{3-8}

\cite{nerini2024dual} & \multicolumn{1}{c|}{} & \multicolumn{1}{c|}{\begin{tabular}[c]{@{}c@{}} Dual-polarized \\BD-RIS \end{tabular}} & \multicolumn{1}{c|}{SU-SISO-DL} & \multicolumn{1}{c|}{Block diagonal} & \multicolumn{1}{c|}{Reflective} & \begin{tabular}[c]{@{}c@{}} Single-/Fully-/Group-/\\Tree-connected \end{tabular} & Received power \\ \cline{1-1} \cline{3-8}

\cite{zhou2024novel} & \multicolumn{1}{c|}{} & \multicolumn{1}{c|}{\begin{tabular}[c]{@{}c@{}} Q-stem connected \\ architecture \end{tabular}} & \multicolumn{1}{c|}{MU-MISO-DL} & \multicolumn{1}{c|}{Block diagonal} & \multicolumn{1}{c|}{Reflective} & Q-stem connected & Channel gain \\ \cline{1-1} \cline{3-8}

\cite{10403525} & \multicolumn{1}{c|}{} & \multicolumn{1}{c|}{\begin{tabular}[c]{@{}c@{}} Channel estimation  \end{tabular}} & \multicolumn{1}{c|}{SU-MISO-DL} & \multicolumn{1}{c|}{Block diagonal} & \multicolumn{1}{c|}{Reflective} & \multicolumn{1}{c|}{Group-connected} & MSE \\ \cline{1-1} \cline{3-8}

\cite{10587164}  & \multicolumn{1}{c|}{} & \multicolumn{1}{c|}{\begin{tabular}[c]{@{}c@{}} Channel estimation \end{tabular}} & \multicolumn{1}{c|}{\begin{tabular}[c]{@{}c@{}} SU-MIMO-DL \\ MU-MISO-DL \end{tabular} } & \multicolumn{1}{c|}{Block diagonal} & \multicolumn{1}{c|}{\begin{tabular}[c]{@{}c@{}} Reflective/\\ Hybrid/\\ Multi-sector \end{tabular} }& \multicolumn{1}{c|}{\begin{tabular}[c]{@{}c@{}} Group-connected \\ and cell-wise \\ group-connected \end{tabular}} & MSE, SE \\ \cline{1-1} \cline{3-8}

\cite{de2024channel}  & \multicolumn{1}{c|}{} & \multicolumn{1}{c|}{\begin{tabular}[c]{@{}c@{}} Channel estimation \end{tabular}} & \multicolumn{1}{c|}{SU-MISO-DL} & \multicolumn{1}{c|}{Block diagonal} & \multicolumn{1}{c|}{Reflective} & \multicolumn{1}{c|}{Group-connected} & NMSE \\ \cline{1-1} \cline{3-8}

\cite{sokal2024decoupled}  & \multicolumn{1}{c|}{} & \multicolumn{1}{c|}{\begin{tabular}[c]{@{}c@{}} Channel estimation \end{tabular}} & \multicolumn{1}{c|}{SU-MIMO-DL} & \multicolumn{1}{c|}{Block diagonal} & \multicolumn{1}{c|}{Reflective} & \multicolumn{1}{c|}{Group-connected} & NMSE \\ \cline{1-1} \cline{3-8}

\cite{ginige2024efficient}  & \multicolumn{1}{c|}{} & \multicolumn{1}{c|}{\begin{tabular}[c]{@{}c@{}} Channel estimation \end{tabular}} & \multicolumn{1}{c|}{MU-MIMO-DL} & \multicolumn{1}{c|}{Block diagonal} & \multicolumn{1}{c|}{Reflective} & \multicolumn{1}{c|}{Fully-/Group-connected} & NMSE, Sum-rate \\ \hline

\end{tabular}%
}
\end{table*}

\begin{table*}[!p]
\centering
\vspace{-2em}
\caption{SUMMARY OF EXISTING WORKS IN BD-RIS-BASED SYSTEMS (CONT.)}
\vspace{-1em}
\label{tab: SUMMARY 2}
\resizebox{0.99\textwidth}{!}{%
\begin{tabular}{|c|ccc|ccc|c|}
\hline
\multirow{2}{*}{\textbf{{[}\#{]}}} & \multicolumn{3}{c|}{\textbf{System related content}} & \multicolumn{3}{c|}{\textbf{BD-RIS related content}} & \multirow{2}{*}{\textbf{\begin{tabular}[c]{@{}c@{}}Performance\\ metric(s)\end{tabular}}} \\ \cline{2-7}

& \multicolumn{1}{c|}{\textbf{\begin{tabular}[c]{@{}c@{}}Technical\\ content\end{tabular}}} & \multicolumn{1}{c|}{\textbf{\begin{tabular}[c]{@{}c@{}}Main focus \end{tabular}}} & \multicolumn{1}{c|}{\textbf{\begin{tabular}[c]{@{}c@{}}System  model\end{tabular}}} & \multicolumn{1}{c|}{\textbf{\begin{tabular}[c]{@{}c@{}}Scattering \\ matrix type\end{tabular}}} & \multicolumn{1}{c|}{\textbf{\begin{tabular}[c]{@{}c@{}}Transmission\\ mode\end{tabular}}} & \textbf{Architecture type} &  \\ \hline


\cite{9514409} & \multicolumn{1}{c|}{\multirow{35}{*}{\begin{tabular}[c]{@{}c@{}} BD-RIS \\  evaluation \\(covered in \\ Sec. V-B) \end{tabular}}} & \multicolumn{1}{c|}{\begin{tabular}[c]{@{}c@{}} Modeling and \\ Evaluation \\ of BD-RIS \\ architectures  \end{tabular}} & \multicolumn{1}{c|}{\begin{tabular}[c]{@{}c@{}} SU-SISO-DL \\ SU-MIMO-DL \end{tabular} } & \multicolumn{1}{c|}{Block diagonal} & \multicolumn{1}{c|}{Reflective} & \multicolumn{1}{c|}{\begin{tabular}[c]{@{}c@{}} Single-/Fully-/Group-\\connected \end{tabular}} &  Received Power\\ \cline{1-1} \cline{3-8}

\cite{soleymani2024maximizing} & \multicolumn{1}{c|}{} & \multicolumn{1}{c|}{\begin{tabular}[c]{@{}c@{}} SE and EE\\ maximization  \end{tabular}} & \multicolumn{1}{c|}{MU-MIMO-DL} & \multicolumn{1}{c|}{Block diagonal} & \multicolumn{1}{c|}{Multi-sector} & \multicolumn{1}{c|}{\begin{tabular}[c]{@{}c@{}} Cell-wise\\ single-connected \end{tabular}} &  SE, EE \\ \cline{1-1} \cline{3-8}

\cite{10694505} & \multicolumn{1}{c|}{} & \multicolumn{1}{c|}{\begin{tabular}[c]{@{}c@{}} SE and EE \\ maximization  \end{tabular}} & \multicolumn{1}{c|}{MU-SISO-DL} & \multicolumn{1}{c|}{Block diagonal} & \multicolumn{1}{c|}{Multi-sector} & \multicolumn{1}{c|}{\begin{tabular}[c]{@{}c@{}} Cell-wise\\ single-connected \end{tabular}} & SE, EE \\ \cline{1-1} \cline{3-8}

\cite{10787237}  & \multicolumn{1}{c|}{} & \multicolumn{1}{c|}{\begin{tabular}[c]{@{}c@{}} Comprehensive \\performance \\analysis \end{tabular}} & \multicolumn{1}{c|}{MU-SISO-DL} & \multicolumn{1}{c|}{Block diagonal} & \multicolumn{1}{c|}{Multi-sector} & \multicolumn{1}{c|}{\begin{tabular}[c]{@{}c@{}} Cell-wise\\ single-connected \end{tabular}} & \begin{tabular}[c]{@{}c@{}} SNR, OP, SE,\\ EE, SEP,\\ diversity order \end{tabular} \\ \cline{1-1} \cline{3-8}

\cite{10319662}  & \multicolumn{1}{c|}{} & \multicolumn{1}{c|}{\begin{tabular}[c]{@{}c@{}} A low-complexity\\ BF design  \end{tabular}} & \multicolumn{1}{c|}{MU-MISO-DL} & \multicolumn{1}{c|}{Block diagonal} & \multicolumn{1}{c|}{Reflective} & \multicolumn{1}{c|}{\begin{tabular}[c]{@{}c@{}} Single-/Fully-/Group-\\connected \end{tabular}}& \begin{tabular}[c]{@{}c@{}} Channel gain, \\ sum-rate \end{tabular} \\ \cline{1-1} \cline{3-8}

\cite{10237233} & \multicolumn{1}{c|}{} & \multicolumn{1}{c|}{\begin{tabular}[c]{@{}c@{}} BD-RIS \\ performance \\ complexity \end{tabular}} & \multicolumn{1}{c|}{SU-SISO-DL} & \multicolumn{1}{c|}{\begin{tabular}[c]{@{}c@{}} Block diagonal \\ and \\ non-diagonal \end{tabular}} & \multicolumn{1}{c|}{Reflective} & \multicolumn{1}{c|}{\begin{tabular}[c]{@{}c@{}} Single-/Fully-/Group-\\/Tree-/Forest-connected \end{tabular}}  & Received power \\ \cline{1-1} \cline{3-8}

\cite{10155675} & \multicolumn{1}{c|}{} & \multicolumn{1}{c|}{\begin{tabular}[c]{@{}c@{}} Closed-form\\global \\optimization \end{tabular}} & \multicolumn{1}{c|}{\begin{tabular}[c]{@{}c@{}} SU-SISO-DL \\ SU-MISO-DL \\ SU-MIMO-DL \\ MU-MISO-DL \end{tabular}} & \multicolumn{1}{c|}{Block diagonal} & \multicolumn{1}{c|}{Reflective} & \multicolumn{1}{c|}{\begin{tabular}[c]{@{}c@{}} Fully-/Group-connected \end{tabular}} & Received power \\ \cline{1-1} \cline{3-8}

\cite{10364738}  & \multicolumn{1}{c|}{} & \multicolumn{1}{c|}{\begin{tabular}[c]{@{}c@{}} Power \\minimization and \\ EE maximization \end{tabular}} & \multicolumn{1}{c|}{MU-MISO-DL} & \multicolumn{1}{c|}{Block diagonal} & \multicolumn{1}{c|}{Reflective} & \multicolumn{1}{c|}{\begin{tabular}[c]{@{}c@{}} Fully-/Group-connected \end{tabular}} & \begin{tabular}[c]{@{}c@{}} Transmit power, \\ EE, sum-rate \end{tabular} \\ \cline{1-1} \cline{3-8}

\cite{wu2024optimization} & \multicolumn{1}{c|}{} & \multicolumn{1}{c|}{\begin{tabular}[c]{@{}c@{}} Power, sum-rate, \\ max-min rate, \\ max-min EE \\ optimization \end{tabular}} & \multicolumn{1}{c|}{\begin{tabular}[c]{@{}c@{}} MU-MISO-DL \\ MU-MIMO-DL \end{tabular}} & \multicolumn{1}{c|}{Block diagonal} & \multicolumn{1}{c|}{Reflective} & \multicolumn{1}{c|}{\begin{tabular}[c]{@{}c@{}} Fully-/Group-/\\Tree-connected \end{tabular}} & \begin{tabular}[c]{@{}c@{}} Transmit power, \\ EE, sum-rate \end{tabular} \\ \cline{1-1} \cline{3-8}

\cite{10187688} & \multicolumn{1}{c|}{} & \multicolumn{1}{c|}{\begin{tabular}[c]{@{}c@{}} SNR \\ maximization \end{tabular}} & \multicolumn{1}{c|}{\begin{tabular}[c]{@{}c@{}} SU-SISO-DL \\ SU-MISO-DL \\ SU-SIMO-UL \\ MU-SISO-UL \end{tabular}} & \multicolumn{1}{c|}{Block diagonal} & \multicolumn{1}{c|}{Reflective} & \multicolumn{1}{c|}{\begin{tabular}[c]{@{}c@{}} Fully-/Group-connected \end{tabular}} & SNR, sum-rate \\ \cline{1-1} \cline{3-8}

\cite{10694491} & \multicolumn{1}{c|}{} & \multicolumn{1}{c|}{\begin{tabular}[c]{@{}c@{}} Rate \\ maximization \end{tabular}} & \multicolumn{1}{c|}{SU-MIMO-DL} & \multicolumn{1}{c|}{Block diagonal} & \multicolumn{1}{c|}{Reflective} & \multicolumn{1}{c|}{Fully-connected} & Rate \\ \cline{1-1} \cline{3-8}

\cite{bjornson2024capacity} & \multicolumn{1}{c|}{} & \multicolumn{1}{c|}{\begin{tabular}[c]{@{}c@{}} Capacity \\ maximization \end{tabular}} & \multicolumn{1}{c|}{SU-MIMO-DL} & \multicolumn{1}{c|}{Block diagonal} & \multicolumn{1}{c|}{Reflective} & \multicolumn{1}{c|}{Fully-connected} & Capacity \\ \cline{1-1} \cline{3-8}

\cite{10694582} & \multicolumn{1}{c|}{} & \multicolumn{1}{c|}{\begin{tabular}[c]{@{}c@{}} Sum-rate \\ maximization  \end{tabular}} & \multicolumn{1}{c|}{MU-MISO-DL} & \multicolumn{1}{c|}{Block diagonal} & \multicolumn{1}{c|}{Reflective} & \multicolumn{1}{c|}{Fully-connected} & Sum-rate \\ \cline{1-1} \cline{3-8}

\cite{10839400} & \multicolumn{1}{c|}{} & \multicolumn{1}{c|}{\begin{tabular}[c]{@{}c@{}} Weighted \\ sum-rate \\ maximization  \end{tabular}} & \multicolumn{1}{c|}{MU-MISO-DL} & \multicolumn{1}{c|}{Block diagonal} & \multicolumn{1}{c|}{Reflective} & \multicolumn{1}{c|}{Fully-connected} & \begin{tabular}[c]{@{}c@{}} Weighted \\ sum-rate \end{tabular} \\ \cline{1-1} \cline{3-8}

\cite{10755162}  & \multicolumn{1}{c|}{} & \multicolumn{1}{c|}{Max-min rate} & \multicolumn{1}{c|}{MU-MISO-DL} & \multicolumn{1}{c|}{Block diagonal} & \multicolumn{1}{c|}{Reflective} & \multicolumn{1}{c|}{Fully-connected} & Max-min rate \\ \cline{1-1} \cline{3-8}

\cite{10694177}  & \multicolumn{1}{c|}{} & \multicolumn{1}{c|}{Max-min EE} & \multicolumn{1}{c|}{MU-MISO-DL} & \multicolumn{1}{c|}{Block diagonal} & \multicolumn{1}{c|}{Reflective} & \multicolumn{1}{c|}{Fully-connected} & Max-min EE \\ \hline


\cite{10302331} & \multicolumn{1}{c|}{\multirow{39}{*}{\begin{tabular}[c]{@{}c@{}} BD-RIS \\  integration with \\ emerging \\ technologies/\\schemes \\(covered in \\ Sec. V-C) \end{tabular}}} & \multicolumn{1}{c|}{mmWave} & \multicolumn{1}{c|}{MU-MIMO-DL} & \multicolumn{1}{c|}{Non-diagonal} & \multicolumn{1}{c|}{Reflective} & \multicolumn{1}{c|}{Fully-connected} & \begin{tabular}[c]{@{}c@{}} Worst-case \\ user-rate \end{tabular} \\ \cline{1-1} \cline{3-8}

\cite{10571253} & \multicolumn{1}{c|}{} & \multicolumn{1}{c|}{mmWave} & \multicolumn{1}{c|}{MU-MISO-DL} & \multicolumn{1}{c|}{Non-diagonal} & \multicolumn{1}{c|}{Reflective} & \multicolumn{1}{c|}{Fully-connected} & SE \\ \cline{1-1} \cline{3-8}

\cite{10817282}  & \multicolumn{1}{c|}{} & \multicolumn{1}{c|}{THz} & \multicolumn{1}{c|}{SU-MISO-DL} & \multicolumn{1}{c|}{Block diagonal} & \multicolumn{1}{c|}{Reflective} & \multicolumn{1}{c|}{Fully-connected} & SE \\ \cline{1-1} \cline{3-8}

\cite{10817342}  & \multicolumn{1}{c|}{} & \multicolumn{1}{c|}{THz} & \multicolumn{1}{c|}{MU-MISO-DL} & \multicolumn{1}{c|}{Block diagonal} & \multicolumn{1}{c|}{Hybrid} & \multicolumn{1}{c|}{Fully-connected} & Rate \\ \cline{1-1} \cline{3-8}

\cite{10693959} & \multicolumn{1}{c|}{} & \multicolumn{1}{c|}{ISAC} & \multicolumn{1}{c|}{MU-MISO-DL} & \multicolumn{1}{c|}{Block diagonal} & \multicolumn{1}{c|}{Transmissive} & \multicolumn{1}{c|}{Fully-connected} & Sum-rate, CRB \\ \cline{1-1} \cline{3-8}

\cite{10495009}  & \multicolumn{1}{c|}{} & \multicolumn{1}{c|}{ISAC} & \multicolumn{1}{c|}{MU-MISO-DL} & \multicolumn{1}{c|}{Block diagonal} & \multicolumn{1}{c|}{Reflective} & \multicolumn{1}{c|}{Fully-connected} & Throughput \\ \cline{1-1} \cline{3-8}

\cite{10493847}  & \multicolumn{1}{c|}{} & \multicolumn{1}{c|}{ISAC} & \multicolumn{1}{c|}{MU-MISO-DL} & \multicolumn{1}{c|}{Block diagonal} & \multicolumn{1}{c|}{Reflective} & \multicolumn{1}{c|}{Fully-connected} & Transmit power \\ \cline{1-1} \cline{3-8}

\cite{10777522} & \multicolumn{1}{c|}{} & \multicolumn{1}{c|}{ISAC} & \multicolumn{1}{c|}{MU-MISO-DL} & \multicolumn{1}{c|}{Block diagonal} & \multicolumn{1}{c|}{Reflective} & \multicolumn{1}{c|}{Fully-connected}  & SNR \\ \cline{1-1} \cline{3-8}

\cite{10333560}  & \multicolumn{1}{c|}{} & \multicolumn{1}{c|}{UAV} & \multicolumn{1}{c|}{MU-MISO-DL} & \multicolumn{1}{c|}{Block diagonal} & \multicolumn{1}{c|}{Reflective} & \multicolumn{1}{c|}{Fully-connected} & Sum-rate \\ \cline{1-1} \cline{3-8}

\cite{khan2024integration} & \multicolumn{1}{c|}{} & \multicolumn{1}{c|}{UAV} & \multicolumn{1}{c|}{MU-MISO-DL} & \multicolumn{1}{c|}{Block diagonal} & \multicolumn{1}{c|}{Transmissive} & \multicolumn{1}{c|}{Fully-connected} & SE \\ \cline{1-1} \cline{3-8}

\cite{Wali2025transmissive} & \multicolumn{1}{c|}{} & \multicolumn{1}{c|}{NTN} & \multicolumn{1}{c|}{MU-SISO-DL} & \multicolumn{1}{c|}{Block diagonal} & \multicolumn{1}{c|}{Transmissive} & \multicolumn{1}{c|}{Fully-connected} & SE \\ \cline{1-1} \cline{3-8} 

\cite{mahmood2023joint} & \multicolumn{1}{c|}{} & \multicolumn{1}{c|}{MEC} & \multicolumn{1}{c|}{MU-SIMO-UL} & \multicolumn{1}{c|}{Block diagonal} & \multicolumn{1}{c|}{Reflective} & \multicolumn{1}{c|}{\begin{tabular}[c]{@{}c@{}} Fully-/Group-connected \end{tabular}} & Latency, rate \\ \cline{1-1} \cline{3-8}

\cite{10643599} & \multicolumn{1}{c|}{} & \multicolumn{1}{c|}{Radar} & \multicolumn{1}{c|}{MU-MISO-DL/UL} & \multicolumn{1}{c|}{Block diagonal} & \multicolumn{1}{c|}{Hybrid} & \multicolumn{1}{c|}{\begin{tabular}[c]{@{}c@{}} Cell-wise\\ group-/fully-connected  \end{tabular}} & SCNR \\ \cline{1-1} \cline{3-8}

\cite{10693852} & \multicolumn{1}{c|}{} & \multicolumn{1}{c|}{Radar} & \multicolumn{1}{c|}{MU-MISO-DL/UL} & \multicolumn{1}{c|}{Block diagonal} & \multicolumn{1}{c|}{Hybrid} & \multicolumn{1}{c|}{\begin{tabular}[c]{@{}c@{}}  Cell-wise\\ single-/group-/fully-\\connected \end{tabular}} &  Weighted sum-rate \\ \cline{1-1} \cline{3-8}

\cite{10571002} & \multicolumn{1}{c|}{} & \multicolumn{1}{c|}{SWIPT} & \multicolumn{1}{c|}{MU-MISO-DL} & \multicolumn{1}{c|}{Block diagonal} & \multicolumn{1}{c|}{Reflective} & \multicolumn{1}{c|}{Fully-connected} & \begin{tabular}[c]{@{}c@{}} Ergodic SE, \\ harvested energy \end{tabular} \\ \cline{1-1} \cline{3-8}

\cite{azarbahram2025beyond} & \multicolumn{1}{c|}{} & \multicolumn{1}{c|}{WPT} & \multicolumn{1}{c|}{SU-SISO-DL} & \multicolumn{1}{c|}{Block diagonal} & \multicolumn{1}{c|}{Reflective} & \multicolumn{1}{c|}{Fully-connected} & Harvested power \\ \cline{1-1} \cline{3-8}

\cite{10288244}  & \multicolumn{1}{c|}{} & \multicolumn{1}{c|}{URLLC} & \multicolumn{1}{c|}{MU-MISO-DL} & \multicolumn{1}{c|}{Block diagonal} & \multicolumn{1}{c|}{Reflective} & \multicolumn{1}{c|}{Group-connected} & Rate, EE \\ \cline{1-1} \cline{3-8}

\cite{10755162}  & \multicolumn{1}{c|}{} & \multicolumn{1}{c|}{URLLC} & \multicolumn{1}{c|}{MU-MISO-DL} & \multicolumn{1}{c|}{Block diagonal} & \multicolumn{1}{c|}{Reflective} & \multicolumn{1}{c|}{Fully-connected} & Max-min rate \\ \cline{1-1} \cline{3-8}

\cite{9814551} & \multicolumn{1}{c|}{} & \multicolumn{1}{c|}{RSMA integration} & \multicolumn{1}{c|}{MU-MISO-DL} & \multicolumn{1}{c|}{Block diagonal} & \multicolumn{1}{c|}{Reflective} & \multicolumn{1}{c|}{Fully-/Group-connected} & Sum-rate \\ \cline{1-1} \cline{3-8}

\cite{10411856} & \multicolumn{1}{c|}{} & \multicolumn{1}{c|}{RSMA integration} & \multicolumn{1}{c|}{MU-MISO-DL} & \multicolumn{1}{c|}{Block diagonal} & \multicolumn{1}{c|}{Multi-sector} & \multicolumn{1}{c|}{\begin{tabular}[c]{@{}c@{}} Cell-wise\\ single-connected \end{tabular}} & Sum-rate\\ \cline{1-1} \cline{3-8}

\cite{khisa2024gradient} & \multicolumn{1}{c|}{} & \multicolumn{1}{c|}{RSMA integration} & \multicolumn{1}{c|}{MU-SIMO-UL} & \multicolumn{1}{c|}{Block diagonal} & \multicolumn{1}{c|}{Reflective} & \multicolumn{1}{c|}{Group-connected} & Sum-rate\\ \cline{1-1} \cline{3-8}

\cite{10742100} & \multicolumn{1}{c|}{} & \multicolumn{1}{c|}{NOMA integration} & \multicolumn{1}{c|}{\begin{tabular}[c]{@{}c@{}} MU-MISO-DL \\ SU-SISO-DL \end{tabular}} & \multicolumn{1}{c|}{Block diagonal} & \multicolumn{1}{c|}{Reflective} & \multicolumn{1}{c|}{Fully-/Group-connected} & \begin{tabular}[c]{@{}c@{}} Sum-rate, \\ asymptotic SNR \end{tabular} \\ \cline{1-1} \cline{3-8}

\cite{10623689} & \multicolumn{1}{c|}{} & \multicolumn{1}{c|}{\begin{tabular}[c]{@{}c@{}} Wideband system \end{tabular}} & \multicolumn{1}{c|}{SU-SISO-DL} & \multicolumn{1}{c|}{Block diagonal} & \multicolumn{1}{c|}{Reflective} & \multicolumn{1}{c|}{Fully-connected} & Capacity \\ \cline{1-1} \cline{3-8}

\cite{10694393} & \multicolumn{1}{c|}{} & \multicolumn{1}{c|}{\begin{tabular}[c]{@{}c@{}} Wideband system \end{tabular}} & \multicolumn{1}{c|}{SU-SISO-DL} & \multicolumn{1}{c|}{Block diagonal} & \multicolumn{1}{c|}{Reflective} & \multicolumn{1}{c|}{Group-connected} & Average rate \\ \cline{1-1} \cline{3-8}

\cite{10857964}  & \multicolumn{1}{c|}{} & \multicolumn{1}{c|}{\begin{tabular}[c]{@{}c@{}} Wideband system \end{tabular}} & \multicolumn{1}{c|}{SU-SISO-DL} & \multicolumn{1}{c|}{\begin{tabular}[c]{@{}c@{}} Block diagonal \\ and \\ non-diagonal \end{tabular}} & \multicolumn{1}{c|}{Reflective} & \multicolumn{1}{c|}{\begin{tabular}[c]{@{}c@{}} Group-/Forest-connected \end{tabular}} & Average rate \\ \cline{1-1} \cline{3-8}

\cite{soleymani2024maximizing} & \multicolumn{1}{c|}{} & \multicolumn{1}{c|}{\begin{tabular}[c]{@{}c@{}} Wideband system  \end{tabular}} & \multicolumn{1}{c|}{MU-MIMO-DL} & \multicolumn{1}{c|}{Block diagonal} & \multicolumn{1}{c|}{Multi-sector} & \multicolumn{1}{c|}{\begin{tabular}[c]{@{}c@{}} Cell-wise \\ single-connected \end{tabular}} &  SE, EE \\ \cline{1-1} \cline{3-8}

\cite{10694582} & \multicolumn{1}{c|}{} & \multicolumn{1}{c|}{\begin{tabular}[c]{@{}c@{}} Wideband system  \end{tabular}} & \multicolumn{1}{c|}{MU-MISO-DL} & \multicolumn{1}{c|}{Block diagonal} & \multicolumn{1}{c|}{Reflective} & \multicolumn{1}{c|}{Fully-connected} & Sum-rate \\ \cline{1-1} \cline{3-8}

\cite{loli2024meta}  & \multicolumn{1}{c|}{} & \multicolumn{1}{c|}{\begin{tabular}[c]{@{}c@{}} Large-scale systems \end{tabular}} & \multicolumn{1}{c|}{MU-MISO-DL} & \multicolumn{1}{c|}{Block diagonal} & \multicolumn{1}{c|}{Reflective} & \multicolumn{1}{c|}{Fully-connected} & Sum-rate \\ \cline{1-1} \cline{3-8}

\cite{10834443} & \multicolumn{1}{c|}{} & \multicolumn{1}{c|}{\begin{tabular}[c]{@{}c@{}} Vehicular networks  \end{tabular}} & \multicolumn{1}{c|}{MU-SISO-DL} & \multicolumn{1}{c|}{Block diagonal} & \multicolumn{1}{c|}{Reflective} & \multicolumn{1}{c|}{Fully-/Group-connected} & Sum-rate \\ \hline 

\end{tabular}%
}
\end{table*}

\section{BD-RIS-Based Wireless Communications} \label{sec: BD-RIS-Based Wireless Communications}

The existing technical works that investigated \gls{BD-RIS} up till the time of preparing this article are discussed in this section. Tables~\ref{tab: SUMMARY 1} and~\ref{tab: SUMMARY 2} provide a comprehensive summary of the various technical contributions that investigated \gls{BD-RIS}. For ease of referencing, the headings of this section are categorized based on the ``Technical content'' column of Tables~\ref{tab: SUMMARY 1} and~\ref{tab: SUMMARY 2}. Specifically, one can classify all the technical contributions that investigated \gls{BD-RIS} into five main areas, namely, (i) contributions that delve into the \gls{BD-RIS} modeling (Covered in Section~\ref{sec: Fundamentals of BD-RIS}), (ii) contributions that explore the potential applications of \gls{BD-RIS} (Covered in Section~\ref{sec: Potential Areas of Applications of BD-RIS}), (iii) contributions that develop new and/or enhance existing \gls{BD-RIS} architectures (Covered in Section~\ref{subsection: BD-RIS Architectural Development}), (ii) contributions that evaluate the performance of \gls{BD-RIS} (Covered in Section~\ref{subsection: BD-RIS Evaluation}), and (iii) contributions that integrate \gls{BD-RIS} in emerging technologies/schemes (Covered in Section~\ref{subsection: BD-RIS Integration with Emerging Technologies/Schemes}).

\subsection{BD-RIS Architectural Development}
\label{subsection: BD-RIS Architectural Development}

Beyond the discussion, mentioned in Section~\ref{sec: Classification of BD-RIS}, concerning the default \gls{BD-RIS} architectures, researchers have (i) developed new \gls{BD-RIS} architectures (e.g., new multi-sector \gls{BD-RIS} architectures shown in Table~\ref{tab: Layer charts 2}(j) and (k) and Table~\ref{tab: Layer charts 3}(b) and (c)), (ii) investigated \gls{BD-RIS} with \gls{RF} impairments (e.g., \gls{BD-RIS} with mutual coupling and lossy interconnections), and (iii) proposed solutions to facilitate the \gls{BD-RIS} operation (i.e., channel estimation-related solutions). A detailed discussion of such aspects is provided in this subsection.

When it comes to new \gls{BD-RIS} architectures, several implementations were investigated by researchers to improve the performance of \gls{BD-RIS}. First, the implementation of \gls{BD-RIS} at the transmitter was proposed in~\cite{10693959} and~\cite{10530995}, as shown in Table~\ref{tab: Layer charts 2}(a) and (b), respectively. The use of \gls{BD-RIS} at the transmitter helps alleviate the need for multiple fully digital \gls{RF} chains, which are typically required for high \gls{BF} gain in \gls{mmWave} systems~\cite{10308579}. This reduction lowers both hardware complexity and power consumption of wireless transmitters. 

Second, four new multi-sector \gls{BD-RIS} architectures were proposed in~\cite{10643263} (as shown in Table~\ref{tab: Layer charts 2}(j) and (k) and Table~\ref{tab: Layer charts 3}(b) and (c)). The first architecture is called ``Opposite-link with horizontally-connected sectors \gls{BD-RIS}'' and is shown in Table~\ref{tab: Layer charts 2}(j). In this architecture, the \gls{RIS} elements from sectors opposite to each other are linked. For instance, if there are six sectors, sector $1$ is connected to sector $4$, sector $2$ is linked to sector $5$, and sector $3$ is associated with sector $6$. This design ensures that users can be served even when they are located in a sector opposite to the \gls{BS}. The main merit of this architecture is that it simplifies the circuit while ensuring connectivity for users located in sectors not directly facing the \gls{BS}. The second architecture is called ``Full-link with horizontally-connected sectors \gls{BD-RIS}'' and is shown in Table~\ref{tab: Layer charts 2}(k). In this architecture, opposite to the first architecture, all sectors are connected, allowing any sector to communicate with another. This ensures that users can be served irrespective of their location within the coverage area. The merit of this architecture lies in its flexibility and higher performance as it can cover a wider range of user locations compared to the first architecture, though it comes with increased circuit complexity. The third architecture is called ``Opposite-link with horizontally-connected sectors and vertically connected elements in each sector \gls{BD-RIS}'' and is shown in Table~\ref{tab: Layer charts 3}(b). Compared to the first one, this architecture introduces vertical non-diagonal connections within groups of \gls{RIS} elements. This non-diagonal setup allows for more flexibility as the signal impinging on one element can be reflected from another. The merit of this architecture is that it enhances performance by leveraging intra-group flexibility while maintaining a relatively simple circuit design compared to fully connected architectures. The fourth architecture is called ``Full-link with horizontally-connected sectors and vertically connected elements in each sector \gls{BD-RIS}'' and is shown in Table~\ref{tab: Layer charts 3}(c). This architecture combines the full-link approach of the second architecture with non-diagonal vertical connections within groups of \gls{RIS} elements. The non-diagonal nature enables even more sophisticated signal routing, further enhancing performance. The main merit of this architecture is that it achieves nearly the same performance as fully connected architectures but with significantly reduced complexity and energy consumption. The four aforementioned multi-sector \gls{BD-RIS} architectures attain a balance between performance, coverage, and circuit complexity, making them suitable for different deployment scenarios~\cite{10643263}.

Third, a line of research has focused on optimizing the element connection pattern to improve the performance of the \gls{BD-RIS}-aided systems. For instance, in~\cite{10472097}, a coordinated \gls{BD-RIS} architecture was proposed to combat channel fading, where both the configurable impedances and the element connection pattern can be optimized. The coordinated \gls{BD-RIS}-aided system outperforms both single-connected and group-connected \gls{BD-RIS}-aided systems in terms of power gain. Also, as the number of reflecting elements increases, its power gain reaches that of the fully-connected \gls{BD-RIS}, but with a significantly smaller number of adjustable impedance settings. In~\cite{10159457, 10526360}, dynamic and static grouping strategies for group-connected \gls{BD-RIS} were proposed, respectively. Specifically, in~\cite{10159457}, using an impedance-switch network that can alter the interconnections between \gls{BD-RIS} groups, as illustrated in Table~\ref{tab: Layer charts 2}(h), the best grouping strategy was obtained with high circuit complexity and control overhead. In~\cite{10526360}, a static grouping strategy based on the channel statistics was proposed. This grouping strategy aims to enhance the achievable rate of group-connected \gls{BD-RIS} without increasing the circuit complexity or requiring additional control overhead.

Fourth, some researchers have investigated distributed \gls{BD-RIS}-aided systems~\cite{10670007, 10902602}. Usually, conventional \gls{BD-RIS}, also referred to as ``localized \gls{BD-RIS}''~\cite{10902602}, is deployed in the channel between the \gls{Tx} and the \gls{Rx}. Depending on the application, the deployment locations are either closer to the \gls{Tx}, in the middle between the \gls{Tx} and the \gls{Rx}, or closer to the \gls{Rx}. On the contrary, distributed \gls{BD-RIS} has two main implementations, either (i) several \glspl{BD-RIS} are deployed at different locations between the \gls{Tx} and the \gls{Rx} or (ii) a long \gls{BD-RIS} is utilized, which cover all the distance between the \gls{Tx} and the \gls{Rx}, with a spacing between elements that can be much longer than the wavelength (See Table~\ref{tab: Layer charts}(f)). Deploying multiple \glspl{BD-RIS} between \gls{Tx} and \gls{Rx} can provide several benefits in terms of coverage, channel quality, capacity, energy efficiency, security, and flexibility~\cite{10670007}. The distributed \gls{BD-RIS} offers also several benefits over the localized \gls{BD-RIS} by providing superior performance, particularly through its ability to guide electromagnetic signals within the \gls{BD-RIS} circuit, resulting in higher signal gains and better coverage, especially in highly obstructed environments~\cite{10902602}. 

Fifth, \gls{BD-RIS} enables advanced inter-element connections that produce non-diagonal scattering matrices, facilitating sophisticated signal control. A reciprocal \gls{BD-RIS} features symmetric matrices where the surface behavior remains consistent regardless of the signal's direction of incidence. This limits its ability to simultaneously optimize different communications paths, making it effective only when uplink and downlink users are aligned or share similar channel conditions. On the other hand, non-reciprocal \gls{BD-RIS} introduces an asymmetry in the scattering matrix, allowing the surface to respond differently depending on the direction of incoming waves. This unique property enables more flexible and effective handling of full-duplex scenarios where uplink and downlink users are in distinct locations, allowing both to communicate optimally with the same base station. In~\cite{li2024non}, the authors highlighted that the non-reciprocal \gls{BD-RIS} achieves significant performance gains in full-duplex scenarios by overcoming the constraints of symmetric reciprocity, which reciprocal \gls{BD-RIS} cannot address unless specific user alignments are met. However, these advantages come with challenges, such as complex circuit designs (to support non-reciprocity) and the need to account for structural scattering effects (that can impact the signal performance). Through theoretical analysis and simulations, this work demonstrated the superior adaptability of non-reciprocal \gls{BD-RIS}, showcasing its potential for enhancing channel gains and expanding communications coverage in full-duplex systems. Later, the work in~\cite{li2024non} was extended and generalized by~\cite{liu2024non} while focusing on sum-rate maximization perspective of non-reciprocal \gls{BD-RIS} full-duplex communications.

Sixth, the authors of~\cite{nerini2024dual} explored the dual-polarized \gls{BD-RIS}-aided systems compared to the literature that primarily analyzed \gls{BD-RIS}-aided uni-polarized systems. Modern \gls{MIMO} wireless systems employ dual-polarized antenna arrays to maximize the number of antennas in constrained spaces and enhance diversity by leveraging the polarization dimension~\cite{5626934}. In~\cite{nerini2024dual}, the authors studied the scenarios where the \gls{Tx}-\gls{Rx} pair has the same polarization or an opposite polarization and assumed that half of the \gls{BD-RIS} elements have vertical polarizations and the other half have horizontal polarizations. It is concluded that the \gls{BD-RIS} architecture always provides a better gain than the \gls{D-RIS} counterpart regardless of whether or not the \gls{Tx}-\gls{Rx} pair has the same polarization or opposite polarization.

Seventh, in~\cite{zhou2024novel}, a new circuit topology for \gls{BD-RIS} structures that strikes a favorable complexity-performance trade-off in multi-user systems was proposed and named `Q-stem connected' topology. This new topology attains the performance of the fully connected topology with a reduced circuit complexity. Also, Q-stem connected topology is a general \gls{BD-RIS} topology, which can be reduced to fully connected \gls{BD-RIS}, tree connected \gls{BD-RIS}, and single connected \gls{RIS} by adjusting the parameter `Q'.

To ensure the reliable operation of \gls{BD-RIS} in practical systems, researchers have investigated the performance of \gls{BD-RIS} under \gls{RF} impairments (i.e., under mutual coupling~\cite{10418928,nerini2024global} and lossy impedance interconnections~\cite{10902602} between \gls{RIS} elements) and with discrete-value scattering matrix. Mutual coupling refers to the phenomenon where the electromagnetic field of one \gls{RIS} element influences the electromagnetic field of another \gls{RIS} element. Subsequently, the mutual coupling depends on the distances between the \gls{RIS} elements. Specifically, the smaller the inter-element distance, the higher the effect of the mutual coupling. Mutual coupling can have both positive and negative effects on the overall performance of \gls{BD-RIS}-aided systems. On the positive side, small spacing between the individual reflecting elements allows for packing a large number of elements in \gls{BD-RIS} structure, potentially resulting in enhanced \gls{BF} capabilities. However, accounting for mutual coupling can significantly increase the computational complexity of system design~\cite{10418928,nerini2024global}. The lossy impedance interconnections refer to the electrical connections between the individual \gls{RIS} elements that introduce energy loss. This loss can be due to factors such as resistance, inductance, or capacitance within the interconnection network. Considering the effect of lossy interconnections can lead to a more realistic and practical model for BD-RIS systems. The lossy interconnections introduce additional challenges in designing and optimizing the RIS, as they can reduce the overall efficiency and limit the achievable \gls{BF} performance~\cite{10902602}. In~\cite{10197228}, a challenge of practical implementation where the impedance matrix values must be discretized, rather than treated as continuous, was investigated. Two primary solutions were proposed to achieve this, namely, scalar-discrete \glspl{RIS} and vector-discrete \glspl{RIS}. The scalar-discrete \gls{RIS} solution discretizes each entry of the \gls{RIS} impedance matrix independently, using an offline-designed codebook to minimize distortion from the discretization process. While this method is simple and has low optimization complexity, it requires more resolution bits to approach the performance of continuous-value \glspl{RIS}. On the other hand, the vector-discrete \gls{RIS} solution jointly discretizes groups of impedance matrix entries, resulting in higher performance with fewer resolution bits, but at the cost of increased computational complexity.

To facilitate the operation of \gls{BD-RIS}, several researchers have proposed channel estimation solutions for \gls{BD-RIS}-aided systems~\cite{10403525, 10587164, de2024channel, sokal2024decoupled, ginige2024efficient}. In~\cite{10403525}, a \gls{LS}-based channel estimation method was proposed for a \gls{MISO} system with a single-antenna receiver assisted by a group-connected \gls{BD-RIS}. The proposed \gls{LS}-based channel estimation method relies solely on the variation of \gls{BD-RIS} matrix. Later, the work of~\cite{10403525} is extended in~\cite{10587164} to a multi-user \gls{MIMO} system while relying also on the \gls{LS}-based channel estimation method. The authors of~\cite{10587164} introduced a \gls{BF} design based on the channel estimates and showed the channel estimation training overhead and transmission performance trade-off. In~\cite{de2024channel}, novel channel estimation methods that effectively exploit the tensor structure of the received pilot signals were proposed. By formulating the combined channel as a Tucker tensor decomposition, the proposed methods decouple the estimation of individual channel matrices, leading to accurate \gls{CSI} acquisition with minimal training overhead. The first solution provides a closed-form solution by solving parallel rank-one matrix approximation problems. The second solution employs an iterative estimation procedure to estimate the individual channels directly. Both methods significantly outperform conventional \gls{LS} estimation while requiring significantly less training data, demonstrating the advantages of leveraging the tensor signal structure for channel estimation in \gls{BD-RIS} systems. In~\cite{sokal2024decoupled}, the authors proposed a decoupled channel estimation method that first utilized the \gls{LS} estimate of the combined \gls{BD-RIS} channel from the method in~\cite{10587164}, and the individual \gls{BD-RIS} channels are revealed by exploiting the Kronecker decomposition structure of the combined \gls{BD-RIS} channel and some rank-one matrix approximation. In~\cite{ginige2024efficient}, joint channel estimation and prediction strategies for \gls{BD-RIS}-aided \gls{MIMO} system with both channel aging and correlated fast-fading environments were proposed. To decompose the \gls{BD-RIS} cascaded channels into effective channels of reduced dimension, a Tucker2 decomposition with bilinear alternative \gls{LS} was utilized. With the aid of \gls{CNN} and an autoregressive predictor, a channel prediction framework was advised. Then, a sum-rate maximization problem that optimizes the \gls{BD-RIS} phase shifts while utilizing the estimated/predicted \gls{CSI} was formulated and solved. The authors demonstrated through simulations the robustness of the proposed approach to channel aging with low pilot overhead and high estimation accuracy.

\subsection{BD-RIS Evaluation}
\label{subsection: BD-RIS Evaluation}

The researchers evaluated the performance of \gls{BD-RIS} from different perspectives, namely, from (i) an optimization perspective, (ii) a performance analysis perspective, and (iii) a circuit complexity perspective. A detailed discussion of such aspects is provided in this subsection.

In the literature, the performance of \gls{BD-RIS} was optimized under different performance  metrics, namely, the \gls{SNR} maximization~\cite{10187688}, the rate maximization~\cite{10694491}, the capacity maximization~\cite{bjornson2024capacity}, the sum-rate maximization~\cite{10694582,10319662,wu2024optimization}, the weighted sum-rate maximization~\cite{10839400}, the \gls{SE} and \gls{EE} maximization~\cite{soleymani2024maximizing,10694505}, the power minimization and \gls{EE} maximization~\cite{10364738, wu2024optimization}, the max-min rate~\cite{10755162, wu2024optimization}, and the max-min \gls{EE}~\cite{10694177, wu2024optimization}. For almost all these metrics and under various setups, there is consensus that the performance of \gls{BD-RIS} outperforms the performance of \gls{D-RIS} at the expanse of high circuit complexity. Conversely, in~\cite{10694177}, the authors found an exception to this rule for the minimum \gls{EE} metric for large \gls{BD-RIS} architectures. Specifically, at a certain point, the static power consumption of the circuitry in large \gls{BD-RIS} architectures overshadow the improvement in the rate performance compared to \gls{D-RIS} architecture, making the \gls{BD-RIS} architecture energy inefficient from the point-of-view of the \gls{EE} metric. Therefore, this is an important design factor that the researchers need to pay attention to while utilizing \gls{BD-RIS} architectures. In another optimization track, the work presented in~\cite{10155675} focused on developing closed-form global optimal solutions for the scattering matrix of the group- and fully-connected architectures. Also,~\cite{10155675} derived tight performance upper bounds for single-user \gls{MIMO} and multi-user \gls{MISO} fully-connected \gls{BD-RIS} systems with negligible direct links.

A comprehensive performance analysis for the multi-sector \gls{BD-RIS} was conducted in~\cite{10787237}. This work explores a multi-user communications system with \gls{BD-RIS} operating in time-switching mode, analyzing critical metrics such as the outage probability, \gls{SE}, \gls{EE}, and \gls{SEP}. The study derives closed-form expressions for the \gls{SNR} distributions and demonstrates that increasing the number of sectors within a \gls{BD-RIS} significantly enhances performance, particularly in terms of the \gls{SE} and the \gls{EE}. However, this sectorization leads to a trade-off, improving performance but reducing diversity order. The results show substantial gains of up to $182\%$ for the \gls{SE} and up to $238\%$ for the \gls{EE} when increasing the number of sectors from $2$ to $6$, highlighting the benefits of sectorization while also considering the complexity and hardware requirements involved in real-world deployments.

A Pareto frontier for the trade-off between performance and complexity in different \gls{BD-RIS} architectures was derived and characterized in~\cite{10237233}. This work aimed to determine the \gls{BD-RIS} architectures that are most suited for bridging the gap between single-connected \gls{RIS} and tree-connected \gls{BD-RIS}.

\subsection{BD-RIS Integration with Emerging Technologies/Schemes}
\label{subsection: BD-RIS Integration with Emerging Technologies/Schemes}

\Gls{BD-RIS}-assisted communications has emerged as a promising technology for improving the performance of wireless communications systems. Integrating \gls{BD-RIS} technology into wireless systems offers significant potential to support robust, reliable, and long-range data transmission, helping meet the requirements of \gls{B5G} and \gls{6G} wireless networks. This subsection outlines the integration of \gls{BD-RIS} with a range of emerging wireless technologies and schemes, namely, \gls{mmWave}/\gls{THz} communications~\cite{9264161,9569475}, \gls{ISAC}~\cite{9705498}, \gls{MEC}~\cite{10093902}, radar communications~\cite{8828016}, \gls{SWIPT} systems~\cite{8214104}, \gls{URLLC}~\cite{9826826}, the \gls{RSMA} and \gls{NOMA} schemes~\cite{10375270, 9154358}, wideband systems~\cite{7763821}, and large-scale systems~\cite{9330752}.

\subsubsection{mmWave/THz Communications} 

The integration of \gls{BD-RIS} in \gls{mmWave}/\gls{THz} communications were explored in~\cite{10302331, 10571253, 10817282, 10817342}. In~\cite{10302331}, a max-min fairness optimization problem was investigated for \gls{BD-RIS}-assisted \gls{mmWave} \gls{MIMO} system while assuming a hybrid \gls{BF} at both the \gls{BS} and the users and a passive \gls{BF} at the \gls{BD-RIS} structure. The obtained results demonstrated that, when deploying a large \gls{BD-RIS} structure, the proposed system outperforms the \gls{D-RIS} counterpart, in terms of the worst-case \gls{UE} rate, and approaches the performance of a half-duplex relay counterpart. Also, the \gls{EE} performance of the proposed system outperforms both the \gls{D-RIS} and half-duplex relay counterparts as well as the full-duplex relay counterpart at high \gls{SNR}. In~\cite{10571253}, a \gls{SE} maximization problem was investigated for \gls{BD-RIS}-assisted \gls{mmWave} \gls{MISO} system while considering both the \gls{BS} \gls{BF} matrix and \gls{BD-RIS} scattering matrix. The obtained results demonstrated that the \gls{SE} performance of the \gls{BD-RIS} structure is superior to the \gls{D-RIS} structure counterpart. A similar \gls{SE} maximization problem was investigated in~\cite{10817282} but for the \gls{THz} communications. In this work, the authors utilize a low-resolution digital beamformer at the \gls{BS} and a discrete phase-shift design at the \gls{BD-RIS}. In \gls{THz} communications, employing the fully digital \gls{BF} method at the \gls{BS} is impractical, due to the need for a large number of \gls{RF} chains for the operation of the massive number of antenna elements at the \gls{BS}. Also, the obtained results in~\cite{10817282} demonstrated that the \gls{SE} performance of the \gls{BD-RIS} structure is superior to the \gls{D-RIS} structure counterpart. In~\cite{10817342}, a hybrid \gls{BD-RIS} architecture employs a \gls{THz} communications system to serve both an indoor user and an outdoor user. A sum-rate maximization problem that jointly optimizes the phase shifts in the hybrid \gls{BD-RIS} architecture and the hybrid \gls{BF} vectors of the \gls{BS} was formulated and solved. The authors demonstrated through simulations the rate superiority of the proposed approach compared to time-division hybrid \gls{RIS}, frequency-division hybrid \gls{RIS}, and \gls{STAR-RIS} counterparts.

\subsubsection{ISAC Systems}

The integration of \gls{BD-RIS} in \gls{ISAC} systems were explored in~\cite{10693959, 10495009, 10493847, 10777522}. The authors of~\cite{10693959} studied a joint optimization problem that maximizes the sum rate for the communications users and minimizes the largest eigenvalue of the \gls{CRB} for the sensing targets. Simulation results showed that the \gls{BD-RIS}-aided system attains a better communications and sensing performance compared to the \gls{D-RIS} counterpart. In~\cite{10495009}, the maximization problem of the networks' throughput while maintaining the sensing quality was analyzed. The provided simulation results revealed the \gls{BD-RIS}-aided system superiority compared to the \gls{D-RIS} counterpart. The authors of~\cite{10493847} considered a \gls{BS}'s transmit power minimization problem while maintaining both the communications and sensing quality. Simulation results illustrated that the \gls{BD-RIS}-aided system can reduce the \gls{BS}'s transmit power by $25$\%-$30$\% compared to the \gls{D-RIS} counterpart. The authors of~\cite{10777522} optimized the weighted sum of the \gls{SNR} at the communications users and the sensing target. Simulation results revealed that, while employing a \gls{BD-RIS} structure, the \gls{SNR} values for both communications users and the sensing target improve by several orders of magnitude compared to the \gls{D-RIS} counterpart.

\subsubsection{UAV Communications}
The integration of \gls{BD-RIS} in \gls{UAV} communications was examined in~\cite{10333560,khan2024integration}. The representative work in~\cite{10333560} considered several \glspl{UAV} transmitting signals to groups of users with \gls{LoS} links and with \gls{NLoS} links through a \gls{BD-RIS} structure. It is assumed that each \gls{UAV} transmitter is serving a unique user group using \gls{OFDMA} and the users within each group are served using the \gls{RSMA} scheme. A sum-rate maximization problem that jointly considers \glspl{UAV} \gls{BF} design, \gls{BD-RIS} elements allocation to groups, \gls{BD-RIS} elements phase shifts, and the \gls{RSMA} common-rate allocation was investigated. Simulation results demonstrated that the sum-rate performance of the proposed \gls{BD-RIS}-aided \gls{RSMA}-assisted \gls{UAV} system outperforms the counterpart schemes of (i) an \gls{RSMA}-assisted \gls{D-RIS} scheme, (ii) a power-domain \gls{NOMA}-assisted \gls{D-RIS} scheme, and (iii) an \gls{RSMA}-assisted \gls{UAV} system without the \gls{RIS} technology. In~\cite{khan2024integration}, the authors highlighted the possible deployment locations of the \gls{BD-RIS} structures in \gls{UAV}-based systems. Notably, \gls{BD-RIS} structures can be deployed into; (i) terrestrial fixed structures such as buildings to enhance the signal strength and coverage between \gls{UAV} transmitters and their respective users, mainly in urban areas, (ii) aerial mobile units to resolve the \gls{NLoS} blockages and maintain optimal communications links between the ground stations and their respective users, and (iii)  a dual terrestrial-based and aerial-based configuration to achieve a robust system that can be suitable for a variety of use-cases. They also investigated an \gls{SE} maximization problem for a \gls{BD-RIS} mounted \gls{UAV} multi-user \gls{MISO} communications system. Through simulations, the authors illustrated the superiority of the investigated system in terms of \gls{SE} performance compared to the \gls{D-RIS} counterpart.   

\subsubsection{Non-Terrestrial Networks} 
The integration of \gls{BD-RIS} in \gls{NTN} was examined in~\cite{10716670,Wali2025transmissive}. In~\cite{10716670}, the authors investigated a multi-user \gls{NOMA}-enabled \gls{LEO} satellite communications system while employing a \gls{BD-RIS} structure to enhance the communications links between the \gls{LEO} satellite and its ground users. A sum-rate maximization problem was studied while ensuring the \gls{QoS} of the ground users. Simulation results revealed the superiority of the proposed \gls{BD-RIS}-aided system compared to the \gls{D-RIS} counterpart for different satellite transmit powers and different numbers of elements in the employed \gls{BD-RIS} structure. In~\cite{Wali2025transmissive}, a transmissive \gls{BD-RIS} was mounted on an \gls{LEO} satellite to aid the downlink \gls{NOMA} transmission to two ground \gls{IoT} devices. A \gls{SE} maximization problem that optimizes both the \gls{LEO} \gls{NOMA} power allocation and the phase shift of the transmissive \gls{BD-RIS} was formulated and solved. Through simulations, the authors demonstrated that the proposed optimized scheme outperforms the fixed \gls{NOMA} power allocation counterpart in terms of the system's \gls{SE}.

\subsubsection{MEC Networks}

In \gls{MEC} networks, numerous users offload their duties to the \gls{MEC} server due to restricted resources and short battery life, then the \gls{MEC} server performs the heavy computations and sends back the results to the users to take actions. The authors of~\cite{mahmood2023joint} developed a \gls{BD-RIS} structure mounted on \gls{UAV} to help the remote users in offloading their tasks to the \gls{MEC} server that is mounted at the \gls{BS}. A minimization problem for the utility function of task computational delay and \gls{UAV}'s hovering time was investigated. The optimization problem takes into account the location of the \gls{UAV}, the \gls{BD-RIS} phase shifts, the \gls{BS} receive \gls{BF} vector, the task segmentation variable, users transmission power, and the computational resources of the \gls{MEC} server and users. The results demonstrated that (i) fully-connected and group-connected architectures exhibit superior worst-case rate performance compared to the single-connected architecture, (ii) the optimal \gls{BD-RIS} deployment significantly enhances the system's utility function relative to the conventional building-mounted \gls{RIS}, (iii) the optimized offloading strategy achieves a 13.44\% improvement in the system's utility function over the binary offloading strategy.

\subsubsection{Radar Communications} 

\gls{DFRC} technology emerged to enable spectrum sharing between radar and communication, as radar systems compete for the same limited resources. The integration of communications and radar operations into a single platform in \gls{DFRC} systems results in increased spectrum efficiency, reduced power consumption, and lower hardware costs. As a result, it is anticipated that \gls{DFRC} will be crucial to the development of new environment-aware applications~\cite{9656537}, including smart homes, environmental monitoring, and automotive networks. The \gls{LoS} connectivity between the \gls{BS} and communications users/sensing targets are essential to the current designs of \gls{DFRC}. Unfortunately, this led to two challenges: (i) barriers can readily obstruct the \gls{LoS} link toward sensing targets or communications users; and (ii) considerable path-loss may occur in the \gls{LoS} channels, particularly at high frequencies. The deployment of \gls{BD-RIS} has the potential to resolve these challenges because \gls{BD-RIS} not only achieves full-space coverage but also offers better \gls{BF} capabilities than \gls{D-RIS}. 

A generic \gls{BD-RIS}-assisted \gls{DFRC} system was proposed in~\cite{10643599}, comprising a hybrid \gls{BD-RIS} structure, numerous users, and various sensing targets corrupted by multiple clutter sources. A maxmin radar output \gls{SCNR} optimization problem is formulated while considering the transmit waveform at the dual-function \gls{BS}, the transmissive and reflective \gls{BF} at the \gls{BD-RIS}, and matched filters at the radar sensing receivers. Simulation results showed that group- and fully-connected \gls{BD-RIS} structure exhibit superior radar \gls{SCNR} performance compared to the \gls{STAR-RIS} structure under identical communications constraints. Another \gls{BD-RIS}-assisted \gls{DFRC} system was proposed in~\cite{10693852} comprising also a hybrid \gls{BD-RIS} structure, indoor and outdoor users, and multiple far-located sensing targets. This work investigated different \gls{BD-RIS} structures, namely, the cell-wise single-/group-/fully connected structures. A maximization of users weighted sum-rate problem was considered while taking into account the transmit waveform at the dual-function \gls{BS} and the transmissive and reflective \gls{BF} at the \gls{BD-RIS}. The provided simulation results illustrated that the cell-wise fully connected structure achieves a higher weighted sum-rate performance compared to the \gls{STAR-RIS} structure for different numbers of \gls{RIS} elements and \gls{BS} transmit powers.

\subsubsection{SWIPT/WPT Systems}

In energy-constrained \gls{IoT} networks, \gls{SWIPT} presents a viable method for cost-effective power delivery~\cite{9775078}. A BS with a continuous power supply simultaneously broadcasts wireless signals to \glspl{IR} and \glspl{ER}. The main difficulty with \gls{SWIPT} systems is that different power supplies are needed for the \glspl{ER} and \glspl{IR} to function. \Glspl{ER} explicitly demand received power at a far higher order than \glspl{IR}. Because of the signal attenuation, the \glspl{ER}' practical operating range is limited. Hence to harvest enough power, they need to be deployed closer to the \gls{BS} than \glspl{IR}. The operational range of \glspl{ER} can be increased by deploying a \gls{BD-RIS} structure between the \gls{BS} and the \glspl{ER}. In~\cite{10571002}, a \gls{BD-RIS}-aided \gls{CF-mMIMO} \gls{SWIPT} system was proposed. An analytical framework that focuses on the ergodic \gls{SE} of the \glspl{IR} and the average harvested energy of the \glspl{ER} was advised. The conducted simulation results illustrated the superiority of the designed \gls{BD-RIS}-aided system compared to systems (i) with a random \gls{BD-RIS} design and (ii) without \gls{BD-RIS} structure. In~\cite{azarbahram2025beyond}, a multi-carrier \gls{SISO} \gls{WPT} system that employed a fully-connected \gls{BD-RIS} structure was proposed to mitigate the low end-to-end power transfer efficiency challenge. A receiver harvested power maximization problem that optimizes both the \gls{BS} \gls{BF} and the multi-carrier waveforms was formulated and solved. The authors revealed through simulations that the \gls{BD-RIS}-aided system achieved the same receiver harvested power performance as the \gls{D-RIS}-aided system under far-field \gls{LoS} conditions, while it outperforms \gls{D-RIS}-aided system under Rician pure \gls{NLoS} conditions.

\subsubsection{URLLC Systems} 

The integration of \gls{BD-RIS} in \gls{URLLC} systems was inspected in~\cite{10288244,10755162}. In~\cite{10288244}, the role of \gls{BD-RIS} and \gls{RSMA} in a \gls{MISO} multi-cell \gls{URLLC} system was studied. This work demonstrates that the impact of \gls{BD-RIS} on the \gls{RSMA} scheme varied with network load. Specifically, \gls{BD-RIS} can amplify the gains of the \gls{RSMA} scheme in overloaded regimes, whereas the benefits of the \gls{RSMA} scheme can be attenuated or even nullified by \gls{BD-RIS} optimization in underloaded regimes. Also, this work investigated the influence of packet length on the \gls{RSMA} scheme performance. Specifically, the performance gains of the \gls{RSMA} scheme are enhanced as packet lengths decrease or as the maximum tolerable decoding failure probability diminishes. Moreover, the combination of the \gls{RSMA} scheme and the group-connected \gls{BD-RIS} structure (group size $2$) in \gls{URLLC} systems can substantially improve the \gls{SE} and the \gls{EE} of the system. 

In~\cite{10755162}, the authors compared the performance of three \gls{RIS} architectures, namely, locally passive \gls{D-RIS}, globally passive \gls{D-RIS}, and globally passive \gls{BD-RIS}, in a multi-user \gls{MISO} \gls{URLLC} system. In locally passive \gls{D-RIS} architecture, i.e., conventional \gls{D-RIS} architecture, the \gls{D-RIS} reflection constraints are imposed independently on each individual reflecting element, while in globally passive \gls{D-RIS}/\gls{BD-RIS} architectures, the \gls{D-RIS} is defined by a unified reflection constraint that simultaneously affects all reflecting elements. This expanded the set of possible reflection matrices, which led to a better performance at the expanse of a higher computational complexity~\cite{10268023}. This work solved an average max-min rate optimization problem and its findings demonstrated that the max-min rate performance of globally passive \gls{BD-RIS} architecture outperforms (i) globally passive \gls{D-RIS}, (ii) locally passive \gls{D-RIS}, (iii) an un-optimized \gls{D-RIS}, i.e., random \gls{D-RIS}, and (iv) No \gls{D-RIS} architectures. 

\subsubsection{RSMA and NOMA Schemes} 

The integration of \gls{BD-RIS} with the \gls{RSMA} scheme was investigated in~\cite{9814551,10411856,khisa2024gradient}. A fully-connected \gls{BD-RIS} aided downlink multi-antenna multi-user \gls{RSMA} transmission paradigm was proposed in~\cite{9814551}. A sum-rate maximization problem that considers both the transmit beamformer of \gls{BS} and the scattering matrix of \gls{BD-RIS} was investigated. Through simulations, the authors demonstrated that the sum-rate performance of the fully-connected \gls{BD-RIS} outperforms both the counterparts with the \gls{D-RIS} structure and without \gls{RIS}. In~\cite{10411856}, the integration of multi-sector \gls{BD-RIS} and \gls{RSMA} in a multiuser-\gls{MISO} communications system was considered. A stochastic average sum-rate maximization problem was formulated while jointly considering the \gls{BS} transmit beamformer and the \gls{BD-RIS} matrix under the imperfect \gls{CSI} conditions. The simulation results exhibited an improved sum-rate performance compared to the \gls{SDMA} scheme. Furthermore, the results showed that the numbers of active and passive antennas, at the \gls{BS} and the \gls{BD-RIS} structure, respectively, can be effectively reduced with the combination of multi-sector \gls{BD-RIS} with the \gls{RSMA} scheme. In~\cite{khisa2024gradient}, a \gls{BD-RIS}-aided uplink \gls{RSMA} system was investigated. A sum-rate optimization problem that jointly considers the \gls{BS} receive beamformer, the \gls{BD-RIS} scattering matrix, and the \glspl{UE} transmit power was formulated and solved. Simulation results showed that the sum-rate performance of the proposed \gls{BD-RIS}-aided \gls{RSMA} system outperforms an \gls{RIS}-aided \gls{RSMA} counterpart system.

The integration of \gls{BD-RIS} with the power-domain \gls{NOMA} scheme was investigated in~\cite{10742100}. In~\cite{10742100}, an asymptotic received \gls{SNR} formula for a single-user \gls{SISO} \gls{NOMA}-enabled \gls{BD-RIS}-aided system was derived. Also, a sum-rate maximization problem for a multi-user \gls{MISO} \gls{NOMA}-enabled \gls{BD-RIS}-aided system was formulated and solved. Simulations results indicated that the \gls{BD-RIS} offers a superior square-law \gls{SNR} than \gls{D-RIS} and a higher sum-rate performance than both \gls{D-RIS} and traditional \gls{OMA}-enabled systems.

\subsubsection{Wideband Systems}

The integration of \gls{BD-RIS} in wideband systems was considered in~\cite{10623689,10694393,10857964,10694582,soleymani2024maximizing}. It should be noted that the modeling of the \gls{BD-RIS} structure in wideband applications is investigated in two directions. In~\cite{10623689,soleymani2024maximizing}, the authors modeled the \gls{BD-RIS} structure using a frequency-independent modeling method. While in~\cite{10857964,10694393,10694582}, the authors modeled the \gls{BD-RIS} using a frequency-dependent modeling method. The authors of~\cite{10857964,10694393} justified this by the fact that, in wideband applications, the scattering matrix of the \gls{BD-RIS} is better modeled using a reconfigurable admittance network, where the admittance of each tunable element has different values at different frequencies. The findings of~\cite{10857964,10694393} demonstrated that the frequency-independent model facilitates the \gls{BD-RIS} design, but at the expense of causing some performance loss to the \gls{BD-RIS}-aided wideband systems. Also, the importance of taking into account the frequency-dependency becomes more important when the circuit complexity of the \gls{BD-RIS} increases.

\subsubsection{Large-Scale Systems}

The integration of \gls{BD-RIS} with the large-scale systems was investigated in~\cite{loli2024meta}. This work introduced a meta-learning-based optimization framework aimed at optimizing \gls{BD-RIS}-aided large-scale wireless systems. Traditional optimization techniques are computationally impractical at large scales, so the proposed framework leverages unsupervised meta-learning to reduce complexity without needing extensive training data. Specifically, a sum-rate maximization problem that jointly takes into account the \gls{BS} transmit \gls{BF} and the phase-shifts of the \gls{BD-RIS} scattering matrix for a multi-user \gls{MISO} system was formulated and solved using the meta-learning-based optimization framework. Through simulations, the authors illustrated the performance gains of the evaluated \gls{BD-RIS}-aided system compared to the \gls{D-RIS} counterpart. For example, in an extreme case, the sum-rate performance of the evaluated \gls{BD-RIS}-aided system with $15$ antennas at the \gls{BS} outperforms a \gls{RIS}-aided system with $100$ antennas at the \gls{BS}. This shows the potential of the \gls{BD-RIS} structure in reducing the complexity and energy consumption at the wireless transmitters.

\subsubsection{Vehicular Networks}
The integration of \gls{BD-RIS} with vehicular networks was studied in~\cite{10834443}. This study investigated a multi-cell transportation network that was aided by a \gls{BD-RIS} to improve the \gls{SE} of the network. Also, the authors formulated and solved a sum-rate maximization problem that optimized both the \gls{BS} allocation power and the \gls{BD-RIS} phase shift design in each cell. Through simulations, the authors illustrated the superiority of the proposed \gls{BD-RIS}-aided transportation network in terms of sum-rate performance compared to the \gls{D-RIS}, \gls{STAR-RIS}, and sub-optimal optimization counterpart.

\subsection{Lessons Learnt}

There has been progress made in the research activities that focus on the \gls{BD-RIS}-based systems. Below are some key takeaways from the reviewed literature.
\begin{itemize}
    \item There are several architectural development investigations for \gls{BD-RIS} structures. These developments start by proposing different transmission modes, such as reflective, hybrid, or multi-sector to enhance the coverage in wireless systems. They also employed different circuit topologies, such as fully-/group-/tree-/forest-connected with and without cells to expand the wave-domain processing capabilities by increasing the number of optimization variables and the possible scattering matrices. Moreover, non-diagonal \gls{BD-RIS} structures that allow for precise control over the direction of the reflected signal, enhancing the overall performance and efficiency of the antenna system can be adopted. Furthermore, the location of deployment of \gls{BD-RIS} structures is not limited to the channel between the \gls{Tx} and \gls{Rx}. \Gls{BD-RIS} can also be deployed at the transmitter to reduce its hardware complexity and power consumption. Additionally, distributed \gls{BD-RIS} structures were proposed while either utilizing several \glspl{BD-RIS} between the \gls{Tx} and \gls{Rx} or a long \gls{BD-RIS} structure to achieve several benefits in terms of coverage, channel quality, capacity, energy efficiency, security, and flexibility. Knowing which structure to adopt for different use cases is a key design factor to benefit the most from the enhanced capabilities that \gls{BD-RIS} can provide. 
    \item To increase the practicality of \gls{BD-RIS} structures, researchers have investigated the performance of \gls{BD-RIS} under systems with non-ideal aspects, such as mutual coupling, lossy impedance interconnections, and discretized impedance values. To fully understand the potential and challenges of \gls{BD-RIS} structures in wireless systems, there is a need for examining the \gls{BD-RIS} performance under additional practical scenarios, for example, with imperfect \gls{CSI}, with transceiver hardware impairments, under generalized fading models that better fit real-time experimental data. 
    \item Compared to the \gls{D-RIS} structure, the optimization of \gls{BD-RIS} structures is subject to additional constraints, for example, the need for the scattering matrix of the \gls{BD-RIS} to be unitary, making the solution process more complex than that of the \gls{D-RIS} structure. A potential approach involves relaxing the strict unitary constraint, thereby simplifying the optimization problem. However, this relaxation can compromise performance, as the derived solution may deviate from the optimal solution of the original problem. Another approach involves employing the manifold optimization algorithm that represents all feasible solutions as a geometric shape (manifold). By doing this, the original constrained problem is converted into an unconstrained problem and subsequently can be efficiently solved through standard searching techniques with minor adjustments to account for the manifold's specific properties.
    \item \gls{BD-RIS} structures offer significant advantages in various wireless communications networks. By introducing additional degrees of freedom in phase shift control, \gls{BD-RIS} can enhance signal propagation, improve channel quality, and mitigate interference. This leads to improved spectral efficiency, energy efficiency, and link reliability. In \gls{mmWave}/\gls{THz} communications, \gls{BD-RIS} can compensate for severe path loss and \gls{BF} challenges. In \gls{ISAC} systems, \gls{BD-RIS} can enable reliable simultaneous sensing and communications by intelligently controlling the reflected signals. In \gls{UAV} communications, \gls{BD-RIS} can provide flexible and dynamic coverage, while in \gls{MEC} networks, it can enhance network capacity and offloading capabilities. In radar communications, \gls{BD-RIS} can improve target detection and tracking performance. In \gls{SWIPT} systems, \gls{BD-RIS} can enhance power transfer and information reception. In \gls{URLLC} systems, it can improve reliability and latency. Wideband systems benefit from \gls{BD-RIS}'s ability to mitigate inter-carrier interference. In large-scale systems, \gls{BD-RIS} can reduce the number of active antennas and improve energy efficiency. Overall, this synergy \gls{BD-RIS} structures and the aforementioned emerging technologies/schemes leads to significant gains in terms of coverage, capacity, and spectral efficiency. 
\end{itemize}

\section{Challenges and Future Research Directions in BD-RIS} 
\label{sec: Challenges And Future Research Directions In BD-RIS}

In this section, we discuss some key challenges and future research directions of \gls{BD-RIS}-aided wireless communications systems.

\subsection{Challenges}

Despite the benefits and potential applications of \gls{BD-RIS}, there exist some challenges with respect to its implementation, such as:

\subsubsection{CSI Estimation Acquisitions}
To take full advantage of the benefits provided by the \gls{BD-RIS} and to ensure coherent signal detection in the \gls{BD-RIS}-based wireless communications networks, there is a need for the provision of accurate \gls{CSI}. The semi-passive channel estimation strategy in which \gls{D-RIS} is equipped with a few low-power \gls{RF} chains to enable the pilot transmission/reception for channel estimation can still be employed in some \gls{BD-RIS}-based systems. However, this will be at the expense of additional power consumption due to the introduced \gls{RF} chains. Besides, there is a need for redesigning the \gls{BD-RIS} pattern for uplink training because of the different constraints on the scattering matrix, which yields different dimensions and structures of the cascaded channel. Thus, it is important to develop new channel estimation strategies with smaller estimation errors, lower overhead, and reduced power consumption for \gls{BD-RIS} in the near future.

\subsubsection{Joint Consideration of Non-ideal Aspects}
Suppose there are mismatches and mutual coupling of antennas, combined with lossy impedance components which is a reality in practical scenarios, then the channel model in the \gls{BD-RIS}-based system will no longer be a linear function of the scattering matrix as a result of these \gls{RF} impairments. Consequently, the \gls{BF} design will get more complicated.

\subsubsection{Continuous, Discrete, and Quantized Phase-shift Designs}

A promising avenue for future research involves a comprehensive investigation of phase-shift design strategies in \gls{BD-RIS}-aided systems. While continuous phase-shift design offers optimal performance, it is impractical due to hardware limitations. Discrete phase-shift design provides a trade-off between performance and complexity, but it may still require precise control. Quantized phase-shift design, on the other hand, can significantly reduce hardware complexity but may lead to performance degradation. Future research should explore hybrid approaches that combine the advantages of these strategies, such as using continuous phase-shift design for critical elements and quantized phase-shift design for less sensitive elements. Additionally, with the aid of advanced \gls{ML} algorithms, adaptive phase-shift design algorithms can be developed to dynamically adjust phase-shift values based on channel conditions and system requirements.

\subsubsection{Near-field vs. Far-field Propagation}

Integrating \gls{BD-RIS} in the near-field poses significant challenges due to the rapid variation of the electromagnetic field in this region. Accurate channel estimation and precise control of the \gls{BD-RIS} elements are crucial to effectively manipulate the wavefronts. Additionally, the design of compact and reconfigurable \gls{BD-RIS} elements is essential to accommodate the near-field's unique characteristics. For example, in a near-field \gls{WPT} scenario, \gls{BD-RIS} could be used to focus the energy towards the receiver, but this requires precise control of the phase shifts to compensate for the rapid phase variations. In contrast, \gls{BD-RIS} implementation in the far-field is relatively simpler as the wavefronts are planar and the channel characteristics are more predictable. Here, \gls{BD-RIS} can be used to create virtual \gls{LoS} paths, enhance signal coverage, and improve the network \gls{EE}.

\subsubsection{Modeling the Inter-Sector Interference of Multi-sector BD-RIS-based Systems}

The representative work in~\cite{10411856} assumes that every \gls{BD-RIS} antenna has a perfect unidirectional radiation pattern with no overlaps between the various \gls{BD-RIS} sectors. In real-world applications, nevertheless, the \gls{BD-RIS} antennas may have strong sidelobes that result in overlaps, covering users by more than one \gls{BD-RIS} sector altogether. It is necessary to explore fresh perspectives for imperfect \gls{BD-RIS} models since the \gls{BF} design and analysis based on the ideal model will no longer hold when this effect is taken into consideration.

\subsection{Future Research Directions}

The recently proposed \gls{BD-RIS}-based system has been the subject of decent research activities. However, to optimize the system's benefits, further research areas could be investigated. This subsection describes some of these potential future research areas.

\subsubsection{Active vs. Passive BD-RIS Structures}

A research area to be considered will be the comparative performance analyses between the passive elements-based \gls{BD-RIS} and active elements-based \gls{BD-RIS} systems. The fundamental tradeoffs in terms of performance, \gls{EE}, and computational complexity cost for these two \gls{BD-RIS} alternatives need further investigation.

\subsubsection{Integration with Additional Emerging Wireless Technologies}

Exploring the potential uses of \gls{BD-RIS} models through integration with additional emerging technologies (Other than the ones discussed in Section~\ref{subsection: BD-RIS Integration with Emerging Technologies/Schemes}). In \gls{OWC}, \gls{BD-RIS} can enhance signal quality and extend coverage by precisely controlling the reflection and refraction of light beams. For \gls{D2D} communications, \gls{BD-RIS} can facilitate efficient resource allocation and interference mitigation, leading to improved throughput and reduced latency. In \gls{CRN}, \gls{BD-RIS} can enable dynamic spectrum access and interference avoidance, optimizing spectrum utilization. Vehicular Communications can benefit from \gls{BD-RIS}-assisted signal reflection and \gls{BF}, improving connectivity and enabling advanced applications like autonomous driving. Integrated terrestrial-satellite networks can seamlessly leverage \gls{BD-RIS} to bridge the gap between terrestrial and satellite networks, enhancing coverage and capacity. \Gls{HAPS} networks can utilize \gls{BD-RIS} for flexible \gls{BF} and coverage extension, especially in remote or disaster-affected regions. Underwater Communications can overcome the challenges of signal propagation and attenuation through \gls{BD-RIS}-assisted signal reflection and focusing. \Gls{CoMP} schemes can leverage \gls{BD-RIS} for coordinated \gls{BF} and interference suppression, improving the \gls{SE} and user experience. Finally, Backscatter Communications can benefit from \gls{BD-RIS}-assisted energy harvesting and backscattering, enabling low-power devices to communicate efficiently.

\subsubsection{Utilizing Advanced Signal Processing Tools to Optimize BD-RIS-based Systems}

There are some new advanced signal processing tools, whose applications on \gls{BD-RIS} technology will make a difference in terms of performance and implementation. Two prominent of these include both advanced \gls{ML} and \gls{QC}. \gls{ML} techniques, unlike traditional model-based approaches, rely on data-driven models, eliminating the need for precise system-level modeling. Advanced \gls{ML} techniques, such as federated learning, meta-learning, graph learning, transfer learning, and quantum learning, can be applied to \gls{BD-RIS}-based systems to efficiently optimize the configuration of \gls{BD-RIS} elements due to their massive number. In addition, advanced \gls{ML} can be used to acquire \gls{CSI}, perform signal detection, and other signal processing requirements for the optimum performance of \gls{BD-RIS}. Significant advancements in \gls{QC} algorithms and hardware in recent years offer a new paradigm for tackling challenging computational issues. In \gls{BD-RIS}-based wireless communications systems, \gls{QC} techniques can be used to get around the anticipated high computational optimization complexity related to the \gls{BD-RIS}-assisted smart radio environment. This can be achieved by recasting the \gls{BD-RIS}-aided wireless and/or electromagnetic problem into a physical formulation that can be effectively addressed with newly developed \gls{QC} hardware~\cite{gradoni2021smart}.

\subsubsection{Deploying a Transmissive BD-RIS at the Receiver}
\glspl{BD-RIS} offer enhanced flexibility in wave manipulation compared to \gls{D-RIS}, enabling improved wireless communications performance. While \gls{BD-RIS} structures are commonly deployed in the channel or at the transmitter side, positioning it at the receiver side presents additional advantages. Strategically placing a transmissive \gls{BD-RIS} structure at the receiver would enhance the signal reception and empower the receiver to actively shape the incoming signals, leading to significant improvements in overall system performance and user experience. To realize these benefits, the \gls{BD-RIS} structure must be compact enough to be integrated into receiver devices without compromising their form factor or functionality. However, deploying \gls{BD-RIS} at the receiver also presents certain challenges. These include the need for precise control and synchronization between the transmitter and the receiver, as well as the potential for increased hardware complexity and power consumption at the receiver. Overcoming these challenges will be essential to fully harness the potential of \gls{BD-RIS} in future wireless communications systems. Also, the joint deployment of \gls{BD-RIS} structures in both the channel
and in the receiver is an interesting area of future research, similar to the research efforts that investigated such a joint deployment with the \gls{D-RIS} technology~\cite{10183987}.

\subsubsection{Prototyping/Experimentation}
Thus far, none of the \gls{BD-RIS} architectures has been physically realized. Important open research topics include the physical realization of \gls{BD-RIS} architectures and the associated experimental measurements. Different practical limits of the various \gls{BD-RIS} architectures (i.e., group-/fully-/tree-/forest-connected and cell-wise single-/group-/dynamically group-/fully-connected) need to be documented, and this will require further investigation while considering real-time scenario issues. Therefore, to experimentally evaluate their benefits in terms of performance enhancement, prototyping of the \gls{BD-RIS}  should be considered. Similarly, the designing of efficient reconfigurable impedance networks with reduced circuit complexity and costs should be explored.

\subsubsection{Development of Standards for the BD-RIS Technology}

According to the international mobile telecommunications-$2030$ ``Future Technology Trends'' report~\cite{M.2516-0} developed by the international telecommunications union - radio communications sector, \gls{RIS} is included as technology to enhance radio interface. In September 2021, the \gls{ISG} on \gls{RIS} technology was established by the \gls{ETSI} to facilitate the formal standardization of \glspl{RIS}.  According to~\cite{liu2024sustainable}, the analysis of the \gls{RIS} technology potential, validation, maturity deadlines, and standardization requirements were carried out in \gls{ISG}'s first phase, which ran from $2022$ to $2023$. Within $2024$ and $2025$, preliminary guidelines for the \gls{RIS} technology together with potential group specifications for its functional architecture are scheduled. In the third and final phase of the \gls{ISG}, which will run from $2026$ to $2027$, the various \gls{RIS} solutions will be considered mature. Since consideration regarding standards for the traditional \gls{RIS} is still ongoing, it may be worthwhile to incorporate \gls{BD-RIS} into any standard that will eventually be approved for the traditional \gls{RIS}. In this regard, it is worth noting that the \gls{ETSI} on one of their reports about the traditional \gls{RIS} briefly discussed \gls{BD-RIS} in Section $5.2.1.1$ that entitled ``Impedance-based structures''~\cite{ETSI.GR.RIS.002}.

\section{Conclusion} \label{sec: Conclusion}

\Glspl{RIS}, a well-known candidate technology under investigation for \gls{6G}, offers the potential to enhance the efficiency of wireless communications in ways that are both cost-effective and energy-efficient. The analysis of the \gls{RIS} technology potential, validation, maturity deadlines, and standardization requirements are still under consideration by different interest groups. Its simple architecture and limitations have given rise to a brand-new \gls{RIS} super-set called \gls{BD-RIS}. We have started by providing the differences between the \gls{D-RIS} structure and the \gls{BD-RIS} structures in terms of design, modeling, limitations, and merits. Then, we have classified the \gls{BD-RIS} according to its scattering matrix types, modes of transmission, inter-cell architectures, and circuit topologies. Afterward, we have explored some potential applications of \gls{BD-RIS} that can help in addressing contemporary communications challenges. Next, we have focused on the overview of research progress made in \gls{BD-RIS} in terms of architectural development, evaluation, and its integration with emerging wireless technologies. Finally, we have discussed some key challenges and future research directions of \gls{BD-RIS}-aided wireless systems. This survey aims to provide an up-to-date comprehensive review of \gls{BD-RIS} development and to aid researchers and professionals in accelerating the theoretical investigation and practical implementation of this technology in cutting-edge wireless networks.

\bibliographystyle{IEEEtran}
\bibliography{Main.bib}

\begin{thebibliography}{100}
\providecommand{\url}[1]{#1}
\csname url@samestyle\endcsname
\providecommand{\newblock}{\relax}
\providecommand{\bibinfo}[2]{#2}
\providecommand{\BIBentrySTDinterwordspacing}{\spaceskip=0pt\relax}
\providecommand{\BIBentryALTinterwordstretchfactor}{4}
\providecommand{\BIBentryALTinterwordspacing}{\spaceskip=\fontdimen2\font plus
\BIBentryALTinterwordstretchfactor\fontdimen3\font minus \fontdimen4\font\relax}
\providecommand{\BIBforeignlanguage}[2]{{%
\expandafter\ifx\csname l@#1\endcsname\relax
\typeout{** WARNING: IEEEtran.bst: No hyphenation pattern has been}%
\typeout{** loaded for the language `#1'. Using the pattern for}%
\typeout{** the default language instead.}%
\else
\language=\csname l@#1\endcsname
\fi
#2}}
\providecommand{\BIBdecl}{\relax}
\BIBdecl

\bibitem{8766143}
Z.~Zhang, Y.~Xiao, Z.~Ma, M.~Xiao, Z.~Ding, X.~Lei, G.~K. Karagiannidis, and P.~Fan, ``{6G} wireless networks: Vision, requirements, architecture, and key technologies,'' \emph{IEEE Veh. Technol. Mag.}, vol.~14, no.~3, pp. 28--41, Sep. 2019.

\bibitem{8869705}
W.~Saad, M.~Bennis, and M.~Chen, ``A vision of {6G} wireless systems: Applications, trends, technologies, and open research problems,'' \emph{IEEE Netw.}, vol.~34, no.~3, pp. 134--142, May 2020.

\bibitem{10054381}
C.-X. Wang, X.~You, X.~Gao, X.~Zhu, Z.~Li, C.~Zhang, H.~Wang, Y.~Huang, Y.~Chen, H.~Haas, J.~S. Thompson, E.~G. Larsson, M.~D. Renzo, W.~Tong, P.~Zhu, X.~Shen, H.~V. Poor, and L.~Hanzo, ``On the road to {6G}: Visions, requirements, key technologies, and testbeds,'' \emph{IEEE Commun. Surv. Tutor.}, vol.~25, no.~2, pp. 905--974, 2nd Quart. 2023.

\bibitem{9140329}
M.~Di~Renzo, A.~Zappone, M.~Debbah, M.-S. Alouini, C.~Yuen, J.~de~Rosny, and S.~Tretyakov, ``Smart radio environments empowered by reconfigurable intelligent surfaces: How it works, state of research, and the road ahead,'' \emph{IEEE J. Sel. Areas Commun.}, vol.~38, no.~11, pp. 2450--2525, Nov. 2020.

\bibitem{8811733}
Q.~Wu and R.~Zhang, ``Intelligent reflecting surface enhanced wireless network via joint active and passive beamforming,'' \emph{IEEE Trans. Wirel. Commun.}, vol.~18, no.~11, pp. 5394--5409, Nov. 2019.

\bibitem{9424177}
Y.~Liu, X.~Liu, X.~Mu, T.~Hou, J.~Xu, M.~Di~Renzo, and N.~Al-Dhahir, ``Reconfigurable intelligent surfaces: Principles and opportunities,'' \emph{IEEE Commun. Surv. Tutor.}, vol.~23, no.~3, pp. 1546--1577, 3rd Quart. 2021.

\bibitem{9148781}
N.~S. Perovic, M.~D. Renzo, and M.~F. Flanagan, ``Channel capacity optimization using reconfigurable intelligent surfaces in indoor {mmWave} environments,'' in \emph{IEEE Int. Conf. Commun. (ICC), Dublin, Ireland}, Jun. 2020, pp. 1--7.

\bibitem{9079457}
S.~Zhou, W.~Xu, K.~Wang, M.~Di~Renzo, and M.-S. Alouini, ``Spectral and energy efficiency of {IRS}-assisted {MISO} communication with hardware impairments,'' \emph{IEEE Wirel. Commun. Lett.}, vol.~9, no.~9, pp. 1366--1369, Sep. 2020.

\bibitem{9136592}
C.~Huang, S.~Hu, G.~C. Alexandropoulos, A.~Zappone, C.~Yuen, R.~Zhang, M.~D. Renzo, and M.~Debbah, ``Holographic {MIMO} surfaces for {6G} wireless networks: Opportunities, challenges, and trends,'' \emph{IEEE Wirel. Commun.}, vol.~27, no.~5, pp. 118--125, Oct. 2020.

\bibitem{10598369}
Q.~Ding, J.~Yang, Y.~Luo, and C.~Luo, ``Intelligent reflecting surface vs. conventional full-duplex relay in {mmWave} {MIMO} networks: A comprehensive performance comparison,'' \emph{IEEE Trans. Veh. Technol.}, vol.~73, no.~11, pp. 17\,231--17\,246, Nov. 2024.

\bibitem{9122596}
S.~Gong, X.~Lu, D.~T. Hoang, D.~Niyato, L.~Shu, D.~I. Kim, and Y.-C. Liang, ``Toward smart wireless communications via intelligent reflecting surfaces: A contemporary survey,'' \emph{IEEE Commun. Surv. Tutor.}, vol.~22, no.~4, pp. 2283--2314, 4th Quart. 2020.

\bibitem{khan2024integration}
W.~U. Khan, E.~Lagunas, A.~Mahmood, M.~Asif, M.~Ahmed, and S.~Chatzinotas, ``Integration of beyond diagonal {RIS} and {UAVs} in {6G} {NTNs}: Enhancing aerial connectivity,'' \emph{arXiv preprint arXiv:2409.06073}, Sep. 2024.

\bibitem{8796365}
E.~Basar, M.~Di~Renzo, J.~De~Rosny, M.~Debbah, M.-S. Alouini, and R.~Zhang, ``Wireless communications through reconfigurable intelligent surfaces,'' \emph{IEEE Access}, vol.~7, pp. 116\,753--116\,773, 2019.

\bibitem{10584518}
M.~Ahmed, S.~Raza, A.~A. Soofi, F.~Khan, W.~U. Khan, S.~Z.~U. Abideen, F.~Xu, and Z.~Han, ``Active reconfigurable intelligent surfaces: Expanding the frontiers of wireless communication-{A} survey,'' \emph{IEEE Commun. Surv. Tutor.}, pp. 1--1, early access, 2024.

\bibitem{8910627}
Q.~Wu and R.~Zhang, ``Towards smart and reconfigurable environment: Intelligent reflecting surface aided wireless network,'' \emph{IEEE Commun. Mag.}, vol.~58, no.~1, pp. 106--112, Jan. 2020.

\bibitem{10268023}
R.~K. Fotock, A.~Zappone, and M.~D. Renzo, ``Energy efficiency optimization in {RIS}-aided wireless networks: Active versus nearly-passive {RIS} with global reflection constraints,'' \emph{IEEE Trans. Commun.}, vol.~72, no.~1, pp. 257--272, Jan. 2024.

\bibitem{9475160}
C.~Pan, H.~Ren, K.~Wang, J.~F. Kolb, M.~Elkashlan, M.~Chen, M.~Di~Renzo, Y.~Hao, J.~Wang, A.~L. Swindlehurst, X.~You, and L.~Hanzo, ``Reconfigurable intelligent surfaces for 6{G} systems: Principles, applications, and research directions,'' \emph{IEEE Commun. Mag.}, vol.~59, no.~6, pp. 14--20, Jun., 2021.

\bibitem{9530403}
M.~H. Khoshafa, T.~M.~N. Ngatched, M.~H. Ahmed, and A.~R. Ndjiongue, ``Active reconfigurable intelligent surfaces-aided wireless communication system,'' \emph{IEEE Commun. Lett.}, vol.~25, no.~11, pp. 3699--3703, Nov. 2021.

\bibitem{9998527}
Z.~Zhang, L.~Dai, X.~Chen, C.~Liu, F.~Yang, R.~Schober, and H.~V. Poor, ``Active {RIS} vs. passive {RIS}: Which will prevail in {6G}?'' \emph{IEEE Trans. Commun.}, vol.~71, no.~3, pp. 1707--1725, Mar. 2023.

\bibitem{9370097}
A.~Taha, M.~Alrabeiah, and A.~Alkhateeb, ``Enabling large intelligent surfaces with compressive sensing and deep learning,'' \emph{IEEE Access}, vol.~9, pp. 44\,304--44\,321, Jan. 2021.

\bibitem{9053976}
G.~C. Alexandropoulos and E.~Vlachos, ``A hardware architecture for reconfigurable intelligent surfaces with minimal active elements for explicit channel estimation,'' in \emph{IEEE Int. Conf. Acoust. Speech Signal Process. (ICASSP), Barcelona, Spain}, May 2020, pp. 9175--9179.

\bibitem{9690478}
Y.~Liu, X.~Mu, J.~Xu, R.~Schober, Y.~Hao, H.~V. Poor, and L.~Hanzo, ``{STAR}: Simultaneous transmission and reflection for $360^{\circ}$ coverage by intelligent surfaces,'' \emph{IEEE Wirel. Commun.}, vol.~28, no.~6, pp. 102--109, Dec. 2021.

\bibitem{9722826}
H.~Zhang, S.~Zeng, B.~Di, Y.~Tan, M.~Di~Renzo, M.~Debbah, Z.~Han, H.~V. Poor, and L.~Song, ``Intelligent omni-surfaces for full-dimensional wireless communications: Principles, technology, and implementation,'' \emph{IEEE Commun. Mag.}, vol.~60, no.~2, pp. 39--45, Feb. 2022.

\bibitem{khoshafa2025rsma}
M.~H. Khoshafa, T.~M. Ngatched, M.~H. Ahmed, and Y.~Gadallah, ``{RIS}-empowered rate-splitting multiple access towards {6G} and beyond wireless communications networks: A comprehensive survey,'' \emph{Authorea Preprints}, Feb. 2025.

\bibitem{khoshafa2025ris}
M.~H. Khoshafa, F.~Bueno, T.~M. Ngatched, and M.~Di~Renzo, ``{RIS}-empowered secured space-air-ground integrated networks: Opportunities and challenges,'' \emph{Authorea Preprints}, Feb. 2025.

\bibitem{8466374}
C.~Liaskos, S.~Nie, A.~Tsioliaridou, A.~Pitsillides, S.~Ioannidis, and I.~Akyildiz, ``A new wireless communication paradigm through software-controlled metasurfaces,'' \emph{IEEE Commun. Mag.}, vol.~56, no.~9, pp. 162--169, Sep. 2018.

\bibitem{10663469}
M.~H. Khoshafa, Y.~Gadallah, T.~M.~N. Ngatched, and M.~H. Ahmed, ``Aerial reconfigurable intelligent surface-assisted {LPWAN}s for {I}o{T}: A cross-layer analysis,'' \emph{IEEE Wirel. Commun. Lett.}, vol.~13, no.~10, pp. 2912--2916, Oct. 2024.

\bibitem{10771739}
H.~Yahya, H.~Li, M.~Nerini, B.~Clerckx, and M.~Debbah, ``Beyond diagonal {RIS}: Passive maximum ratio transmission and interference nulling enabler,'' \emph{IEEE Open J. Commun. Soc.}, vol.~5, pp. 7613--7627, 2024.

\bibitem{10499196}
W.~Sun, S.~Sun, T.~Shi, X.~Su, and R.~Liu, ``A new model of beyond diagonal reconfigurable intelligent surfaces ({BD-RIS}) for the corresponding quantization and optimization,'' \emph{IEEE Trans. Wirel. Commun.}, vol.~23, no.~9, pp. 11\,521--11\,534, Sep. 2024.

\bibitem{9941040}
H.~Alakoca, M.~Namdar, S.~Aldirmaz-Colak, M.~Basaran, A.~Basgumus, L.~Durak-Ata, and H.~Yanikomeroglu, ``Metasurface manipulation attacks: Potential security threats of {RIS}-aided 6{G} communications,'' \emph{IEEE Commun. Mag.}, vol.~61, no.~1, pp. 24--30, Jan. 2023.

\bibitem{10316535}
H.~Li, S.~Shen, M.~Nerini, and B.~Clerckx, ``Reconfigurable intelligent surfaces {2.0}: Beyond diagonal phase shift matrices,'' \emph{IEEE Commun. Mag.}, vol.~62, no.~3, pp. 102--108, Mar. 2024.

\bibitem{9514409}
S.~Shen, B.~Clerckx, and R.~Murch, ``Modeling and architecture design of reconfigurable intelligent surfaces using scattering parameter network analysis,'' \emph{IEEE Trans. Wirel. Commun.}, vol.~21, no.~2, pp. 1229--1243, Feb. 2022.

\bibitem{10574199}
M.~Nerini, S.~Shen, H.~Li, M.~D. Renzo, and B.~Clerckx, ``A universal framework for multiport network analysis of reconfigurable intelligent surfaces,'' \emph{IEEE Trans. Wirel. Commun.}, vol.~23, no.~10, pp. 14\,575--14\,590, Oct. 2024.

\bibitem{10694015}
P.~Del~Hougne, ``Physics-compliant diagonal representation of beyond-diagonal {RIS},'' in \emph{IEEE Workshop Signal Process. Adv. Wirel. Commun. (SPAWC), Lucca, Italy}, Sep. 2024, pp. 931--935.

\bibitem{del2024physics}
P.~del Hougne, ``A physics-compliant diagonal representation for wireless channels parametrized by beyond-diagonal reconfigurable intelligent surfaces,'' \emph{arXiv preprint arXiv:2409.20509}, Sep. 2024.

\bibitem{10453384}
M.~Nerini, S.~Shen, H.~Li, and B.~Clerckx, ``Beyond diagonal reconfigurable intelligent surfaces utilizing graph theory: Modeling, architecture design, and optimization,'' \emph{IEEE Trans. Wirel. Commun.}, vol.~23, no.~8, pp. 9972--9985, Aug. 2024.

\bibitem{5446312}
M.~T. Ivrlac and J.~A. Nossek, ``Toward a circuit theory of communication,'' \emph{IEEE Trans. Circuits Syst. I: Regul. Pap.}, vol.~57, no.~7, pp. 1663--1683, Jul. 2010.

\bibitem{10694393}
H.~Li, M.~Nerini, S.~Shen, and B.~Clerckx, ``Wideband modeling and beamforming for beyond diagonal reconfigurable intelligent surfaces,'' in \emph{IEEE Workshop Signal Process. Adv. Wirel. Commun. (SPAWC), Lucca, Italy}, Sep. 2024, pp. 926--930.

\bibitem{10857964}
H.~Li, M.~Nerini, S.~Shen, and B.~Clerckx, ``Beyond diagonal reconfigurable intelligent surfaces in wideband {OFDM} communications: Circuit-based modeling and optimization,'' \emph{IEEE Trans. Wirel. Commun.}, pp. 1--1, early access, 2025.

\bibitem{xu2008divide}
W.~Xu and S.~Qiao, ``A divide-and-conquer method for the {Takagi} factorization,'' \emph{SIAM J. Matrix Anal. Appl.}, vol.~30, no.~1, pp. 142--153, 2008.

\bibitem{10716670}
W.~U. Khan, A.~Mahmood, M.~A. Jamshed, E.~Lagunas, M.~Ahmed, and S.~Chatzinotas, ``Beyond diagonal {RIS} for {6G} non-terrestrial networks: Potentials and challenges,'' \emph{IEEE Netw.}, vol.~39, no.~1, pp. 80--89, Jan. 2025.

\bibitem{ahmed2024comprehensive}
M.~Ahmed, F.~Xu, A.~Wahid, K.~Ali, M.~A. Mirza, W.~Khan, K.~Dev, S.~A. Hassan, and Z.~Han, ``A comprehensive survey of artificial intelligence advances in {RIS}-assisted wireless networks,'' \emph{Authorea Preprints}, Aug. 2024.

\bibitem{9913356}
H.~Li, S.~Shen, and B.~Clerckx, ``Beyond diagonal reconfigurable intelligent surfaces: From transmitting and reflecting modes to single-, group-, and fully-connected architectures,'' \emph{IEEE Trans. Wirel. Commun.}, vol.~22, no.~4, pp. 2311--2324, Apr. 2023.

\bibitem{Wali2025transmissive}
W.~U. Khan, E.~Lagunas, and S.~Chatzinotas, ``Transmissive beyond diagonal {RIS}-mounted {LEO} communication for {NOMA} {IoT} networks,'' \emph{arXiv preprint arXiv:2501.02742}, Jan. 2025.

\bibitem{10530995}
M.~Nerini and B.~Clerckx, ``Physically consistent modeling of stacked intelligent metasurfaces implemented with beyond diagonal {RIS},'' \emph{IEEE Commun. Lett.}, vol.~28, no.~7, pp. 1693--1697, Jul. 2024.

\bibitem{10159457}
H.~Li, S.~Shen, and B.~Clerckx, ``A dynamic grouping strategy for beyond diagonal reconfigurable intelligent surfaces with hybrid transmitting and reflecting mode,'' \emph{IEEE Trans. Veh. Tech.}, vol.~72, no.~12, pp. 16\,748--16\,753, Dec. 2023.

\bibitem{10158988}
H.~Li, S.~Shen, and B.~Clerckx, ``Beyond diagonal reconfigurable intelligent surfaces: A multi-sector mode enabling highly directional full-space wireless coverage,'' \emph{IEEE J. Sel. Areas Commun.}, vol.~41, no.~8, pp. 2446--2460, Aug. 2023.

\bibitem{10643263}
Y.~Dong, Q.~Li, S.~X. Ng, and M.~El-Hajjar, ``Reconfigurable intelligent surface relying on low-complexity joint sector non-diagonal structure,'' \emph{IEEE Open J. Veh. Technol.}, vol.~5, pp. 1106--1123, 2024.

\bibitem{9737373}
Q.~Li, M.~El-Hajjar, I.~Hemadeh, A.~Shojaeifard, A.~A.~M. Mourad, B.~Clerckx, and L.~Hanzo, ``Reconfigurable intelligent surfaces relying on non-diagonal phase shift matrices,'' \emph{IEEE Trans. Veh. Tech.}, vol.~71, no.~6, pp. 6367--6383, Jun. 2022.

\bibitem{10693959}
K.~Chen and Y.~Mao, ``Transmitter side beyond-diagonal {RIS} for {mmWave} integrated sensing and communications,'' in \emph{IEEE Workshop Signal Process. Adv. Wirel. Commun. (SPAWC), Lucca, Italy}, Sep. 2024, pp. 951--955.

\bibitem{10694505}
M.~Samy, A.~B.~M. Adam, K.~Ntontin, H.~Al-Hraishawi, S.~Chatzinotas, and B.~Otteresten, ``Enhancing spectral and energy efficiency with multi-sector beyond-diagonal {RIS},'' in \emph{IEEE Workshop Signal Process. Adv. Wirel. Commun. (SPAWC), Lucca, Italy}, Sep. 2024, pp. 941--945.

\bibitem{9200683}
S.~Zhang, H.~Zhang, B.~Di, Y.~Tan, Z.~Han, and L.~Song, ``Beyond intelligent reflecting surfaces: Reflective-transmissive metasurface aided communications for full-dimensional coverage extension,'' \emph{IEEE Trans. Veh. Tech.}, vol.~69, no.~11, pp. 13\,905--13\,909, Nov. 2020.

\bibitem{10736549}
M.~H. Khoshafa, O.~Maraqa, J.~M. Moualeu, S.~Aboagye, T.~M.~N. Ngatched, M.~H. Ahmed, Y.~Gadallah, and M.~D. Renzo, ``{RIS}-assisted physical layer security in emerging {RF} and optical wireless communication systems: A comprehensive survey,'' \emph{IEEE Commun. Surv. Tutor.}, pp. 1--1, early access, 2024.

\bibitem{9305710}
M.~H. Khoshafa, T.~M.~N. Ngatched, and M.~H. Ahmed, ``Reconfigurable intelligent surfaces-aided physical layer security enhancement in {D2D} underlay communications,'' \emph{IEEE Commun. Lett.}, vol.~25, no.~5, pp. 1443--1447, May 2021.

\bibitem{10778572}
M.~H. Khoshafa, G.~Ahmed, T.~M.~N. Ngatched, and M.~D. Renzo, ``Aerial reconfigurable intelligent surfaces-enabled secured wireless communications: Performance analysis and optimization,'' \emph{IEEE Trans. Commun.}, pp. 1--1, early access, 2024.

\bibitem{10285357}
M.~H. Khoshafa, T.~M.~N. Ngatched, Y.~Gadallah, and M.~H. Ahmed, ``Securing lpwans: A reconfigurable intelligent surface ({RIS})-assisted {UAV} approach,'' \emph{IEEE Wireless Comm. Lett.}, vol.~13, no.~1, pp. 158--162, Jan. 2024.

\bibitem{10219539}
M.~H. Khoshafa, T.~M.~N. Ngatched, and M.~H. Ahmed, ``{RIS}-aided physical layer security improvement in underlay cognitive radio networks,'' \emph{IEEE Sys. J.}, vol.~17, no.~4, pp. 6437--6448, Dec. 2023.

\bibitem{9330587}
R.~Borralho, A.~Mohamed, A.~U. Quddus, P.~Vieira, and R.~Tafazolli, ``A survey on coverage enhancement in cellular networks: Challenges and solutions for future deployments,'' \emph{IEEE Commun. Surv. Tutor.}, vol.~23, no.~2, pp. 1302--1341, 2nd Quart. 2021.

\bibitem{10766364}
A.~S. De~Sena, M.~Rasti, N.~H. Mahmood, and M.~Latva-aho, ``Beyond diagonal {RIS} for multi-band multi-cell {MIMO} networks: A practical frequency-dependent model and performance analysis,'' \emph{IEEE Trans. Wirel. Commun.}, vol.~24, no.~1, pp. 749--766, Jan. 2025.

\bibitem{9737357}
F.~Liu, Y.~Cui, C.~Masouros, J.~Xu, T.~X. Han, Y.~C. Eldar, and S.~Buzzi, ``Integrated sensing and communications: Toward dual-functional wireless networks for 6{G} and beyond,'' \emph{IEEE J. Sel. Areas Commun.}, vol.~40, no.~6, pp. 1728--1767, Jun. 2022.

\bibitem{9606831}
Y.~Cui, F.~Liu, X.~Jing, and J.~Mu, ``Integrating sensing and communications for ubiquitous {I}o{T}: Applications, trends, and challenges,'' \emph{IEEE Netw.}, vol.~35, no.~5, pp. 158--167, Oct. 2021.

\bibitem{9705498}
A.~Liu, Z.~Huang, M.~Li, Y.~Wan, W.~Li, T.~X. Han, C.~Liu, R.~Du, D.~K.~P. Tan, J.~Lu, Y.~Shen, F.~Colone, and K.~Chetty, ``A survey on fundamental limits of integrated sensing and communication,'' \emph{IEEE Commun. Surv. Tutor.}, vol.~24, no.~2, pp. 994--1034, 2nd Quart. 2022.

\bibitem{10495009}
Z.~Liu, Y.~Liu, S.~Shen, Q.~Wu, and Q.~Shi, ``Enhancing {ISAC} network throughput using beyond diagonal {RIS},'' \emph{IEEE Wirel. Commun. Lett.}, vol.~13, no.~6, pp. 1670--1674, Jun. 2024.

\bibitem{10493847}
Z.~Guang, Y.~Liu, Q.~Wu, W.~Wang, and Q.~Shi, ``Power minimization for {ISAC} system using beyond diagonal reconfigurable intelligent surface,'' \emph{IEEE Trans. Veh. Tech.}, vol.~73, no.~9, pp. 13\,950--13\,955, Sep. 2024.

\bibitem{10777522}
T.~Esmaeilbeig, K.~V. Mishra, and M.~Soltanalian, ``Beyond diagonal {RIS}: Key to next-generation integrated sensing and communications?'' \emph{IEEE Signal Process. Lett.}, vol.~32, pp. 216--220, 2025.

\bibitem{10902602}
M.~Nerini, G.~Ghiaasi, and B.~Clerckx, ``Localized and distributed beyond diagonal reconfigurable intelligent surfaces with lossy interconnections: Modeling and optimization,'' \emph{IEEE Trans. Commun.}, pp. 1--1, early access, 2025.

\bibitem{10643599}
B.~Wang, H.~Li, S.~Shen, Z.~Cheng, and B.~Clerckx, ``A dual-function radar-communication system empowered by beyond diagonal reconfigurable intelligent surface,'' \emph{IEEE Trans. Commun.}, pp. 1--1, early access, 2024.

\bibitem{raeisi2024efficient}
M.~Raeisi, H.~Chen, H.~Wymeersch, and E.~Basar, ``Efficient localization with base station-integrated beyond diagonal {RIS},'' \emph{arXiv preprint arXiv:2411.13295}, Nov. 2024.

\bibitem{10187717}
S.~Zheng, B.~Lv, T.~Zhang, Y.~Xu, G.~Chen, R.~Wang, and P.~C. Ching, ``On {D}o{F} of active {RIS}-assisted {MIMO} interference channel with arbitrary antenna configurations: When will {RIS} help?'' \emph{IEEE Trans. Veh. Technol.}, vol.~72, no.~12, pp. 16\,828--16\,833, Dec. 2023.

\bibitem{9200661}
M.~A. ElMossallamy, H.~Zhang, R.~Sultan, K.~G. Seddik, L.~Song, G.~Y. Li, and Z.~Han, ``On spatial multiplexing using reconfigurable intelligent surfaces,'' \emph{IEEE Wirel. Commun. Lett.}, vol.~10, no.~2, pp. 226--230, Feb. 2021.

\bibitem{9734015}
A.~H.~A. Bafghi, V.~Jamali, M.~Nasiri-Kenari, and R.~Schober, ``Degrees of freedom of the $k$-user interference channel assisted by active and passive {IRSs},'' \emph{IEEE Trans. Commun.}, vol.~70, no.~5, pp. 3063--3080, May 2022.

\bibitem{10314137}
S.~Meng, W.~Tang, W.~Chen, J.~Lan, Q.~Y. Zhou, Y.~Han, X.~Li, and S.~Jin, ``Rank optimization for {MIMO} channel with {RIS}: Simulation and measurement,'' \emph{IEEE Wirel. Commun. Lett.}, vol.~13, no.~2, pp. 437--441, Feb. 2024.

\bibitem{zhao2024channel}
Y.~Zhao, H.~Li, M.~Franceschetti, and B.~Clerckx, ``Channel shaping using beyond diagonal reconfigurable intelligent surface: Analysis, optimization, and enhanced flexibility,'' \emph{arXiv preprint arXiv:2407.15196}, Jul. 2024.

\bibitem{alegria2024channel}
J.~V. Alegr{\'\i}a, J.~Thunberg, and O.~Edfors, ``Channel orthogonalization with reconfigurable surfaces: General models, theoretical limits, and effective configuration,'' \emph{arXiv preprint arXiv:2403.15165}, Mar. 2024.

\bibitem{10680138}
A.~Papazafeiropoulos, P.~Kourtessis, and S.~Chatzinotas, ``Effect of channel aging on beyond diagonal reconfigurable intelligent surfaces,'' \emph{IEEE Open J. Commun. Soc.}, vol.~5, pp. 6303--6313, 2024.

\bibitem{10379500}
N.~U. Hassan, J.~An, M.~Di~Renzo, M.~Debbah, and C.~Yuen, ``Efficient beamforming and radiation pattern control using stacked intelligent metasurfaces,'' \emph{IEEE Open J. Commun. Society}, vol.~5, pp. 599--611, 2024.

\bibitem{10534211}
A.~Papazafeiropoulos, J.~An, P.~Kourtessis, T.~Ratnarajah, and S.~Chatzinotas, ``Achievable rate optimization for stacked intelligent metasurface-assisted holographic {MIMO} communications,'' \emph{IEEE Trans. Wirel. Commun.}, vol.~23, no.~10, pp. 13\,173--13\,186, Oct. 2024.

\bibitem{10515204}
J.~An, C.~Yuen, C.~Xu, H.~Li, D.~W.~K. Ng, M.~Di~Renzo, M.~Debbah, and L.~Hanzo, ``Stacked intelligent metasurface-aided {MIMO} transceiver design,'' \emph{IEEE Wirel. Commun.}, vol.~31, no.~4, pp. 123--131, Aug. 2024.

\bibitem{9437234}
J.~Xu, Y.~Liu, X.~Mu, and O.~A. Dobre, ``{STAR-RISs}: Simultaneous transmitting and reflecting reconfigurable intelligent surfaces,'' \emph{IEEE Commun. Lett.}, vol.~25, no.~9, pp. 3134--3138, Sep. 2021.

\bibitem{9754364}
J.~Xu, Y.~Liu, X.~Mu, J.~T. Zhou, L.~Song, H.~V. Poor, and L.~Hanzo, ``Simultaneously transmitting and reflecting intelligent omni-surfaces: Modeling and implementation,'' \emph{IEEE Veh. Tech. Mag.}, vol.~17, no.~2, pp. 46--54, Jun. 2022.

\bibitem{10445725}
H.~Wang, Z.~Han, and A.~L. Swindlehurst, ``Channel reciprocity attacks using intelligent surfaces with non-diagonal phase shifts,'' \emph{IEEE Open J. Commun. Soc.}, vol.~5, pp. 1469--1485, 2024.

\bibitem{10694006}
H.~Wang, J.~Nossek, and A.~L. Swindlehurst, ``Beyond-diagonal {RIS} attacks on physical layer key generation,'' in \emph{IEEE Workshop Signal Process. Adv. Wirel. Commun. (SPAWC), Lucca, Italy}, Sep. 2024, pp. 946--950.

\bibitem{zhang2024full}
Y.~Zhang, X.~Shao, H.~Li, B.~Clerckx, and R.~Zhang, ``Full-space wireless sensing enabled by multi-sector intelligent surfaces,'' \emph{arXiv preprint arXiv:2406.15945}, Jun. 2024.

\bibitem{10308579}
A.~Mishra, Y.~Mao, C.~D'Andrea, S.~Buzzi, and B.~Clerckx, ``Transmitter side beyond-diagonal reconfigurable intelligent surface for massive {MIMO} networks,'' \emph{IEEE Wirel. Commun. Lett.}, vol.~13, no.~2, pp. 352--356, Feb. 2024.

\bibitem{10670007}
K.~D. Katsanos, P.~D. Lorenzo, and G.~C. Alexandropoulos, ``Multi-{RIS}-empowered multiple access: A distributed sum-rate maximization approach,'' \emph{IEEE J. Sel. Top. Signal Process.}, vol.~18, no.~7, pp. 1324--1338, Oct. 2024.

\bibitem{10472097}
Q.~Li, M.~El-Hajjar, I.~Hemadeh, A.~Shojaeifard, and L.~Hanzo, ``Coordinated reconfigurable intelligent surfaces: Non-diagonal group-connected design,'' \emph{IEEE Trans. Veh. Tech.}, vol.~73, no.~7, pp. 10\,811--10\,816, Jul. 2024.

\bibitem{10526360}
M.~Nerini, S.~Shen, and B.~Clerckx, ``Static grouping strategy design for beyond diagonal reconfigurable intelligent surfaces,'' \emph{IEEE Commun. Lett.}, vol.~28, no.~7, pp. 1708--1712, Jul. 2024.

\bibitem{10418928}
H.~Li, S.~Shen, M.~Nerini, M.~Di~Renzo, and B.~Clerckx, ``Beyond diagonal reconfigurable intelligent surfaces with mutual coupling: Modeling and optimization,'' \emph{IEEE Commun. Lett.}, vol.~28, no.~4, pp. 937--941, Apr. 2024.

\bibitem{nerini2024global}
M.~Nerini, H.~Li, and B.~Clerckx, ``Global optimal closed-form solutions for intelligent surfaces with mutual coupling: Is mutual coupling detrimental or beneficial?'' \emph{arXiv preprint arXiv:2411.04949}, Nov. 2024.

\bibitem{li2024non}
H.~Li and B.~Clerckx, ``Non-reciprocal beyond diagonal {RIS}: Multiport network models and performance benefits in full-duplex systems,'' \emph{arXiv preprint arXiv:2411.04370}, Nov. 2024.

\bibitem{liu2024non}
Z.~Liu, H.~Li, and B.~Clerckx, ``Non-reciprocal beyond diagonal {RIS}: Sum-rate maximization in full-duplex communications,'' \emph{arXiv preprint arXiv:2411.18523}, Dec. 2024.

\bibitem{10197228}
M.~Nerini, S.~Shen, and B.~Clerckx, ``Discrete-value group and fully connected architectures for beyond diagonal reconfigurable intelligent surfaces,'' \emph{IEEE Trans. Veh. Tech.}, vol.~72, no.~12, pp. 16\,354--16\,368, Dec. 2023.

\bibitem{nerini2024dual}
M.~Nerini and B.~Clerckx, ``Dual-polarized beyond diagonal {RIS},'' \emph{arXiv preprint arXiv:2412.16097}, Dec. 2024.

\bibitem{zhou2024novel}
X.~Zhou, T.~Fang, and Y.~Mao, ``A novel {Q}-stem connected architecture for beyond-diagonal reconfigurable intelligent surfaces,'' \emph{arXiv preprint arXiv:2411.18480}, Nov. 2024.

\bibitem{10403525}
H.~Li, Y.~Zhang, and B.~Clerckx, ``Channel estimation for beyond diagonal reconfigurable intelligent surfaces with group-connected architectures,'' in \emph{IEEE Int. Workshop Comput. Adv. Multi-Sensor Adaptive Process. (CAMSAP), Herradura, Costa Rica}, Dec. 2023, pp. 21--25.

\bibitem{10587164}
H.~Li, S.~Shen, Y.~Zhang, and B.~Clerckx, ``Channel estimation and beamforming for beyond diagonal reconfigurable intelligent surfaces,'' \emph{IEEE Trans. Signal Process.}, vol.~72, pp. 3318--3332, 2024.

\bibitem{de2024channel}
A.~L. de~Almeida, B.~Sokal, H.~Li, and B.~Clerckx, ``Channel estimation for beyond diagonal {RIS} via tensor decomposition,'' \emph{arXiv preprint arXiv:2407.20402}, Jul. 2024.

\bibitem{sokal2024decoupled}
B.~Sokal, A.~L. de~Almeida, H.~Li, B.~Clerckx \emph{et~al.}, ``A decoupled channel estimation method for beyond diagonal {RIS},'' \emph{arXiv preprint arXiv:2412.06683}, Dec. 2024.

\bibitem{ginige2024efficient}
N.~Ginige, A.~S. de~Sena, N.~H. Mahmood, N.~Rajatheva, and M.~Latva-aho, ``Efficient channel prediction for beyond diagonal {RIS}-assisted {MIMO} systems with channel aging,'' \emph{arXiv preprint arXiv:2411.17725}, Nov. 2024.

\bibitem{soleymani2024maximizing}
M.~Soleymani, I.~Santamaria, A.~Sezgin, and E.~Jorswieck, ``Maximizing spectral and energy efficiency in multi-user {MIMO} {OFDM} systems with {RIS} and hardware impairment,'' \emph{arXiv preprint arXiv:2401.11921}, Jan. 2024.

\bibitem{10787237}
M.~Samy, H.~Al-Hraishawi, A.~B.~M. Adam, S.~Chatzinotas, and B.~Otteresten, ``Beyond diagonal {RIS}-aided networks: Performance analysis and sectorization tradeoff,'' \emph{IEEE Open J. Commun. Soc.}, vol.~6, pp. 302--315, 2025.

\bibitem{10319662}
T.~Fang and Y.~Mao, ``A low-complexity beamforming design for beyond-diagonal {RIS} aided multi-user networks,'' \emph{IEEE Commun. Lett.}, vol.~28, no.~1, pp. 203--207, Jan. 2024.

\bibitem{10237233}
M.~Nerini and B.~Clerckx, ``Pareto frontier for the performance-complexity trade-off in beyond diagonal reconfigurable intelligent surfaces,'' \emph{IEEE Commun. Lett.}, vol.~27, no.~10, pp. 2842--2846, Oct. 2023.

\bibitem{10155675}
M.~Nerini, S.~Shen, and B.~Clerckx, ``Closed-form global optimization of beyond diagonal reconfigurable intelligent surfaces,'' \emph{IEEE Trans. Wirel. Commun.}, vol.~23, no.~2, pp. 1037--1051, Feb. 2024.

\bibitem{10364738}
Y.~Zhou, Y.~Liu, H.~Li, Q.~Wu, S.~Shen, and B.~Clerckx, ``Optimizing power consumption, energy efficiency, and sum-rate using beyond diagonal {RIS}-a unified approach,'' \emph{IEEE Trans. Wirel. Commun.}, vol.~23, no.~7, pp. 7423--7438, Jul. 2024.

\bibitem{wu2024optimization}
Z.~Wu and B.~Clerckx, ``Optimization of beyond diagonal {RIS}: A universal framework applicable to arbitrary architectures,'' \emph{arXiv preprint arXiv:2412.15965}, Dec. 2024.

\bibitem{10187688}
I.~Santamaria, M.~Soleymani, E.~Jorswieck, and J.~Gutierrez, ``{SNR} maximization in beyond diagonal {RIS}-assisted single and multiple antenna links,'' \emph{IEEE Signal Process. Lett.}, vol.~30, pp. 923--926, 2023.

\bibitem{10694491}
I.~Santamaria, M.~Soleymani, E.~Jorswieck, and J.~Gutierrez, ``{MIMO} capacity maximization with beyond-diagonal {RIS},'' in \emph{IEEE Workshop Signal Process. Adv. Wirel. Commun. (SPAWC), Lucca, Italy}, Sep. 2024, pp. 936--940.

\bibitem{bjornson2024capacity}
E.~Bj{\"o}rnson and {\"O}.~T. Demir, ``Capacity maximization for {MIMO} channels assisted by beyond-diagonal {RIS},'' \emph{arXiv preprint arXiv:2411.18298}, Nov. 2024.

\bibitem{10694582}
K.~D. Katsanos, P.~Di~Lorenzo, and G.~C. Alexandropoulos, ``The interference broadcast channel with reconfigurable intelligent surfaces: A cooperative sum-rate maximization approach,'' in \emph{IEEE Workshop Signal Process. Adv. Wirel. Commun. (SPAWC), Lucca, Italy}, Sep. 2024, pp. 551--555.

\bibitem{10839400}
X.~Zhou, T.~Fang, and Y.~Mao, ``Joint active and passive beamforming optimization for beyond diagonal {RIS}-aided multi-user communications,'' \emph{IEEE Commun. Lett.}, pp. 1--1, early access, 2025.

\bibitem{10755162}
M.~Soleymani, A.~Zappone, E.~Jorswieck, M.~D. Renzo, and I.~Santamaria, ``Rate region of {RIS}-aided {URLLC} broadcast channels: Diagonal versus beyond diagonal globally passive {RIS},'' \emph{IEEE Wirel. Commun. Lett.}, vol.~14, no.~2, pp. 320--324, Feb. 2025.

\bibitem{10694177}
M.~Soleymani, I.~Santamaria, E.~Jorswieck, M.~Di~Renzo, and J.~Gutierrez, ``Energy efficiency comparison of {RIS} architectures in {MISO} broadcast channels,'' in \emph{IEEE Workshop Signal Process. Adv. Wirel. Commun. (SPAWC), Lucca, Italy}, Sep. 2024, pp. 701--705.

\bibitem{10302331}
J.~Singh, S.~Srivastava, A.~K. Jagannatham, and L.~Hanzo, ``Joint transceiver and reconfigurable intelligent surface design for multiuser {mmWave} {MIMO} systems relying on non-diagonal phase shift matrices,'' \emph{IEEE Open J. Commun. Soc.}, vol.~4, pp. 2897--2912, 2023.

\bibitem{10571253}
S.~Sobhi-Givi, M.~Nouri, H.~Behroozi, and Z.~Ding, ``Joint {BS} and beyond diagonal {RIS} beamforming design with {DRL} methods for {mmWave} {6G} mobile communications,'' in \emph{IEEE Wirel. Commun. Netw. Conf. (WCNC), Dubai, United Arab Emirates}, Apr. 2024, pp. 1--6.

\bibitem{10817282}
W.~U. Khan, C.~K. Sheemar, Z.~Abdullah, E.~Lagunas, and S.~Chatzinotas, ``Beyond diagonal {IRS} assisted ultra massive {THz} systems: A low resolution approach,'' in \emph{IEEE Int. Symp. Pers. Indoor Mob. Radio Commun. (PIMRC), Valencia, Spain}, Sep. 2024, pp. 1--5.

\bibitem{10817342}
A.~Mahmood, T.~X. Vu, S.~Chatzinotas, and B.~Ottersten, ``Enhancing indoor and outdoor {THz} communications with beyond diagonal-{IRS}: Optimization and performance analysis,'' in \emph{IEEE 35th IEEE Int. Symp. Pers. Indoor Mob. Radio Commun. (PIMRC), Valencia, Spain}, Sep. 2024, pp. 1--6.

\bibitem{10333560}
A.~M. Huroon, Y.-C. Huang, and L.-C. Wang, ``Optimized transmission strategy for {UAV-RIS} {2.0} assisted communications using rate splitting multiple access,'' in \emph{IEEE Veh. Technol. Conf. (VTC2023-Fall), Hong Kong, Hong Kong}, Oct. 2023, pp. 1--6.

\bibitem{mahmood2023joint}
A.~Mahmood, T.~X. Vu, W.~U. Khan, S.~Chatzinotas, and B.~Ottersten, ``Joint computation and communication resource optimization for beyond diagonal {UAV-IRS} empowered {MEC} networks,'' \emph{arXiv preprint arXiv:2311.07199}, Nov. 2023.

\bibitem{10693852}
X.~Peng, Z.~Chen, J.~Ye, P.~Zhang, and L.~Huang, ``Beyond-diagonal {RIS} aided {DFRC} systems: A joint beamforming optimization design method,'' in \emph{IEEE/CIC Int. Conf. Commun. in China (ICCC Workshops), Hangzhou, China}, Aug. 2024, pp. 782--787.

\bibitem{10571002}
T.~D. Hua, M.~Mohammadi, H.~Q. Ngo, and M.~Matthaiou, ``Cell-free massive {MIMO} {SWIPT} with beyond diagonal reconfigurable intelligent surfaces,'' in \emph{IEEE Wirel. Commun. Netw. Conf. (WCNC), Dubai, United Arab Emirates}, Apr. 2024, pp. 1--6.

\bibitem{azarbahram2025beyond}
A.~Azarbahram, O.~L. L{\'o}pez, B.~Clerckx, M.~Di~Renzo, and M.~Latva-aho, ``Beyond diagonal reconfigurable intelligent surfaces for multi-carrier {RF} wireless power transfer,'' \emph{arXiv preprint arXiv:2501.01787}, Jan. 2025.

\bibitem{10288244}
M.~Soleymani, I.~Santamaria, E.~A. Jorswieck, and B.~Clerckx, ``Optimization of rate-splitting multiple access in beyond diagonal {RIS}-assisted {URLLC} systems,'' \emph{IEEE Trans. Wirel. Commun.}, vol.~23, no.~5, pp. 5063--5078, May 2024.

\bibitem{9814551}
T.~Fang, Y.~Mao, S.~Shen, Z.~Zhu, and B.~Clerckx, ``Fully connected reconfigurable intelligent surface aided rate-splitting multiple access for multi-user multi-antenna transmission,'' in \emph{IEEE Int. Conf. Commun. Workshops (ICC Workshops), Seoul, Korea}, May 2022, pp. 675--680.

\bibitem{10411856}
H.~Li, S.~Shen, and B.~Clerckx, ``Synergizing beyond diagonal reconfigurable intelligent surface and rate-splitting multiple access,'' \emph{IEEE Trans. Wirel. Commun.}, vol.~23, no.~8, pp. 8717--8729, Aug. 2024.

\bibitem{khisa2024gradient}
S.~Khisa, A.~Amhaz, M.~Elhattab, C.~Assi, and S.~Sharafeddine, ``Gradient-based meta learning for uplink {RSMA} with beyond diagonal {RIS},'' \emph{arXiv preprint arXiv:2410.17896}, Oct. 2024.

\bibitem{10742100}
Q.~Zhang, G.~Luo, Z.~Dong, F.~Sun, X.~Wang, and J.~Liu, ``Beyond-diagonal reconfigurable intelligent surface enhanced {NOMA} systems,'' \emph{IEEE Wirel. Commun. Lett.}, vol.~14, no.~1, pp. 118--122, Jan. 2025.

\bibitem{10623689}
O.~T. Demir and E.~Bjornson, ``Wideband channel capacity maximization with beyond diagonal {RIS} reflection matrices,'' \emph{IEEE Wirel. Commun. Lett.}, vol.~13, no.~10, pp. 2687--2691, Oct. 2024.

\bibitem{loli2024meta}
R.~C. Loli and B.~Clerckx, ``Meta-learning based optimization for large scale wireless systems,'' \emph{arXiv preprint arXiv:2407.01823}, Jul. 2024.

\bibitem{10834443}
C.~Zhang, W.~U. Khan, A.~K. Bashir, A.~K. Dutta, A.~U. Rehman, and M.~M.~A. Dabel, ``Sum rate maximization for {6G} beyond diagonal {RIS}-assisted multi-cell transportation systems,'' \emph{IEEE Trans. Intell. Transp. Syst.}, pp. 1--11, early access, 2025.

\bibitem{5626934}
T.~Kim, B.~Clerckx, D.~J. Love, and S.~J. Kim, ``Limited feedback beamforming systems for dual-polarized {MIMO} channels,'' \emph{IEEE Trans. Wirel. Commun.}, vol.~9, no.~11, pp. 3425--3439, Nov. 2010.

\bibitem{9264161}
A.~S. Rajasekaran, O.~Maraqa, H.~U. Sokun, H.~Yanikomeroglu, and S.~Al-Ahmadi, ``User clustering in {mmWave}-{NOMA} systems with user decoding capability constraints for b5g networks,'' \emph{IEEE Access}, vol.~8, pp. 209\,949--209\,963, 2020.

\bibitem{9569475}
O.~Maraqa, A.~S. Rajasekaran, H.~U. Sokun, S.~Al-Ahmadi, H.~Yanikomeroglu, and S.~M. Sait, ``Energy-efficient coverage enhancement of indoor {THz}-{MISO} systems: An {FD-NOMA} approach,'' in \emph{IEEE Int. Symp. Pers. Indoor Mob. Radio Commun. (PIMRC), Helsinki, Finland}, Sep. 2021, pp. 483--489.

\bibitem{10093902}
O.~Maraqa, S.~Al-Ahmadi, A.~S. Rajasekaran, H.~U. Sokun, H.~Yanikomeroglu, and S.~M. Sait, ``Energy-efficient optimization of multi-user {NOMA}-assisted cooperative {THz}-{SIMO} {MEC} systems,'' \emph{IEEE Trans. Commun.}, vol.~71, no.~6, pp. 3763--3779, 2023.

\bibitem{8828016}
L.~Zheng, M.~Lops, Y.~C. Eldar, and X.~Wang, ``Radar and communication coexistence: An overview,'' \emph{IEEE Signal Process. Mag.}, vol.~36, no.~5, pp. 85--99, Sep. 2019.

\bibitem{8214104}
T.~D. Ponnimbaduge~Perera, D.~N.~K. Jayakody, S.~K. Sharma, S.~Chatzinotas, and J.~Li, ``Simultaneous wireless information and power transfer ({SWIPT}): Recent advances and future challenges,'' \emph{IEEE Commun. Surv. Tutor.}, vol.~20, no.~1, pp. 264--302, 1st Quart. 2018.

\bibitem{9826826}
B.~S. Khan, S.~Jangsher, A.~Ahmed, and A.~Al-Dweik, ``{URLLC} and {eMBB} in {5G} industrial {IoT}: A survey,'' \emph{IEEE Open J. Commun. Soc.}, vol.~3, pp. 1134--1163, 2022.

\bibitem{10375270}
O.~Maraqa, S.~Aboagye, and T.~M.~N. Ngatched, ``Optical {STAR}-{RIS}-aided {VLC} systems: {RSMA} versus {NOMA},'' \emph{IEEE Open J. Commun. Soc.}, vol.~5, pp. 430--441, 2024.

\bibitem{9154358}
O.~Maraqa, A.~S. Rajasekaran, S.~Al-Ahmadi, H.~Yanikomeroglu, and S.~M. Sait, ``A survey of rate-optimal power domain {NOMA} with enabling technologies of future wireless networks,'' \emph{IEEE Commun. Surv. Tutor.}, vol.~22, no.~4, pp. 2192--2235, 4th Quart. 2020.

\bibitem{7763821}
V.~Niemela, J.~Haapola, M.~Hamalainen, and J.~Iinatti, ``An ultra wideband survey: Global regulations and impulse radio research based on standards,'' \emph{IEEE Commun. Surv. Tutor.}, vol.~19, no.~2, pp. 874--890, 2nd Quart. 2017.

\bibitem{9330752}
M.~Kordestani, A.~A. Safavi, and M.~Saif, ``Recent survey of large-scale systems: Architectures, controller strategies, and industrial applications,'' \emph{IEEE Syst. J.}, vol.~15, no.~4, pp. 5440--5453, Dec. 2021.

\bibitem{9656537}
K.~Wu, J.~A. Zhang, X.~Huang, and Y.~J. Guo, ``Frequency-hopping {MIMO} radar-based communications: An overview,'' \emph{IEEE Aerosp. Electron. Syst. Mag.}, vol.~37, no.~4, pp. 42--54, Apr. 2022.

\bibitem{9775078}
H.~Zhang, B.~Di, K.~Bian, Z.~Han, H.~V. Poor, and L.~Song, ``Toward ubiquitous sensing and localization with reconfigurable intelligent surfaces,'' \emph{Proc. IEEE}, vol. 110, no.~9, pp. 1401--1422, Sep. 2022.

\bibitem{gradoni2021smart}
G.~Gradoni, M.~Di~Renzo, A.~Diaz-Rubio, S.~Tretyakov, C.~Caloz, Z.~Peng, A.~Alu, G.~Lerosey, M.~Fink, V.~Galdi \emph{et~al.}, ``Smart radio environments,'' \emph{arXiv preprint arXiv:2111.08676}, Nov. 2021.

\bibitem{10183987}
O.~Maraqa and T.~M.~N. Ngatched, ``Optimized design of joint mirror array and liquid crystal-based {RIS}-aided {VLC} systems,'' \emph{IEEE Photonics J.}, vol.~15, no.~4, pp. 1--11, Aug. 2023.

\bibitem{M.2516-0}
\BIBentryALTinterwordspacing
{ITU-R, Report M.2516-0}, \emph{{Future Technology Trends of Terrestrial International Mobile Telecommunications Systems Towards 2030 and Beyond}}, Nov. 2022. [Online]. Available: \url{https://www.itu.int/pub/R-REP-M.2516-2022}
\BIBentrySTDinterwordspacing

\bibitem{liu2024sustainable}
R.~Liu, S.~Zheng, Q.~Wu, Y.~Jiang, N.~Zhang, Y.~Liu, M.~Di~Renzo \emph{et~al.}, ``Sustainable wireless networks via reconfigurable intelligent surfaces ({RISs}): Overview of the {ETSI} {ISG} {RIS},'' \emph{arXiv preprint arXiv:2406.05647}, Jun. 2024.

\bibitem{ETSI.GR.RIS.002}
\BIBentryALTinterwordspacing
{ETSI, Report ETSI.GR.RIS.002 V1.1.1}, \emph{{Reconfigurable Intelligent Surfaces ({RIS}); Technological Challenges, Architecture and Impact on Standardization}}, Aug. 2023. [Online]. Available: \url{https://www.etsi.org/deliver/etsi_gr/RIS/001_099/002/01.01.01_60/gr_ris002v010101p.pdf}
\BIBentrySTDinterwordspacing

\end{thebibliography}

\ifCLASSOPTIONcaptionsoff
  \newpage
\fi

\end{document}